\documentclass[
prd,
preprint,
superscriptaddress,
showpacs,
showkeys,
nofootinbib,
aps,
amsfonts,
amssymb,
]{revtex4-1}

\usepackage{color}
\usepackage{multirow} 
\usepackage{amsmath}
\usepackage{graphicx}
\usepackage{slashed}
\usepackage{epic}
\usepackage{eepic}
\usepackage{epsfig}
\usepackage{latexsym}
\usepackage{float}
\usepackage{amsfonts}
\usepackage{amssymb} 
\usepackage{color}
\usepackage{multirow,array}
\usepackage{tikz}
\usepackage[colorlinks, urlcolor=blue, citecolor=red, link
color=blue]{hyperref}

\usepackage[caption=false]{subfig}

\usepackage{geometry}
\newcolumntype{L}[1]{>{\raggedright\arraybackslash}p{#1}}
\newcolumntype{C}[1]{>{\centering\arraybackslash}p{#1}}
\newcolumntype{R}[1]{>{\raggedleft\arraybackslash}p{#1}}
                                                                          
\begin{document}

\preprint{IPPP/15/36, DCPT/15/72, HRI-RECAPP-2015-012}

\title{Searching for a Heavy Higgs boson in a Higgs-portal B-L Model}

\author{Shankha Banerjee}
\email{shankha@hri.res.in}
\affiliation{Regional Centre for Accelerator-based Particle Physics,
Harish-Chandra Research Institute, Chhatnag Road, Jhusi, Allahabad  211019, India.}

\author{Manimala Mitra}
\email{manimala@iisermohali.ac.in}
\affiliation{Indian Institute of Science Education and Research Mohali, Knowledge city, Sector 81, SAS Nagar, Manauli PO 140306.}

\author{Michael Spannowsky}
\email{michael.spannowsky@durham.ac.uk }
\affiliation{Institute for Particle Physics Phenomenology, Durham University, Durham DH1 3LE, United Kingdom.}
  
\begin{abstract}

We study the discovery prospects of a heavy neutral scalar arising from a $U(1)_{B-L}$ extension of the Standard Model (SM) during the 
Large Hadron Collider's high luminosity runs (HL-LHC). This heavy neutral scalar mixes with the SM Higgs boson through a Higgs portal 
and interacts with the SM particles with an interaction strength proportional to the sine of the mixing angle. The mixing between the 
two Higgs bosons is constrained by direct and indirect measurements. We choose an experimentally viable mixing angle and explore in 
detail the $ZZ$ and $WW$ decay modes of the heavy Higgs boson. For the $ZZ$ case, we focus on the cleanest $4\ell$ and $2\ell 2j$ final 
states and find that a heavy Higgs boson of mass smaller than 500 GeV can be discovered at the HL-LHC. For the $WW$ decay mode, we 
analyze the $\ell jj \slashed{E}_T$ signature. We implement novel background reduction techniques in order to tackle the huge 
background by performing both cut-based and multivariate analyses. However, large backgrounds render this channel challenging. 
We briefly discuss the discovery prospects of the heavy $Z'$-boson arising in this model.

\end{abstract}

\pacs{12.60.-i,12.60.Cn,12.60.Fr}
\keywords{B-L model, Singlet Extension, HL-LHC}

\maketitle

\section{Introduction}

The CMS and ATLAS collaborations at the Large Hadron Collider (LHC) have successfully discovered a new resonance~\cite{Aad:2012tfa,Chatrchyan:2012ufa} 
with a mass of $125$ GeV~\cite{Aad:2015zhl}, which has properties consistent with the Higgs boson predicted by the Standard Model (SM). 
The signal strengths of this boson in its various final states are in good agreement  with the SM expectations at 1$\sigma$. The nominal 
variations in certain production and decay modes could be due to some physics beyond the standard model (BSM) or could simply 
be due to insufficient statistics.

It is well known that the SM cannot be the final theory of nature. The successful explanation of the 
hierarchy problem requires some new physics (NP) near the TeV scale. In addition, the observation of small neutrino masses and their very particular mixing indicates the presence of physics beyond the standard model (BSM). There are a few  well motivated theories, such as supersymmetric extensions of the SM or theories with extra spatial dimensions, that cure the aforementioned limitations of the SM. However, 
neither ATLAS nor CMS have yet conclusively discovered any particle that serves as proof for BSM physics. Now, with the discovery of the Higgs boson, effects of new physics can be searched for in its coupling measurements ~\cite{Masso:2012eq,Corbett:2012ja,Falkowski:2013dza,Corbett:2013pja,Dumont:2013wma,Banerjee:2012xc,Gainer:2013rxa,
Corbett:2013hia,Elias-Miro:2013mua,Pomarol:2013zra,Einhorn:2013tja,Banerjee:2013apa,Willenbrock:2014bja,
Ellis:2014dva,Belusca-Maito:2014dpa,Gupta:2014rxa,Masso:2014xra,Biekoetter:2014jwa,Englert:2014cva,
Ellis:2014jta,Edezhath:2015lga,Gorbahn:2015gxa,Han:2004az,Ciuchini:2013pca,Blas:2013ana,Chen:2013kfa,
Alonso:2013hga,Englert:2014uua,Trott:2014dma,Falkowski:2014tna,Henning:2014wua,deBlas:2014mba,
Berthier:2015oma,Efrati:2015eaa}.

In this paper, we consider the simplest manifestation of a BSM extension through an extra singlet scalar. As a first 
step, we would like to see how the addition of just an additional neutral Higgs boson fares with the discovery prospects at 
the high-luminosity run at LHC (HL-LHC) with a final integrated luminosity of 3000 fb$^{-1}$.

The presence of a heavy Higgs-like neutral scalar is innate in various models, such as, the minimal supersymmetric standard model 
(MSSM), two Higgs doublet models (2HDMs), models with extra spatial dimensions, etc. However, the simplest among 
these models is the SM augmented with a gauge singlet. This can originate very naturally from a  $U(1)_{B-L}$ model with an extra 
$U(1)$ local gauge symmetry~\cite{Carlson:1986cu}, where $B$ and $L$ represents the baryon number and lepton number respectively. In particular, we focus on 
a TeV scale $B-L$ model, that can further be embedded in a TeV scale Left-Right symmetric 
model~\cite{Mohapatra:1974hk,Mohapatra:1974gc,Senjanovic:1975rk,Gu:2012in,Khachatryan:2014dka}. The $B-L$ symmetry group
is a part of a Grand Unified Theory (GUT) as described by a $SO(10)$ group~\cite{Buchmuller:1991ce}. Besides, the $B-L$ symmetry 
breaking scale is  related to the masses of the heavy right-handed Majorana neutrinos, which participate in  the celebrated 
\textit{seesaw} mechanism~\cite{Minkowski:1977sc,Mohapatra:1979ia,Yanagida:1980xy,Schechter:1980gr} and  generate   the light neutrino masses. 

Another important theoretical motivation of this model is 
that  the right handed neutrinos, that are an essential ingredient of this model participate in generating the baryon asymmetry of the universe via leptogenesis \cite{Fukugita:1986hr}. Hence, the  $B-L$ breaking scale is  strongly linked to  leptogenesis via sphaleron interactions that preserve $B-L$. It is important to note that in the 
$U(1)_{B-L}$ model, the symmetry breaking can take place at scales much lower than that of any GUT scale, e.g. the 
electroweak (EW) scale or TeV scale. Because the $B+L$ symmetry is broken due to sphaleron interactions, baryogenesis or leptogenesis 
cannot occur above the $B-L$ breaking scale. Hence, the $B-L$ breaking around the TeV scale naturally implies TeV scale baryogenesis. 

The presence of heavy neutrinos, a TeV scale extra neutral gauge boson and an additional heavy neutral Higgs,
makes the model phenomenologically rich, testable at the LHC as well as future $e^{+} e^{-}$ 
colliders~\cite{Gluza:2002vs,Emam:2007dy,Basso:2008iv,Basso:2009hf,Perez:2009mu,Huitu:2008gf,Basso:2010si,Pruna:2011me,Basso:2010yz,Englert:2011yb,Khoze:2013oga,Lopez:2013hsa,Basso:2013vla,Gluza:2015goa}. The Majorana nature of the 
heavy neutrinos can be probed for example through same-sign 
dileptonic signatures at the LHC~\cite{Keung:1983uu}. On the other hand, the extra gauge boson $Z'$ in this model interacts with SM leptons and 
quarks. Non-observation of an excess in dilepton and di-jet signatures by ATLAS and CMS have placed stringent constraints on the $Z'$ mass~\cite{CMS-PAS-EXO-12-023,Aad:2014cka,Khachatryan:2014fba,Khachatryan:2015sja,Chatrchyan:2012ku,Aad:2015osa}. 

In this work,  we examine in detail the discovery prospects of the second Higgs at the HL-LHC for a TeV scale $U(1)_{B-L}$ model. The
vacuum expectation value ($vev$) of the gauge singlet Higgs breaks the $U(1)_{B-L}$ symmetry and generates the masses of the
right handed neutrinos. We consider the $B-L$ breaking scale to be of the order of a few TeVs, for which the right handed neutrino 
masses can naturally be in 
the TeV range. The physical second Higgs state mixes with the SM Higgs boson with a mixing angle $\theta$, constrained by
electroweak precision measurements from LEP~\cite{Robens:2015gla,Lopez-Val:2014jva,Alcaraz:2006mx}, as well as from 
Higgs coupling measurements at LHC~\cite{Khachatryan:2014jba,ATLAS-CONF-2015-007}. The second Higgs is dominantly produced
by gluon fusion with subsequent decay into heavy particles. The largest branching ratios are into $W$, $Z$ and 
Higgs bosons. We discuss in detail the different channels through which the second Higgs state can be probed at the HL-LHC. 

Note that there are other possible $B-L$ extensions of the SM, that have been studied in Refs.~\cite{Feldman:2011ms}, Refs.~\cite{Heeck:2014zfa} and Refs.~\cite{FileviezPerez:2011kd}. 
In Refs.~\cite{Feldman:2011ms}, the $B-L$ gauge boson $Z^{\prime}$ acquires mass through the Stueckelberg mechanism \cite{Kors:2004dx,Kors:2005uz}. In this case, 
 the $B-L$  symmetry is unbroken, even after $Z^{\prime}$  acquires mass. Hence, the neutrinos in this model are necessarily 
of Dirac nature. To generate the mass of $Z^{\prime}$ via Stueckelberg mechanism, the presence of an axionic scalar is required. 
In addition to the $U(1)_{B-L}$, an additional $U(1)_X$ symmetry is imposed, that brings down the scale of the $Z^{\prime}$ around TeV. 
 As a second option \cite{FileviezPerez:2011kd}, the $B-L$ symmetry is broken spontaneously by an SM gauge singlet Higgs field and as a consequence the $Z^{\prime}$ acquires mass. 
However, due to non-trivial $B-L$ charge assignment of the gauge singlet Higgs, this scenario does not contain any Majorana mass term 
of the heavy right handed neutrinos. The light neutrinos in this model are again necessarily of Dirac nature. The collider signatures 
of these models are very different compared to the $B-L$ extension where the light and heavy neutrinos are of Majorana nature.

The paper is organised as follows: in section~\ref{model}, we review the basics of the  $U(1)_{B-L}$ model. 
We discuss the constraints on the heavy neutrino sector and the limits on $Z'$ in section~\ref{NZ'}. Following this, 
in section~\ref{mix}, we outline the different constraints on the mixing angle between the SM-like Higgs and the second Higgs state. 
We study in detail the collider signatures of the heavy Higgs in section~\ref{col}. We briefly discuss non-standard 
production of the heavy Higgs in section~\ref{non-standard-prod}. Decay of the heavy Higgs to a pair of heavy neutrinos is discussed 
in section~\ref{decay-neutrino}. Eventually we offer conclusions in section~\ref{conclu}.

\section{ Brief Review of the $U(1)_{B-L}$ model \label{model}}

The $U(1)_{B-L}$ model is one of the simplest extensions of the 
SM~\cite{Marshak:1979fm,Mohapatra:1980qe,Wetterich:1981bx,Masiero:1982fi,Mohapatra:1982xz,Buchmuller:1991ce}. In addition to the symmetry group of the SM, it has an additional  
$U(1)$ gauge symmetry, that is identified as $B-L$ symmetry. The full group structure of this model is therefore 
\begin{equation}
 SU(3)_C\times SU(2)_L \times U(1)_Y \times U(1)_{B-L},
\end{equation}
where $U(1)_{\rm{B-L}}$ represents the additional gauge symmetry. The Lagrangian of this model is as follows: 
\begin{equation}
 \mathcal{L}=\mathcal{L}_s+\mathcal{L}_{YM}+\mathcal{L}_f+\mathcal{L}_Y,
\end{equation}
where $\mathcal{L}_s,\mathcal{L}_{YM},\mathcal{L}_f$ and $\mathcal{L}_Y$ are the scalar, Yang-Mills, fermion and Yukawa terms
respectively. The different terms in the Lagrangian are explained in detail in Refs.~\cite{Emam:2007dy,Basso:2008iv,Pruna:2011me}.
The Yang-Mills Lagrangian can be expressed as
\begin{equation}
 \mathcal{L}_{YM}=-\frac{1}{4}G_{\mu\nu}^a G^{a,\mu\nu}-\frac{1}{4}W_{\mu\nu}^b W^{b,\mu\nu}-\frac{1}{4}F_{\mu\nu}F^{\mu\nu}-\frac{1}{4}F'_{\mu\nu}F'^{\mu\nu},
\end{equation}
where the first three terms represent the kinetic terms of the $SU(3)_C,SU(2)_L$ and $U(1)_Y$ gauge groups respectively. $a,b$ 
are the colour and $SU(2)$ indices respectively. The fourth term is the kinetic term for the $U(1)_{B-L}$ gauge group and is 
represented by
\begin{equation}
 F'_{\mu\nu}=\partial_{\mu}B'_{\nu}-\partial_{\nu}B'_{\mu},
\end{equation}
where $B'$ is the $U(1)_{B-L}$ field strength.

In addition to the standard particle contents of SM, the fermion sector of this model has  three   right-handed neutrinos  
$N_R$, that are singlets under SM gauge group. This is required for  anomaly cancellation  and 
these right handed neutrinos generate Majorana 
masses of the light neutrinos through the \textit{seesaw} mechanism, as discussed in section~\ref{NZ'}. Analogous to the 
SM, the covariant derivative for this model is defined as
\begin{equation}
 D_{\mu}=\partial_{\mu}+i g_s t^a G_{\mu}^a + i g T^b W_{\mu}^b + i g_1 Y B_{\mu} + i g' Y_{B-L} B'_{\mu},
\end{equation}
where $g_s,g,g_1$ and $g'$ are  the $SU(3)_C,SU(2)_L,U(1)_Y$ and $U(1)_{B-L}$ couplings with $t^a,T^b,Y$ and $Y_{B-L}$ 
being their respective group generators. In the present study, we explicitly assume that there is no direct mixing between the two $U(1)$ fields $B$ and $B'$. This corresponds to the minimal version of the $B-L$ model. The fermion sector of the Lagrangian is expressed by
\begin{align}
 \mathcal{L}_f=\sum_{i=1,2,3} (i \, \overline{(Q_{L})_i} \gamma^{\mu} D_{\mu} (Q_{L})_i + i \, \overline{(u_{R})_i}\gamma^{\mu} D_{\mu} (u_{R})_i
 + i \, \overline{(d_{R})_i}\gamma^{\mu} D_{\mu} (d_{R})_i \nonumber \\ + i \, \overline{(L_{L})_i}\gamma^{\mu} D_{\mu} (L_{L})_i
 + i \, \overline{(e_{R})_i}\gamma^{\mu} D_{\mu} (e_{R})_i + i \, \overline{(N_{R})_i}\gamma^{\mu} D_{\mu} (N_{R})_i),
\end{align}
where the electromagnetic charges on the fields are the same as the SM ones and the $B-L$ charges are $Y_{B-L}^{quarks}=\frac{1}{3}$ 
and $Y_{B-L}^{leptons}=-1$.

In order to break the $B-L$ gauge symmetry and to generate the mass of the additional $Z'$ boson corresponding to this extra gauge symmetry, 
one needs to introduce  an 
extra complex  Higgs field $\chi$.  
The field $\chi$ is  a singlet under the SM gauge group with non-zero $B-L$ charge $Y_{B-L}^{\chi}=+2$. The $B-L$ symmetry is broken spontaneously by the {\it vev} of the Higgs field  $\chi$. The SM Higgs field is 
neutral under the $B-L$  gauge group, hence it has $Y_{B-L}^{H}=0$. This particular choice preserves gauge invariance.

The most general and renormalisable scalar Lagrangian of this model can be expressed as
\begin{eqnarray}
\mathcal{L}_s=(D^{\mu}H)^{\dagger}D_{\mu}H+(D^{\mu}\chi)^{\dagger}(D_{\mu}\chi)-V(\chi, H),
\end{eqnarray}
where the scalar potential $V(\chi,H)$ has the following form, 
\begin{eqnarray}
V(\chi, H)=M_H^2 H^{\dagger}H+ m^2_{\chi} |\chi|^2 + \lambda_1 (H^{\dagger} H)^2+ \lambda_2 |\chi|^4 + \lambda_3 
(H^{\dagger} H)|\chi|^2.
\label{scpot}
\end{eqnarray}

To complete the discussion on the Lagrangian, we write down the Yukawa term, which in addition to the SM terms has interactions involving the right-handed neutrinos $N_R$,
\begin{align}
 \mathcal{L}_Y=&-y^d_{ij} \overline{(Q_{L})_i} (d_{R})_j H - y^u_{ij} \overline{(Q_{L})_i} (u_{R})_j \widetilde{H} 
 - y^e_{ij} \overline{(L_{L})_i} (e_{R})_j H \nonumber \\ &- y^{\nu}_{ij} \overline{(L_{L})_i} (N_{R})_j \widetilde{H}
 - y^M_{ij} \overline{(N_{R})_i^c} (N_{R})_j \chi + h.c., 
\end{align}
where $\widetilde{H}=i\sigma^2 H^*$ and $i,j$ runs from $1$-$3$. The $vev$ of the $\chi$ field breaks the $B-L$ symmetry and generates the Majorana masses of the right handed neutrinos $N_R$ where  $M_{N_R} = y^M v^{\prime}$.  On the other hand, the  Dirac  masses for the light neutrinos are governed by the Yukawa couplings $y^{\nu}$s.

Next, we turn our attention to spontaneous electroweak symmetry breaking (SSB) in this model. Further details are discussed in 
Ref.~\cite{Pruna:2011me}. In order for the potential to be bounded from below, the couplings $\lambda_{1,2,3}$ should be related as
\begin{align}
 4 \lambda_1 \lambda_2 - \lambda_3^2 > 0, \nonumber \\
 \lambda_{1,2} > 0.
\end{align}
In order to minimise the potential, one requires the above two conditions to hold. 
 
We denote the $vev$s of $H$ and $\chi$ by 
$v$ and $v'$ respectively. On minimising the potential $V(\chi,H)$ with respect to both $vev$s, one obtains~\cite{Emam:2007dy,Basso:2008iv,Pruna:2011me}, 
\begin{eqnarray}
v^2=\frac{4 \lambda_2 M_H^2-2 \lambda_3 M^2_{\chi}  }{\lambda^2_3-4 \lambda_1 \lambda_2}, ~~~ {v^{\prime}}^2=\frac{4 \lambda_1 M_{\chi}^2-2 \lambda_3 M_H^2}{\lambda^2_3-4 \lambda_1 \lambda_2}.
\end{eqnarray}

Since the $B-L$ breaking scale is higher than the electroweak symmetry 
breaking (EWSB) scale, we have  $v'> v$. 

$\chi$ mixes with $H$ due to the $\lambda_3$-term as shown in Eq.~\ref{scpot}. The mass matrix between the two Higgs bosons in the basis 
$(H, \chi)$ is given by
\begin{eqnarray}
{\cal M}(H, \chi) = 2 \left(\begin{array}{cc}
\lambda^2_1 v^2 & \lambda_3 v v'/2 \\ 
\lambda_3 v v'/2 & \lambda_2 {v'}^2
\end{array}\right).
\label{eq:massmatrix}
\end{eqnarray}
Next, we define the mass eigenstates as $(H_1, H_2)$ which are related to the $(H, \chi)$ basis by
\begin{equation}
\left(\begin{array}{c}
H_1 \\
H_2 
\end{array}\right)= \left(\begin{array}{cc}
\cos \theta & -\sin \theta  \\
\sin \theta & \cos \theta
\end{array}\right)
\left(\begin{array}{c}
H \\
\chi 
\end{array}\right),
\label{mixingh}
\end{equation}
where the mixing angle $\theta$ ($-\frac{\pi}{2} < \theta < \frac{\pi}{2}$) satisfies 
\begin{equation}
\tan 2\theta= \frac{ \lambda_3 v' v}{(\lambda_2 v'^2 -\lambda_1 v^2)}.
\label{theta}
\end{equation}
The masses of the physical Higgs bosons, $H_1$ and $H_2$ are 
\begin{eqnarray}
M^2_{H_1}=\lambda_1 v^2+ \lambda_2 v'^2- \sqrt{(\lambda_1 v^2- \lambda_2 v'^2)^2+\lambda^2_3 v'^2 v^2},  \nonumber \\
M^2_{H_2}=\lambda_1 v^2+ \lambda_2 v'^2+ \sqrt{(\lambda_1 v^2- \lambda_2 v'^2)^2+\lambda^2_3 v'^2 v^2}.
\label{mass} 
\end{eqnarray}
After imposing SSB in $\mathcal{L}_s$, one obtains the mass spectrum of the gauge bosons
\begin{align}
 M_{\gamma} &=0, \nonumber \\
 M_{W^{\pm}} &= \frac{1}{2} v g, \nonumber \\
 M_Z &= \frac{v}{2}\sqrt{g^2 + g_1^2}, \nonumber \\
 M_{Z'} &=  2 v' g'.
\end{align}

Note that,  among the two Higgs masses, one chooses $m^2_{H_1}< m^2_{H_2}$, \textit{i.e.}, $H_1$ is chosen to be the 
lightest state. In our subsequent discussion we will consider the case, where $H_1$ is SM-like with a mass around $125$ GeV. 
The other Higgs state, $H_2$ is heavy and we allow its mass to vary in the range 250-900 GeV for our phenomenological studies. 
The interactions of $H_1$ and $H_2$ with the fermions and gauge bosons are expressed in terms of the mixing parameter $\theta$ in 
the following manner 
\begin{eqnarray}
H_1 f \bar{f} : - \frac{e M_f \cos \theta}{2 M_W},  \hspace*{0.5cm} H_2 f \bar{f} : - \frac{e M_f \sin \theta}{2 M_W},  \nonumber \\
H_1 W^{+} W^{-} : \frac{M_W e \cos \theta}{s_w},  \hspace*{0.5cm} H_2 W^{+} W^{-} : \frac{M_W e \sin \theta}{s_w}, \nonumber \\
H_1 Z Z : \frac{M_W e \cos \theta}{c^2_w s_w},  \hspace*{0.5cm} H_2 Z Z :\frac{M_W e \sin \theta}{c^2_w s_w}, \nonumber \\
H_1 Z^{\prime} Z^{\prime} : -8 \sin \theta {g^{\prime}}^2 v^{\prime},  \hspace*{0.5cm} H_2 Z^{\prime} Z^{\prime} : -8 \cos \theta {g^{\prime}}^2 v^{\prime}.
\label{int}
\end{eqnarray}
The scalar self-interactions are given by
\begin{align}
H_1 H_1 H_1 : -3\frac{1}{e}(4\cos^3{\theta} \sin{\theta_w} M_W \lambda_1 - 2 \sin^3{\theta} e \lambda_2 v' -& \nonumber \\
              \cos^2{\theta} \sin{\theta} e \lambda_3 v' + 2 \sin{\theta_w} \sin^2{\theta} \cos{\theta} M_W \lambda_3), \nonumber \\
H_2 H_1 H_1 : -\frac{1}{e}(12\cos^2{\theta}\sin{\theta_w}\sin{\theta} M_W \lambda_1 + 6 \sin^2{\theta} \cos{\theta} e \lambda_2 v' +& \nonumber \\
              (1-3\sin^2{\theta})\cos{\theta}e \lambda_3 v' - 2(2-3\sin^2{\theta}) \sin{\theta_w} \sin{\theta} M_W \lambda_3).
\label{selfc}
\end{align}
In the above expressions, $f$ denotes the SM fermions. We refer the readers to Refs.~\cite{Emam:2007dy,Basso:2008iv,Pruna:2011me} for a 
detailed description of the other interaction terms, arising from this model. Since, there are no extra coloured or 
electromagnetically charged states in this model that can alter the loop functions, we calculate the effective vertices 
$ggH_{1,2}$, $\gamma \gamma H_{1,2}$, $Z \gamma H_{1,2}$ and $Z' \gamma  H_{1,2}$ following the standard loop functions relevant for the SM~(see \cite{Djouadi:2005gi} and references therein). 

\section{Constraints on Heavy Neutrinos and $Z'$ \label{NZ'}} 
Before proceeding to discuss the phenomenological aspects of the heavy Higgs, we briefly discuss the constraints and limits on the
various parameters arising from the heavy neutrinos $N_R$ and $Z'$.
\subsection{Constraints on Heavy Neutrinos}
As we discussed in the previous section, the model consists of three right-handed neutrinos $N_R$, required for anomaly cancellation
in the theory. The low-scale breaking of the $B-L$ gauge symmetry implies the right handed neutrinos to be of the order of a few hundred 
GeVs to a few TeVs. In this work, we consider the heavy right handed neutrinos of TeV scale, that naturally emerge from low scale  $B-L$ breaking, without any unnatural tuning of the Yukawas. This scale will be accessible in the coming runs of the LHC and possibly also at future lepton and hadron collider 
experiments. The right handed neutrinos generate the masses for the light neutrinos via the  \textit{seesaw} mechanism
\begin{eqnarray}
\mathcal{M}_{\nu}=-M^T_D M^{-1}_R M_D.
\label{seesaw}
\end{eqnarray}
In the above,  $M_D$ is the Dirac mass matrix of light neutrinos, whereas  $M_R$ is the  Majorana mass matrix of the heavy neutrinos. By demanding $M_{R} \sim $ TeV and 
$\mathcal{M}_{\nu} \sim $ eV, one is able to constrain the active-sterile neutrino mixing $M_D/M_R \sim 10^{-6}$. In addition to the
constraints from light neutrino masses, the active-sterile mixing can also be constrained from other experimental searches, e.g. the neutrino-less double beta decay ($0\nu 2\beta$), $\beta$-decay, peak searches and kink searches \cite{Atre:2009rg,Gluza:2002vs,Gluza:2015goa}. 
The heavy Majorana neutrino, that mix with the active light neutrino $\nu$ with a mixing angle $\mathcal{\theta_{\nu}}$, participate in  the 
$0\nu 2\beta$-decay, where the amplitude is expressed as 
\begin{eqnarray}
\mathcal{A}_{N} \sim G^2_F \frac{\mathcal{\theta_{\nu}}^2}{M_R}.
\label{0nu2beta}
\end{eqnarray}
The non-observation of any positive signal in this lepton number violating process constrains the active-sterile mixing 
to $\theta_{\nu}^2 < 10^{-5}$, for a heavy neutrino mass $M_R \sim 500$ GeV \cite{Mitra:2011qr}.  A complete discussion on the different 
bounds on the active-sterile neutrino mixing can be found in \cite{Atre:2009rg,Deppisch:2015qwa}. In addition, the collider signatures of the heavy 
neutrinos at LHC has been discussed in details in Ref.~\cite{Perez:2009mu}. A detailed discussion of the like sign dilepton signature from 
right handed neutrino decay has been studied in Ref.~\cite{Perez:2009mu}.

\subsection{Limits on $Z'$ \label{zp}}
\begin{figure}
\centering
\subfloat[]{\includegraphics[width=7.0cm,height=7.0cm]{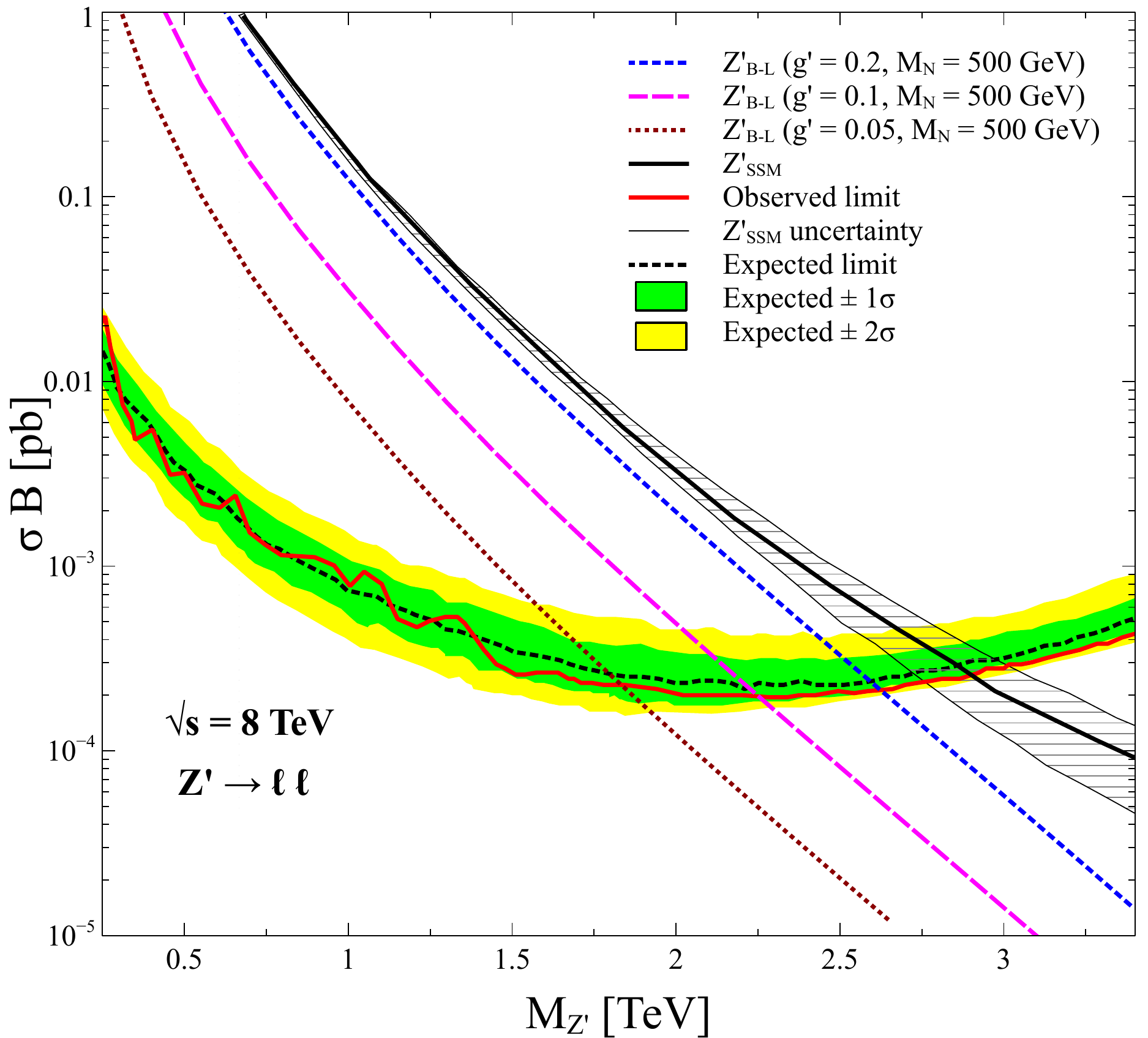}\label{fig:zp}}~~
\subfloat[]{\includegraphics[width=7.0cm,height=7.0cm]{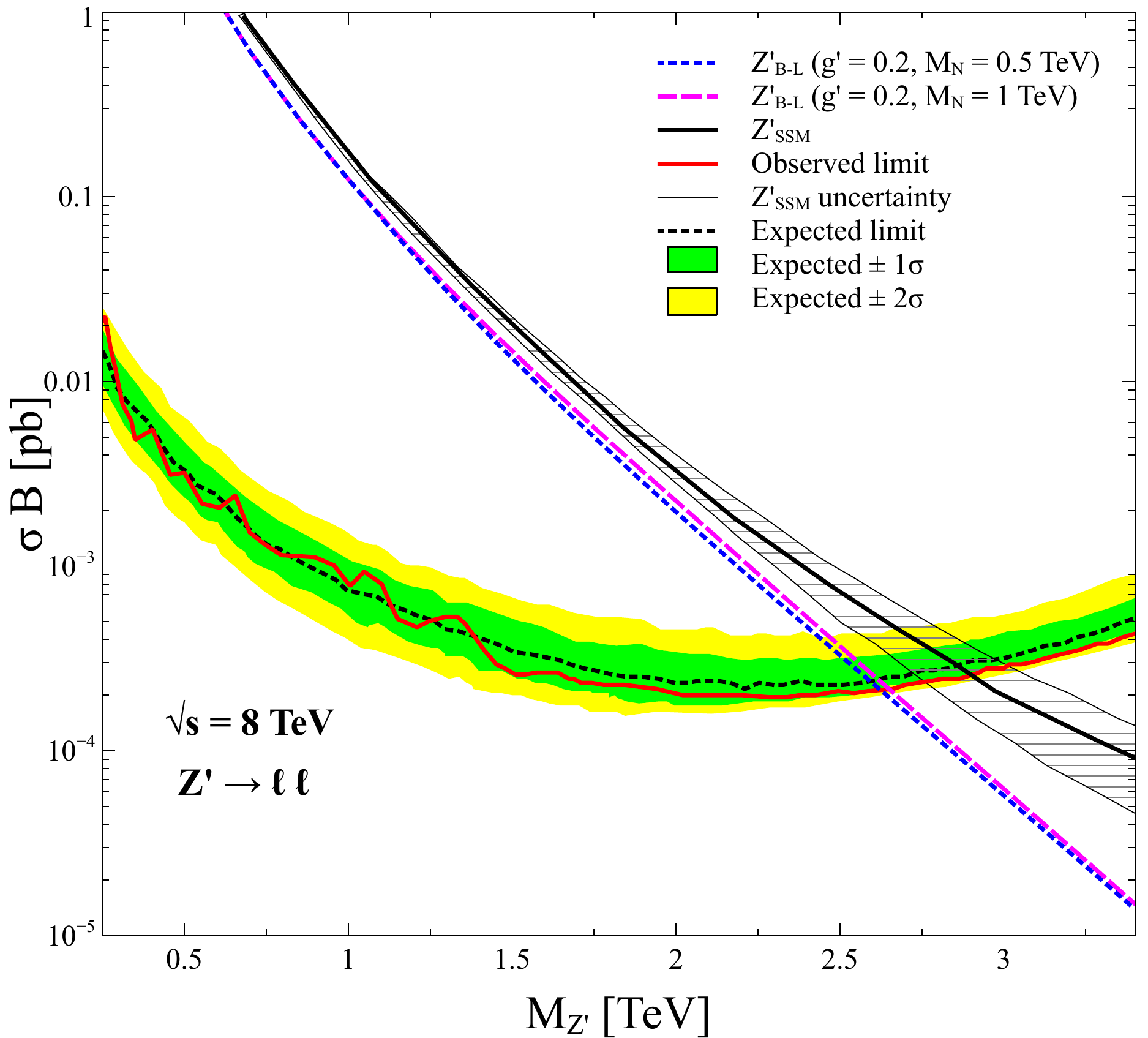}\label{fig:zp1}}
\caption{The comparison between the limits from ATLAS di-lepton search~\cite{Aad:2014cka} with the $B-L$ predictions with 
(a) $M_N=500$ GeV and $g'=0.05$ (brown fine dotted), $g'=0.1$ (magenta dashed) and $g'=0.2$ (blue dotted) and (b) $g'=0.2$ and
$M_N=500$ GeV (blue dotted) and $M_N=1$ TeV (magenta dashed). } 
\label{fig:zp_limits}
\end{figure}

The $B-L$ model has an additional gauge boson $Z'$ of mass $M_Z'=2 v' g_1'$. $Z'$ interacts 
with the leptons, quarks, heavy neutrinos and light neutrinos with interaction strengths proportional to the  $B-L$ gauge coupling 
$g^{\prime}$. The $Z'$ boson can in principle be detected by observing di-leptonic and di-jet signals at colliders. The presence of a
sequential SM-like (SSM) $Z'$ gauge boson has been severely constrained by direct searches at colliders, as well as by indirect searches. The ratio of $Z'$ mass to its coupling is constrained from 
indirect searches to be around~\cite{Carena:2004xs,Cacciapaglia:2006pk,Basso:2008iv,Heeck:2014zfa}
\begin{equation}
\frac{M_{Z'}}{g^{\prime}} \ge 6.9\,  \rm{TeV}. 
\label{indirect}
\end{equation}
Several studies have been carried out by ATLAS and CMS in di-leptonic and di-jet channels to search for this elusive heavy gauge 
boson~\cite{CMS-PAS-EXO-12-023,Chatrchyan:2012ku,Khachatryan:2014fba,Aad:2014cka,Khachatryan:2015sja,Aad:2015osa}.
The $Z'$ can decay to a boosted $t\bar{t}$ pair which provides sensitivity in semi-leptonic or fully hadronic top decays~\cite{Aad:2012raa, Chatrchyan:2012ku}. The cross-section times branching ratio ($\sigma \times B$) has been constrained to be less than 
1-2 pb~\cite{Chatrchyan:2012ku} for a $Z'$ with a width to mass ratio between $\Gamma_{Z'}/M_{Z'}=1\%$ and $\Gamma_{Z'}/M_{Z'}=10\%$. Note that, here and in Table.~\ref{cvalue},  we quote the most conservative  limits of the cross-sections, where for other different masses the cross-sections are  even more stringent. The recent combined analysis by CMS for the di-electron and di-muon mass spectra has further constrained 
the ratio ($R$) of cross-section times branching ratio of a narrow resonance~\cite{Khachatryan:2014fba}. 
In addition, the di-leptonic search by ATLAS has also constrained the sequential $Z'$ \cite{Aad:2014cka}. The other searches correspond 
to 
\begin{enumerate}
 \item The search for a di-jet resonance by CMS~\cite{Khachatryan:2015sja} that constrains $\sigma \times B \times A < 0.2-0.3$ pb 
($A$ being the acceptance for the kinematic requirements) and $M_{Z'_{SSM}} < 1.70$ TeV
\item ATLAS search for $\tau^{+} \tau^{-}$ pair \cite{Aad:2015osa}
\item CMS search for heavy resonance into $b \bar{b}$ pairs that bounds $M_{Z'_{SSM}} $ \cite{CMS-PAS-EXO-12-023}.
\end{enumerate}
In addition, we also show the limits applicable for a $B-L$ model by comparing the limits from the 8 TeV run of the ATLAS di-lepton 
search~\cite{Aad:2014cka}, in Fig.~\ref{fig:zp_limits}. We consider few benchmark values for the free parameter $g^{\prime}$ and also for 
the mass of the heavy neutrino. Note that the production cross-sections of $Z'$ in the $B-L$ model have been computed at leading 
order (LO). The bounds on $M_{Z'}$ from this model are summarised in Table~\ref{dlsearch}. In Fig.~\ref{fig:zp}, we find that by varying 
the coupling parameter $g'$, the bound on $M_{Z'}$ changes considerably by a few 100 GeV, whereas in Fig.~\ref{fig:zp1}, we find that 
by varying the masses of the heavy neutrinos, the bounds shift by $\mathcal{O}$(10) GeV \footnote{A recent study~\cite{Mandal:2015vfa} showed 
the importance of exclusion plots as functions of both masses and couplings.}. Finally, we give an estimate of how the signal fares 
with respect to the SM backgrounds by studying the di-lepton final state using a basic set of trigger cuts on the transverse momentum ($p_T$), pseudo-rapidity ($\eta$) and 
isolation in the pseudo-rapidity-azimuthal angle plane ($\Delta R$), i.e. $p_{T,\ell}>10$ GeV, $|\eta_{\ell}| < 2.5$ and 
$\Delta R_{\ell \ell} > 0.2$. For the 14 TeV run, the benchmark $M_{Z'}=3$ TeV and $g'=0.2$ yields the LO cross-section to be around 1.4 fb. Whereas, for the 
background the LO cross-section is around 1900 pb, several orders of magnitude larger than the signal cross-section. However, by imposing 
a simple invariant mass cut on the dilepton system, $2900 ~ \textrm{GeV} ~ < M_{\ell\ell} < ~ 3100 ~ \textrm{GeV}$, one sees a dramatic reduction in the SM 
background, which amounts to $\sim 0.01$ fb. The signal, however reduces by a nominal amount to $\sim$ 1.3 fb. Hence, a massive $Z'$ boson has a significant discovery potential during the 14 TeV LHC runs.


\begin{table}
\centering
\begin{tabular}{|c|c|c|}
\hline
Searches  & Constraints  & $M_{Z'}(SSM)$ \\ \hline \hline
 Boosted $ t \bar{t}$ \cite{Chatrchyan:2012ku} & $\sigma \times B \le 1-2 \, \rm{pb}$ & - \\ \hline
di-lepton-CMS \cite{Khachatryan:2014fba} & $R < 7 \times 10^{-6}$ & 2.90 TeV \\ \hline
di-lepton-ATLAS \cite{Aad:2014cka} & $\sigma \times B \le 4 \times 10^{-2} \, \rm{pb}$  & $2.90$ TeV \\ \hline
di-jet-ATLAS \cite{Khachatryan:2015sja} & $\sigma \times B \times A \le 0.2-0.3 \, \rm{pb}$ & $1.70$ TeV \\ \hline 
$\tau^{+} \tau^{-}$-ATLAS \cite{Aad:2015osa} & $\sigma \times B \le 0.1 \, \rm{pb}$ & $1.90$ TeV \\ \hline
$b\bar{b}$-CMS \cite{CMS-PAS-EXO-12-023} & -  & $1.20-1.68$ TeV \\ \hline
\end{tabular}
\caption{The recent bounds on $Z'$ production from di-lepton, di-jet and other analyses.}
\label{cvalue}
 \end{table}

\begin{table}
\centering
\begin{tabular}{|c|c|c|}
\hline
$M_N$ (TeV) &  $g^{\prime}$  & $M_{Z'}(B-L)$ (TeV) \\ \hline \hline
0.5 &  0.2 & 2.62 \\ 
1.0 &  0.2 & 2.65 \\ \hline
0.5 &  0.1 & 2.25 \\ \hline
0.5 & 0.05 & 1.83 \\ \hline
\end{tabular}
\caption{The  bounds on $M_{Z'}$ derived from the ATLAS di-lepton search~\cite{Aad:2014cka} relevant for a  $U(1)_{B-L}$ model.}
\label{dlsearch}
 \end{table}

\section{Constraints on Higgs Mixing \label{mix}}

As discussed in the previous section, the light Higgs, $H_1$ (or the SM-like Higgs) and the heavy Higgs, $H_2$ mix with an angle $\theta$. 
Hence, their couplings to the other particles in the model are scaled accordingly. Before discussing the phenomenological aspects of the 
searches, we impose bounds on the mixing parameter from the available experimental results. There are further theoretical bounds on this
parameter which we discuss below.

\begin{itemize}
 \item \textbf{Experimental bounds :} Recent searches from CMS~\cite{Khachatryan:2014jba} and ATLAS~\cite{ATLAS-CONF-2015-007} have already 
 put bounds on a large class of BSM models. We work in the so-called $\kappa$ framework, where the coupling deviations of the SM-like Higgs are parametrized in terms of simple rescalings. The Higgs coupling to two fermions $g_{H_1 ff}$ and 
 two weak bosons $g_{H_1 VV}$ are defined as~\cite{Heinemeyer:2013tqa}, 
\begin{equation}
g_{H_1 ff}=\kappa_f.g^{\rm{SM}}_{Hff} \, \, \, \, \, \textrm{and} \, \, \, \, \, \,  \, g_{H_1 VV}=\kappa_V.g^{\rm{SM}}_{HVV}, 
\label{kappaframe}
\end{equation}
where $\kappa_f$ and $\kappa_V$ are the coupling modifiers and are equal to unity in SM. We quote the $95 \%$ CL intervals on the various 
$\kappa$ parameters from CMS (Fig. 12 in Ref.~\cite{Khachatryan:2014jba}) and ATLAS (Fig. 15 in Ref.~\cite{ATLAS-CONF-2015-007})in 
Table~\ref{kappa-values}.  Here, the experimental collaborations have assumed that the loop level couplings like 
$H_1gg$, $H_1 \gamma \gamma$ and $H_1 Z\gamma$ can be parametrized in terms of the tree level couplings and that no new loop particles 
are involved. They also assume that the invisible branching ratio of $H_1$ is zero. These assumptions agree with the model under 
consideration. We consider the mass of heavy neutrinos to be in the TeV scale, for which the SM like Higgs decay to heavy neutrinos is small. In addition, since the light and heavy sterile mixing is small $\sim 10^{-6}$, hence, the SM like Higgs decaying to two light neutrinos is negligible. In this particular model, the couplings 
$\kappa_t=\kappa_b=\kappa_W=\kappa_Z=\kappa_{\tau}=\cos{\theta}$ and can have a 
maximum value of unity. Thus, all the major production cross-sections for $H_1$, e.g. $ggF$, $VBF$, $VH$ and $t\bar{t}H$ scale
as $\cos^2{\theta}$.
\begin{table}[H]
\begin{tabular}{ccccc}
\hline
$\kappa_W$  & $\kappa_Z$ & $\kappa_t$ & $\kappa_b$ & $\kappa_{\tau}$ \\  \hline 
& & CMS & & \\ \hline
$[0.66,1.24]$ & $[0.69,1.37]$ & $[0.51,1.22]$ &  $[0.07,1.46]$ & $[0.47,1.25]$ \\ \hline
& & ATLAS & &  \\ \hline
$[0.63,1.19]$ & $[-1.20,-0.67] \bigcup [0.67,1.26]$ & $[0.59,1.39]$ & $[-1.29,1.31]$ & $[-1.46,-0.61] \bigcup [0.62.1.47]$ \\ \hline
\end{tabular}
\caption{The 95$\%$ CL ranges on various signal strength modifiers, $\kappa$, as reported by CMS~\cite{Khachatryan:2014jba} 
and ATLAS~\cite{ATLAS-CONF-2015-007}.}
\label{kappa-values}
 \end{table}

Using the ranges in Table~\ref{kappa-values}, we obtain the scale factor of the heavy Higgs, $H_2$ as
$\sin^2{\theta} < 0.31 (0.33)$ for CMS (ATLAS) at 95\% CL.

It is however important to note that the bounds on $\sin{\theta}$ from the coupling measurements of the SM-like 
Higgs are possibly the most desired and robust ones. These bounds are independent of the mass of the heavy higgs and will probably 
get more stringent with more integrated luminosity. As an example, in Ref.~\cite{Peskin:2012we}, the $H \to WW^*$ measurement is shown to constrain 
$\sin{\theta} \sim 0.36$ from the projected study of LHC at 14 TeV with $\int{\mathcal{L} dt} = $300 fb$^{-1}$. The same study also 
projects a smaller $\sin{\theta} \sim 0.25$ at the ILC, running at 250 GeV with an integrated luminosity of 250 fb$^{-1}$. The ILC 
runs with greater centre-of-mass energies and higher integrated luminosities are expected to constrain $\sin{\theta}$ to even smaller values.
In this analysis, we assume $\sin{\theta}=0.2$ which is in sync with the projected study at LHC 14 with 300 fb$^{-1}$.

\item \textbf{Theoretical bounds :} 
\begin{itemize}
 \item \textbf{Constraints from $\boldsymbol{M_W}$ :} One of the strongest constraints on the the mixing angle, $\sin{\theta}$, comes from the
 one-loop correction to the $W$-boson mass, $M_W$, which is required to agree within $2\sigma$ of its experimental 
value, \textit{i.e.}, $M_W=80.385\pm 0.015$ GeV~\cite{Alcaraz:2006mx,Aaltonen:2012bp,D0:2013jba}. This has recently been studied in the context 
of this model in Refs.~\cite{Lopez-Val:2014jva,Robens:2015gla,Martin-Lozano:2015dja}. It has also been shown in Ref.~\cite{Robens:2015gla}, that in the
high mass region, the constraints from the one-loop correction to $M_W$ are stronger than the ones obtained from $S,T ~ \textrm{and} ~ U$ 
parameters~\cite{Altarelli:1990zd,Peskin:1990zt,Peskin:1991sw,Maksymyk:1993zm}. The upper bound on $\sin{\theta}$ decreases from 
$\sim 0.35$ to $\sim 0.20$ as $M_{H_2}$ increases from 250 GeV to 900 GeV~\cite{Robens:2015gla}. In our analysis, we have considered a 
conservative value of $\sin{\theta}=0.20$, throughout, in order to satisfy all the constraints.

\item \textbf{Constraints from perturbative unitarity :} Demanding perturbative unitarity~\cite{Lee:1977eg}, by studying all the $2\to 2$ scattering amplitudes and demanding that the partial wave amplitudes  
$a_{\ell}$s follow
\begin{equation}
 |\mathrm{Re}(a_{\ell})|\leq \frac{1}{2},
\end{equation}
where the subscript $\ell$ denotes the orbital angular momentum, results in an upper 
bound on the Higgs boson mass. The bounds from perturbative unitarity for a model with a scalar 
extension has been derived in Ref.~\cite{Pruna:2013bma}. Perturbative unitarity also poses strong constraints on the ratio 
$\tan{\beta} = v/v'$.

\item \textbf{Perturbativity of the couplings :} All the couplings in the potential are required to conform within perturbative limits, \textit{i.e.,}
$\lambda_{1,2,3} \leq 4 \pi$. These bounds are weaker than the ones obtained from perturbative unitarity at the EW scale.

Besides, constraints from vacuum stability and the renormalisation group evolution of $\lambda_{1,2,3}$ are also studied in 
Refs.~\cite{Lerner:2009xg,Robens:2015gla,Basso:2013vla}.
\end{itemize}
\end{itemize}

\section{Collider searches for the heavy Higgs \label{col}}

The heavy Higgs $H_2$ in the $B-L$ model mixes with the SM-like Higgs, $H_1$, with mixing angle $\theta$, as has been discussed in 
the previous sections. $H_2$ can be produced at the LHC through multiple production processes, e.g. gluon fusion ($ggF$), weak boson fusion ($VBF$), associated $WH_2/ZH_2$ productions and the associated $t \bar{t} H_2$ production mode. Once produced, 
$H_2$ promptly decays into different final states, with $WW$, $H_1 H_1$ and $ZZ$ being the dominant decay modes. In this section, we 
study in detail the collider signatures of $H_2$ produced through its dominant production mode, $ggF$, after including the constraints 
on the mixing angle, $\theta$, as discussed above. In order to study the collider signatures of $H_2$, we implement the model using 
{\tt FeynRules}~\cite{Alloul:2013bka}. The generated {\tt Universal FeynRules Output (UFO)}~\cite{Degrande:2011ua} files are then fed 
to the Monte-Carlo (MC) event generator {\tt MadGraph}~\cite{Alwall:2014hca} for generation of event samples. The parton-showering 
and hadronisation is carried out in the {\tt Pythia 6}~\cite{Sjostrand:2006za} framework. For jet formation, we  use the anti-$k_T$ 
algorithm with a jet parameter of $R=0.4$~\cite{Cacciari:2011ma}. 

\begin{figure}
\centering
\subfloat[]{\includegraphics[width=7.5cm,height=6cm]{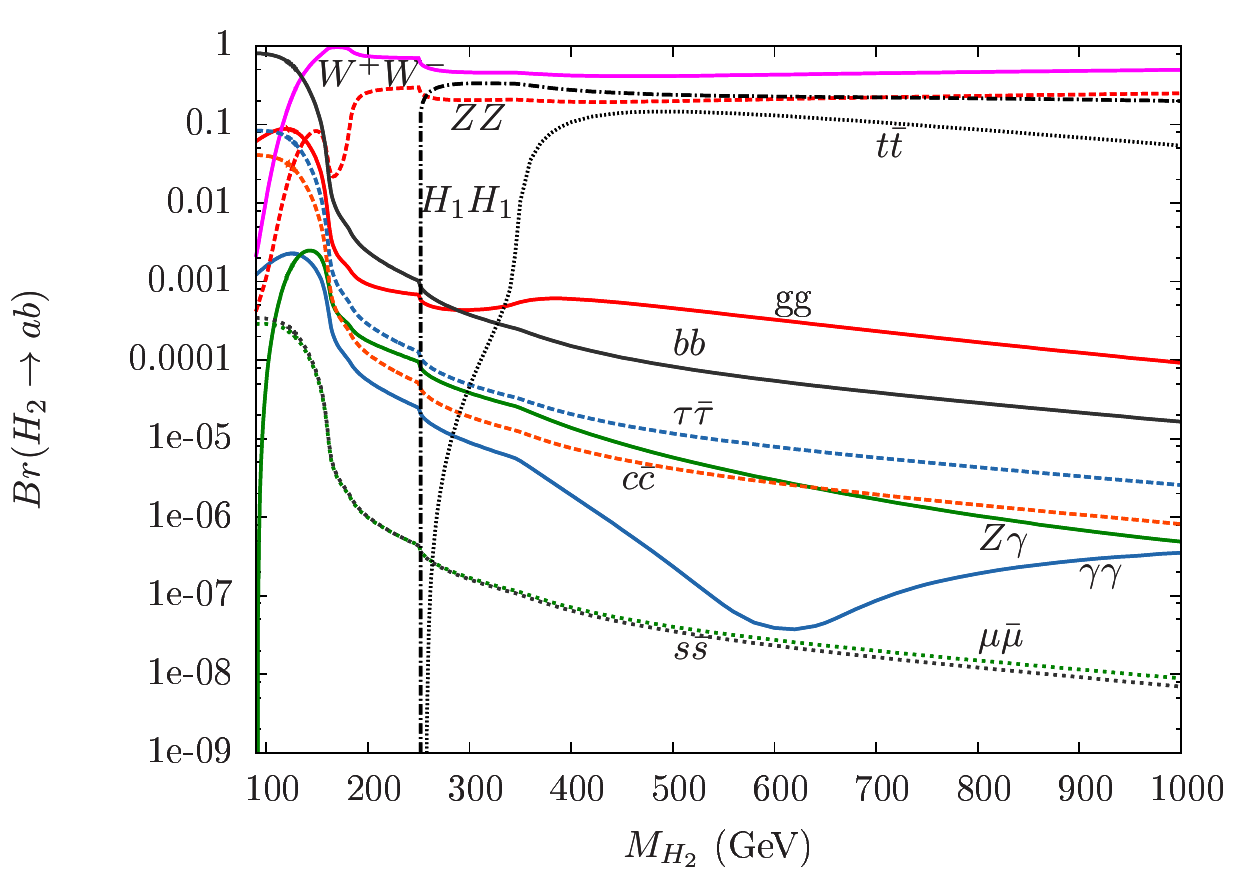}\label{fig:br}}~~
\subfloat[]{\includegraphics[width=7.5cm,height=6cm]{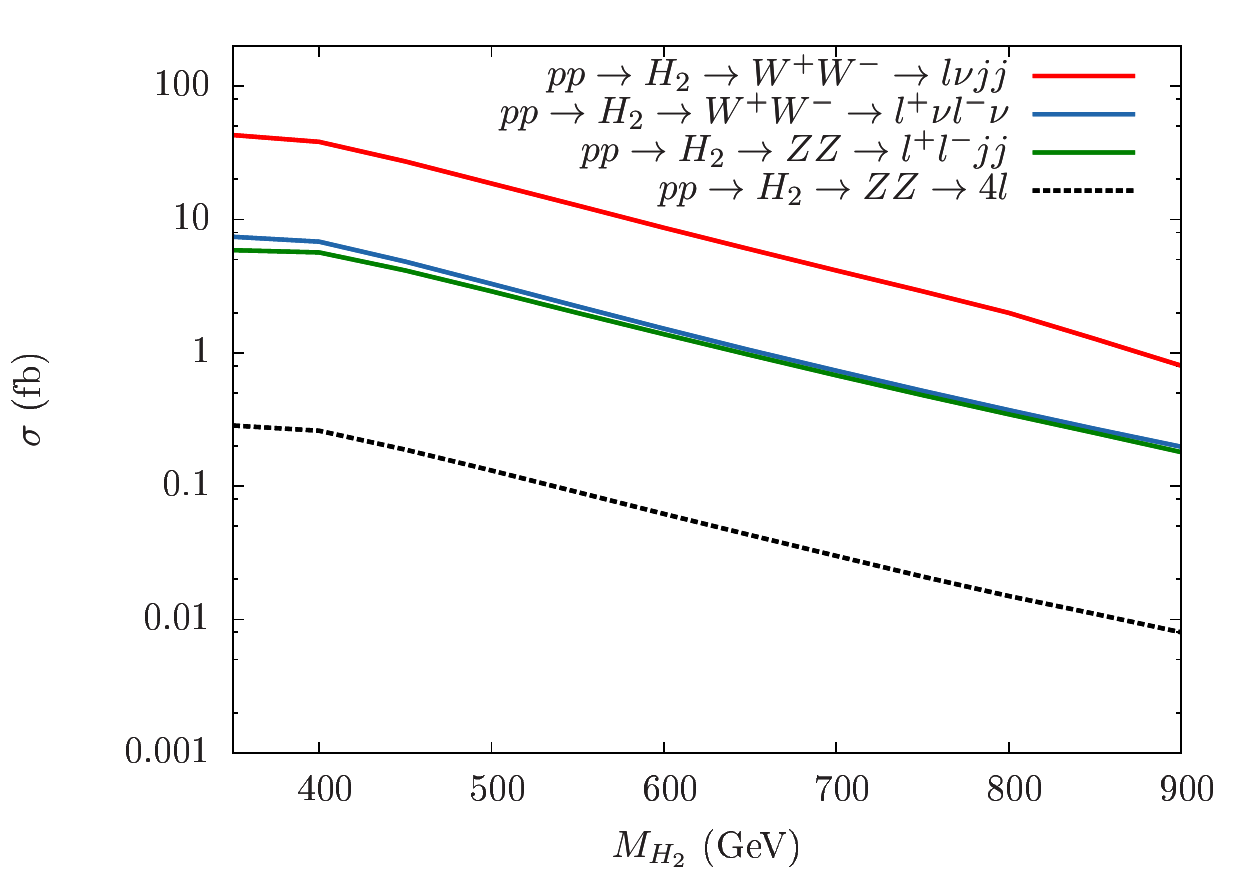}\label{fig:figcrossallx}}~~\\
\caption{Left panel : Branching ratios of $H_2$ (as a function of $M_{H_2}$) to $W^+W^-$ (solid magenta), $ZZ$ (dotted red), $H_1H_1$ (dot-dashed black),
$t\bar{t}$ (dotted black), $b\bar{b}$ (solid black), $gg$ (solid red), $\tau \bar{\tau}$ (dotted blue), $Z\gamma$ (solid green),
$c\bar{c}$ (fine-dotted red), $\gamma \gamma$ (solid blue), $\mu \bar{\mu}$ (fine-dotted green) and $s\bar{s}$ (fine-dotted black).
Right panel : NNLO Cross section (fb) times Branching ratio as functions of $M_{H_2}$ ($pp \to H_2 \to W^+W^- \to \ell \nu j j$ [solid red], 
$pp \to H_2 \to W^+W^- \to \ell^+ \nu \ell^- \nu$ [solid blue], $pp \to H_2 \to ZZ \to \ell^+ \ell^- j j$ [solid green] and 
$pp \to H_2 \to ZZ \to 4 \ell$ [dotted black]). $\sin{\theta}=0.2$ for all the cases.}
\label{fig:brxs}
\end{figure}

In Fig.~\ref{fig:br} we show the branching ratios of $H_2$ to various final states as function of its mass, varying $M_{H_2}$ between 
$250$ GeV and $1$ TeV. As is clear from Fig.~\ref{fig:br}, the three most dominant decay modes of $H_2$ are $WW$, $H_1 H_1$ and $ZZ$.
In Fig.~\ref{fig:figcrossallx}, we show the Next-to-Next-to Leading Order (NNLO) cross-sections of the three different final states mentioned above. Note that, in 
addition to the aforementioned processes, we also show the cross-section for the process $p p \to H_2 \to WW \to 2 \ell \slashed{E}_T$ 
in Fig.~\ref{fig:figcrossallx}. However, we do not consider the phenomenology for the latter process because of a somewhat less amount of 
handle on its kinematics due to fewer visible particles in the final state. The cross-section of $H_2$ decaying to  $\ell \nu 2 j$ is the 
highest, whereas for the $4\ell$ channel, the cross-section is the smallest. We analyze these two channels in considerable details and 
study the discovery prospects of $H_2$ at the HL-LHC. We also briefly mention the $2\ell 2 j$ final state as a potential channel for 
discovering $H_2$. 

\begin{figure}
\centering
\includegraphics[width=8.0cm,height=8.0cm]{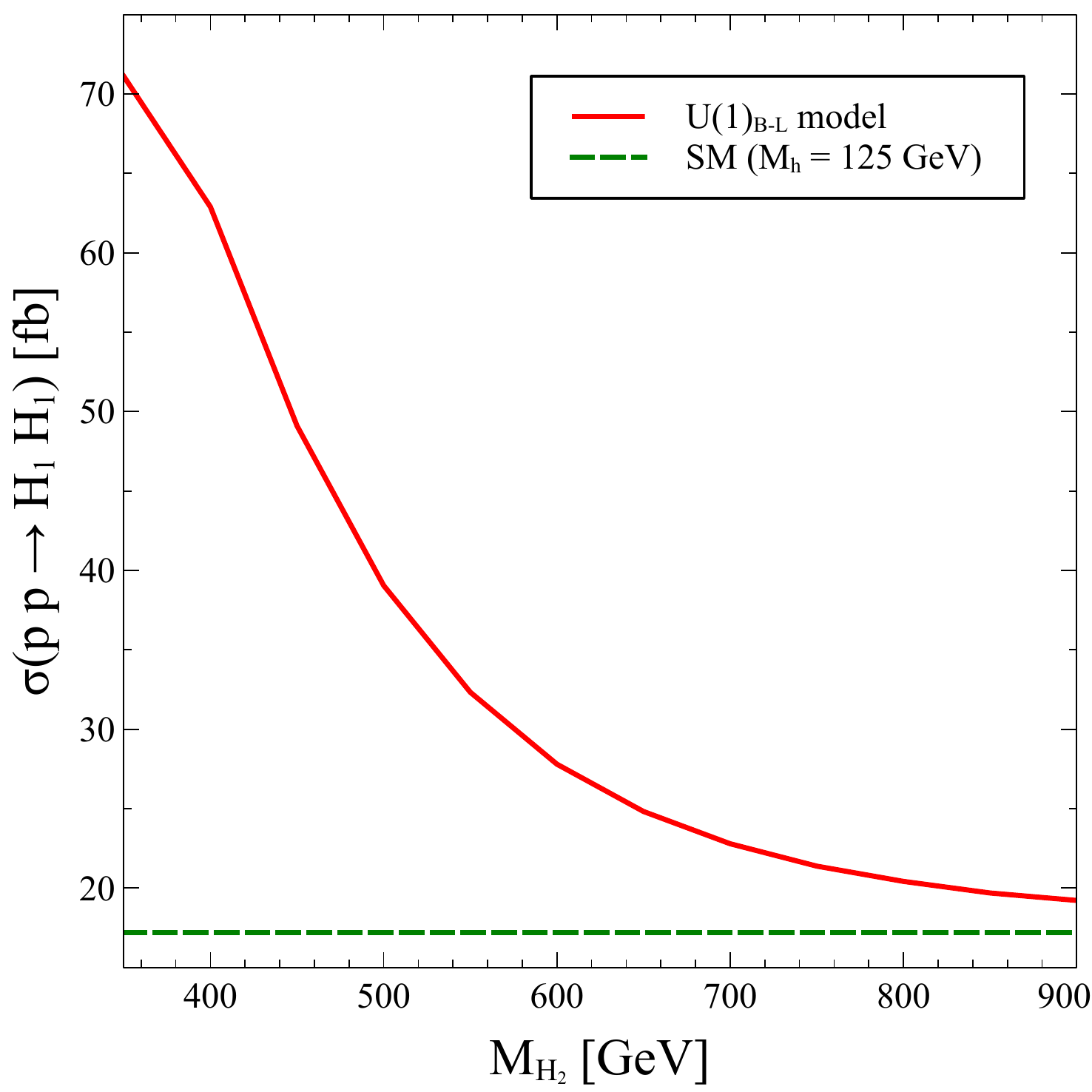}
\caption{LO cross-section for the channel $p p \to H_1 H_1$ for (a) the $U(1)_{B-L}$ model (solid red curve) and (b) the SM 
with $M_h=125$ GeV (green curve). }
\label{fig:H2H1H1}
\end{figure}

Recently, search strategies for 
$H_2 \to H_1 H_1$ have been discussed in Refs.~\cite{Dolan:2012ac,No:2013wsa,Falkowski:2015iwa,Martin-Lozano:2015dja,Buttazzo:2015bka}. CMS~\cite{CMS-PAS-HIG-13-032} and 
ATLAS~\cite{Aad:2014yja,Aad:2015uka} have studied the di-higgs production in the $b\bar{b}b\bar{b}$ and $b\bar{b}\gamma\gamma$ final 
states mostly in the context of models with extra spatial dimensions. The upper limits on $\sigma \times B$ for the resonant and 
non-resonant production of di-higgs in context of such BSM models are found in Refs.~\cite{CMS-PAS-HIG-13-032,Aad:2014yja,Aad:2015uka}.
A naive leading order estimate of the $p p \to H_1 H_1$ cross-sections~\cite{Frederix:2014hta} 
with $v'=3.75$ TeV and $\sin{\theta}=0.2$ reveals that for lower values of $M_{H_2}$, 
\textit{i.e.}, up to $\sim 500$ GeV, the cross-section is substantially enhanced with respect to the SM cross-section. However, with
higher values of $M_{H_2}$, $H_2$ decouples and the cross-section tends to the SM value, see Fig.~\ref{fig:H2H1H1}. Hence, this channel can complement the gauge-boson final states in searches for $H_2$.

In the present work, we focus on the $WW$ and $ZZ$ decay modes and try to devise some search strategies in the context of the 14 TeV run at the HL-LHC. Although, the 
branching ratios of $W/Z$ decaying to di-jet final states are large, still, the leptonic and semi-leptonic decay modes offer the 
cleanest possible signatures because of significantly less backgrounds. Therefore, in our subsequent discussions, we concentrate only
on those channels, that have leptons in the final state, i.e.
\begin{itemize}
\item
$pp \to H_2 \to Z Z \to 4 \ell$,
\item
$pp \to H_2 \to Z Z \to 2 j 2 \ell$ and
\item
 $p p \to H_2 \to WW \to l \nu 2 j$
\end{itemize}

For the search strategies, we adopt two different reconstruction methods which we discuss below.

\begin{itemize}
 \item \textbf{Cut-based analysis (CBA) :}  In this method, we employ rectangular cuts 
 on various kinematic variables in order to optimise the significance $$n=\mathcal{N_S}/\sqrt{\mathcal{N_S}+\mathcal{N_B}}~.$$ 
 \item \textbf{Multivariate analysis (MVA) :} We employ multivariate techniques for better
 signal-to-background discrimination, resulting in better signal significance, $n$. For the present study, we use the Boosted Decision 
 Tree (BDT) algorithm from the TMVA~\cite{2007physics...3039H} framework. In order to perform an MVA, we select the set of kinematic 
 variables that give the maximum discrimination between signal and background. Both the signal 
 and backgrounds are trained by this algorithm and another set of event samples are used to test the BDT output. For any MVA, we must 
 always be alert not to over-train signal and background. The universally accepted Kolmogorov-Smirnov (KS) test can reveal if our choice
 of parameters needs to be changed. The test sample is not over-trained if the KS probability lies in the range $(0.1,0.9)$. For most 
 cases, a critical value of the KS probability greater than 0.01~\cite{KS-test} implies that the samples are not over-trained. For the 
 subsequent studies we ensure that over-training is not an issue over the entire parameter range. 
 In order to estimate the LHC's potential in excluding $H_2$, we use the result of the MVA as input to a binned log-likelihood 
 hypothesis test~\cite{Junk:1999kv}. 
\end{itemize}

In the following subsections we discuss the discovery prospects of the heavy higgs.

\subsection{$pp \to H_2 \to Z Z \to 4\ell$}

\begin{figure}
\centering
\subfloat[]{\includegraphics[width=7.5cm,height=6cm]{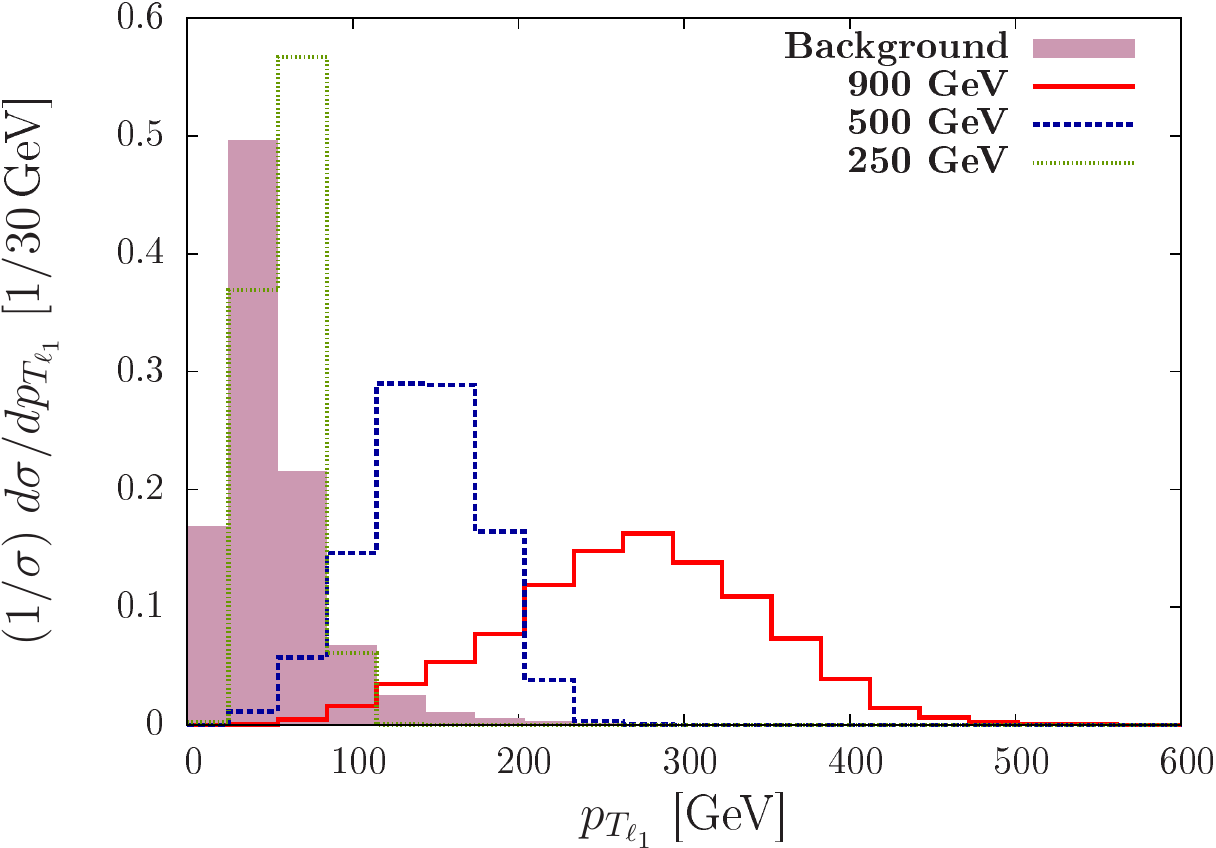}\label{fig:pTl1}}~
\subfloat[]{\includegraphics[width=7.5cm,height=6cm]{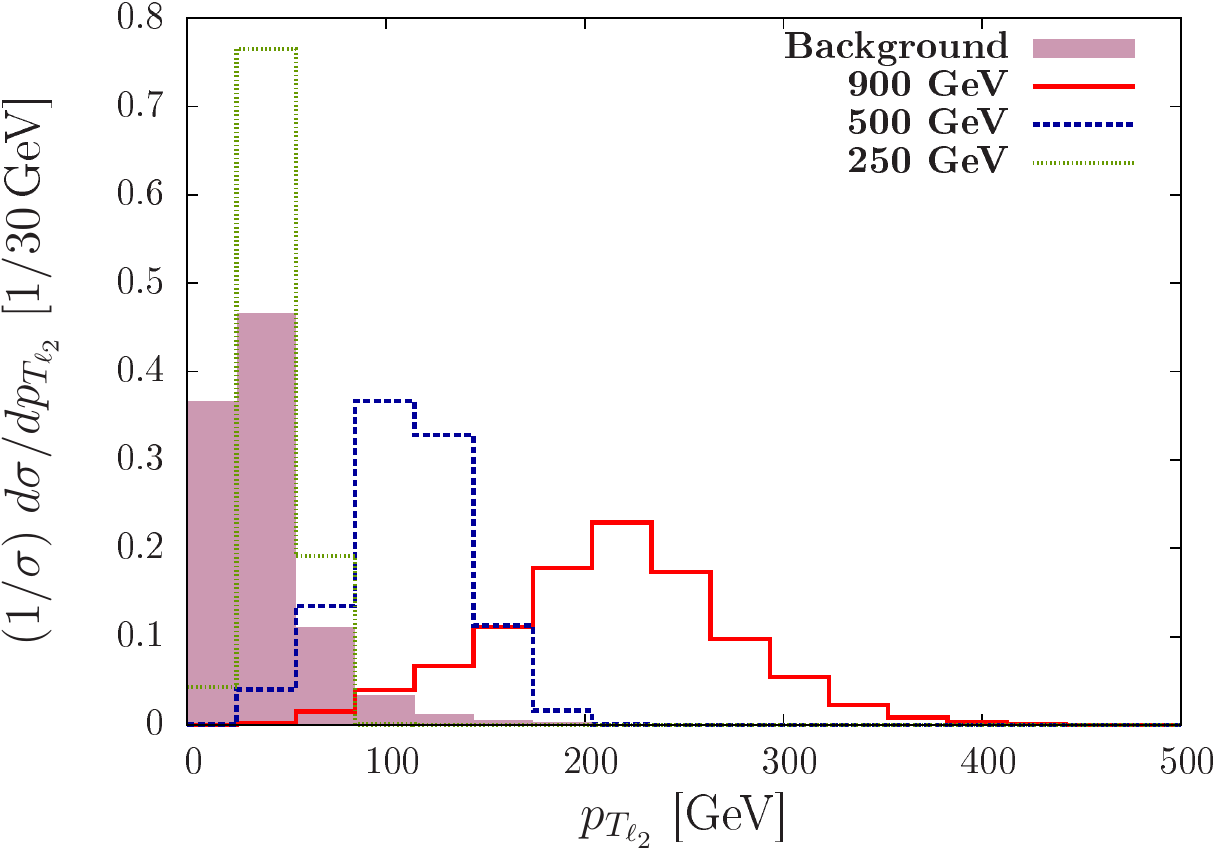}\label{fig:pTl2}}~\\
\subfloat[]{\includegraphics[width=7.5cm,height=6cm]{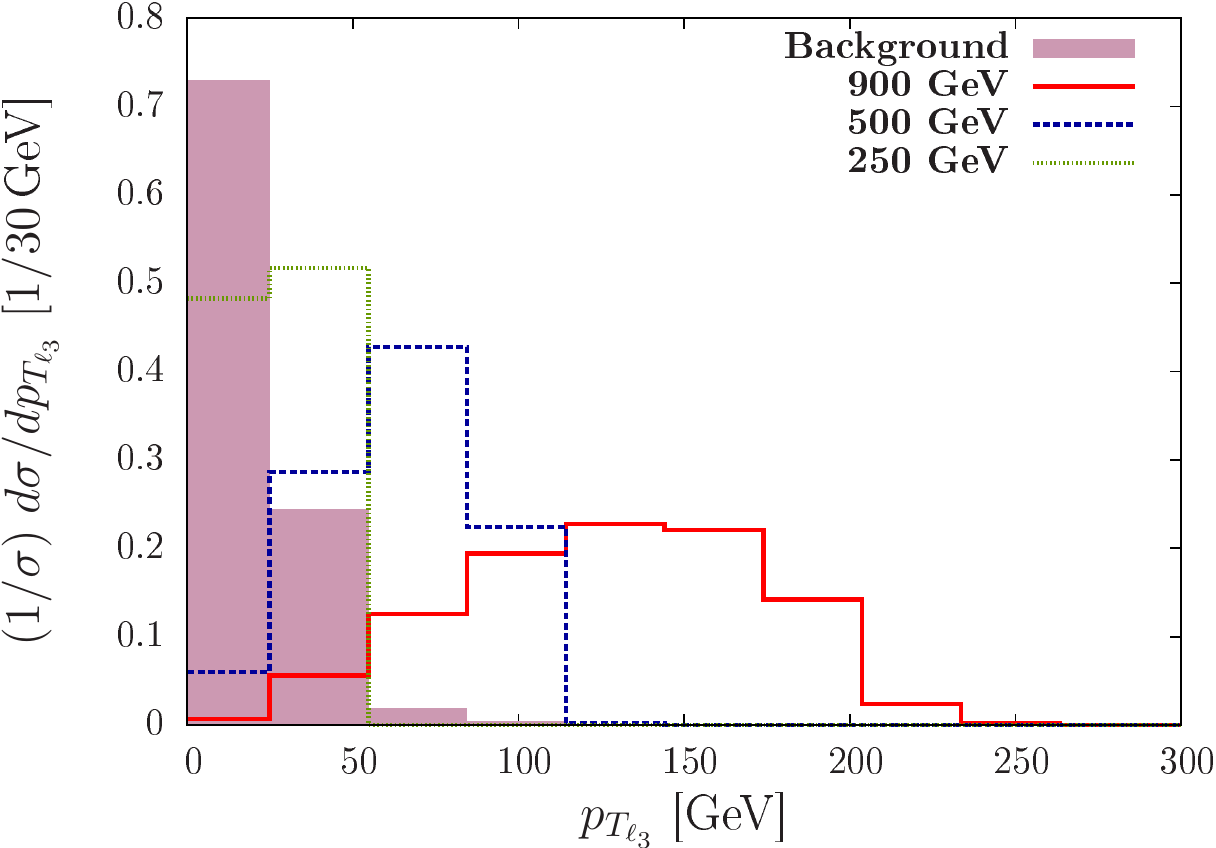}\label{fig:pTl3}}~
\subfloat[]{\includegraphics[width=7.5cm,height=6cm]{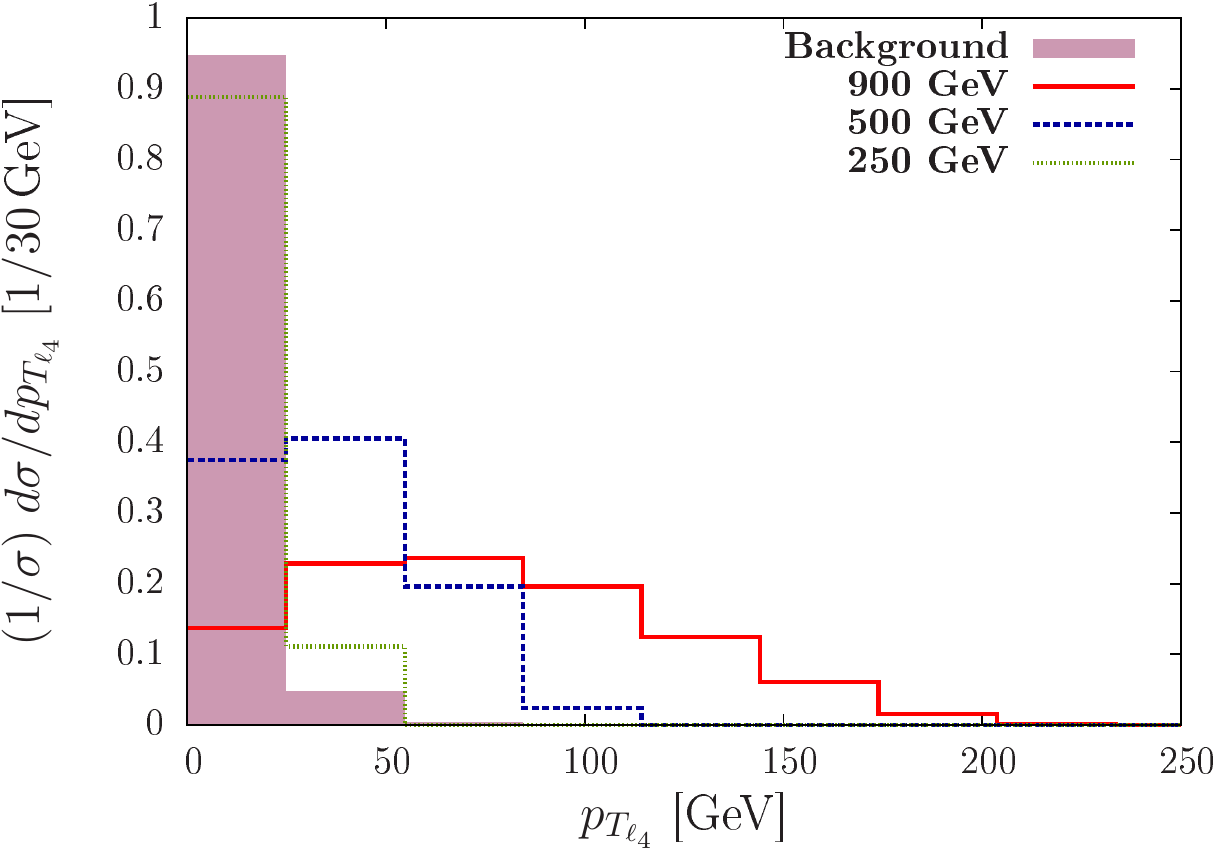}\label{fig:pTl4}}
\caption{$pp\to\ ZZ \to 4 \ell$ channel normalised distributions for $M_{H_2}$ = 250 GeV (green dotted line), 500 GeV (blue dashed line), 
900 GeV (red solid line) and background (purple solid): (a)-(d) $p_T$ distributions of the four leptons, $p_T$ sorted.}
\label{fig:zz1}
\end{figure}

\begin{figure}
\centering
\subfloat[]{\includegraphics[width=7.5cm,height=6cm]{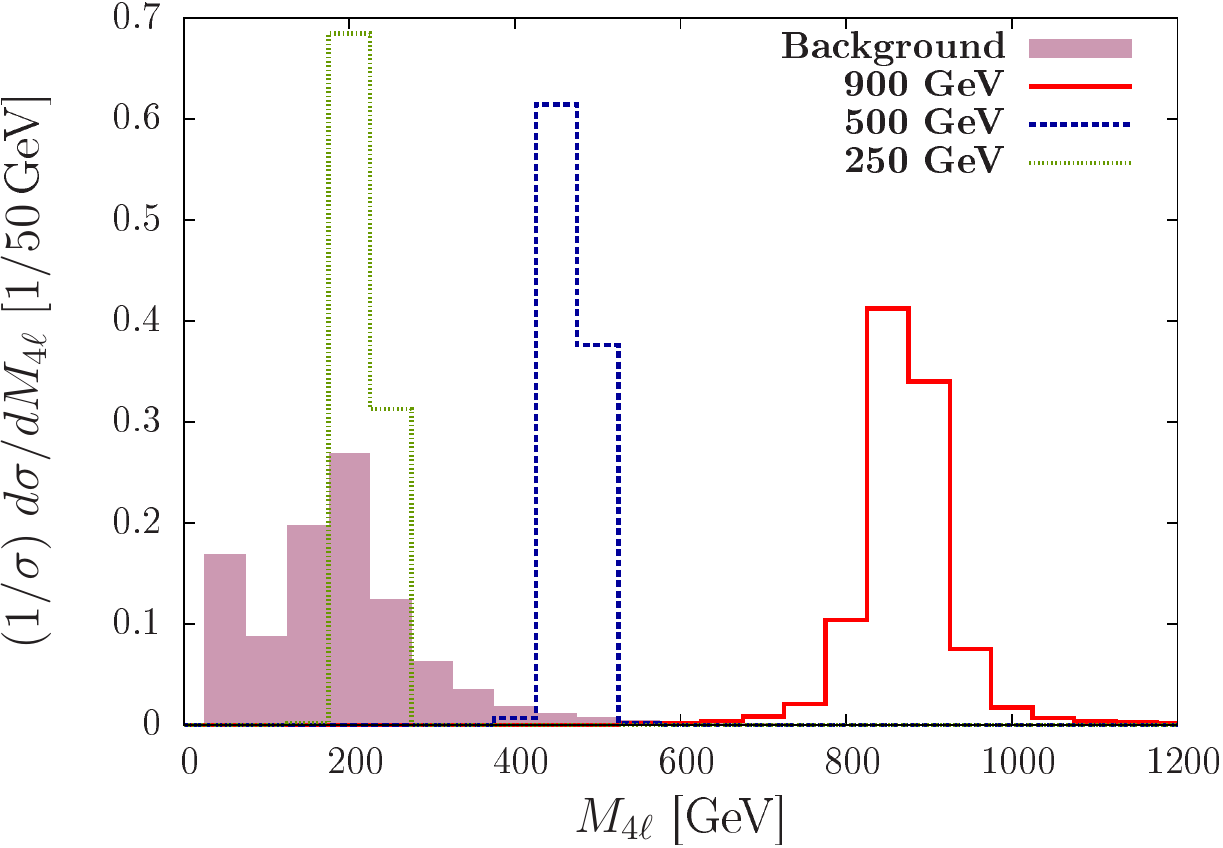}\label{fig:M4l}}~
\subfloat[]{\includegraphics[width=7.5cm,height=6cm]{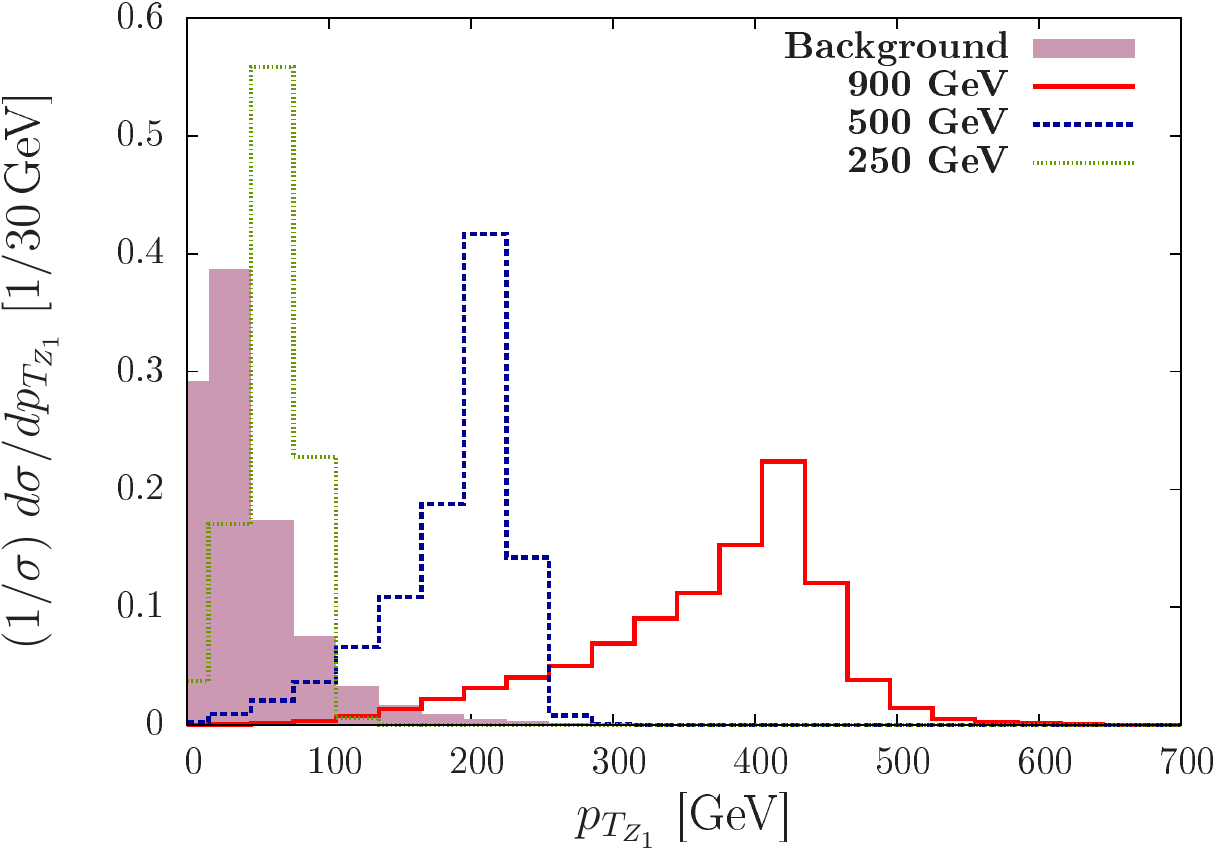}\label{fig:pTZ1}}~\\
\subfloat[]{\includegraphics[width=7.5cm,height=6cm]{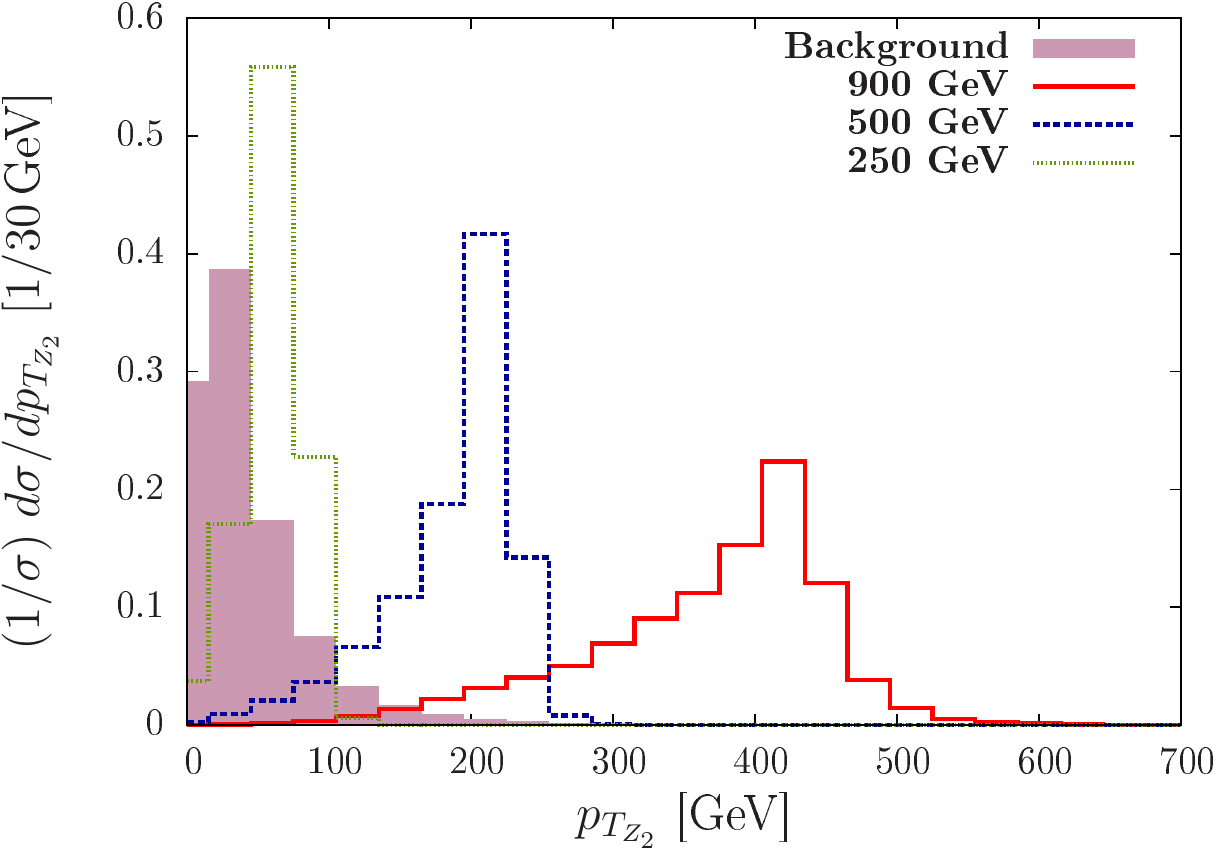}\label{fig:pTZ2}}~
\subfloat[]{\includegraphics[width=7.5cm,height=6cm]{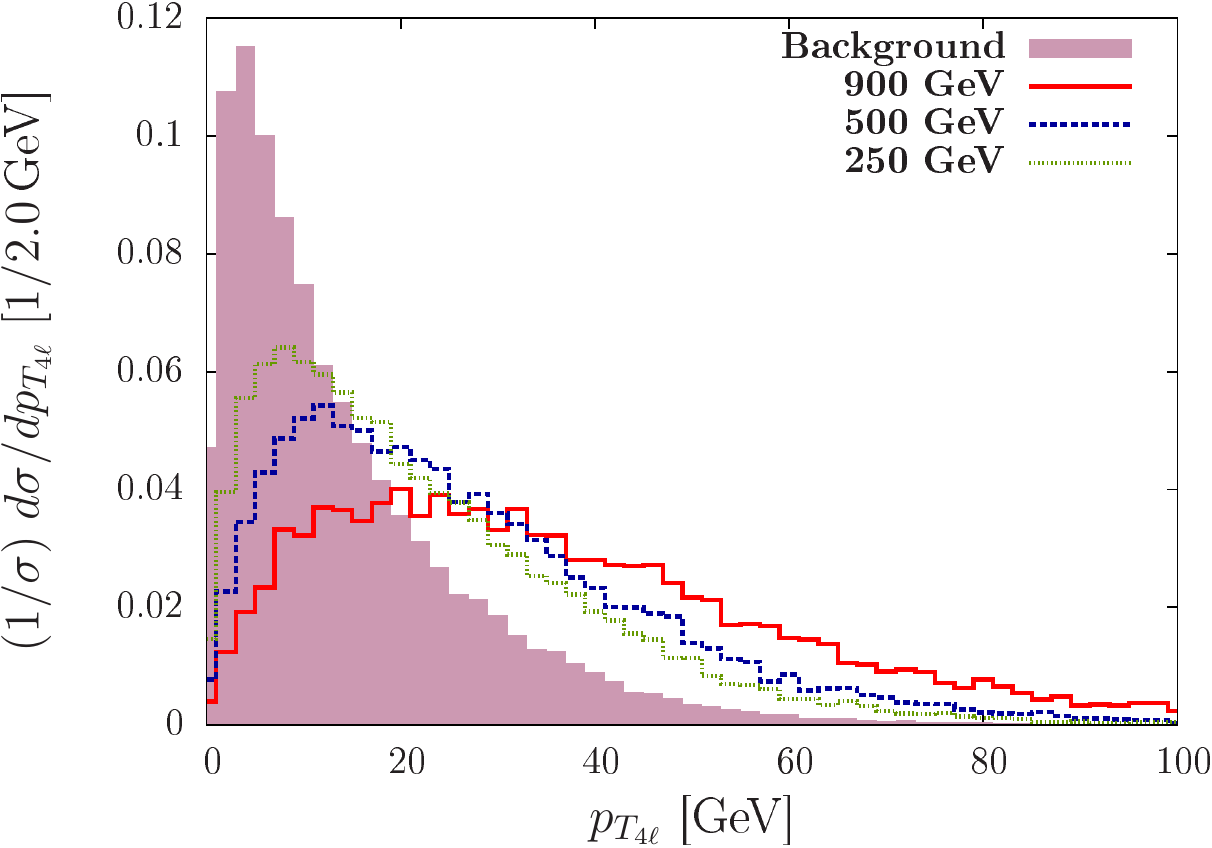}\label{fig:pT4l}}
\caption{$pp\to\ ZZ \to 4 \ell$ channel normalised distributions for $M_{H_2}$ = 250 GeV (green dotted line), 500 GeV (blue dashed line), 
900 GeV (red solid line) and background (purple solid): (a) invariant mass of the four leptons, (b)-(c) 
transverse momenta of the two reconstructed $Z$ bosons and (d) vector sum $p_T$ of the four lepton system.}
\label{fig:zz2}
\end{figure}

In this scenario, $H_2$ is produced on-shell and decays to two $Z$ bosons. The two $Z$ bosons subsequently decay to four leptons. 
The main background for this process is the $ZZ$ production mode that will generate the same final state. To analyse this channel 
we employ the following trigger cuts:
\begin{itemize}
\item
 \underline{\textbf{Trigger Cuts (TC):}} \\
To identify the leptons, we apply the following minimal cuts. 
\begin{enumerate}
\item { Transverse Momentum:} $p_T(l) > 10$ GeV 
\item { Pseudo-rapidity:} $|\eta(l)| < 2.5$
\item { Radial Distance:} $\Delta R(l_i,l_j) > 0.2$ 
\end{enumerate}

We show the normalised distributions of various kinematic variables in Figs.~\ref{fig:zz1} and~\ref{fig:zz2} for $M_{H_2}=250$, $500$ and 
$900$ GeV with respect to the background. It is evident from Fig.~\ref{fig:M4l}, that while for the signal, the invariant mass of 
four leptons peaks at the heavy Higgs mass, $M_{H_2}$, this is not true for the background and hence serves as one of the
better discriminating variables. The leptons originating from the $Z$ decays also 
have higher transverse momentum compared to the background, which peaks at lower $p_T$ values. The two $Z$ bosons also have higher 
$p_T$ with respect to the background, as shown in Figs.~\ref{fig:pTZ1} and~\ref{fig:pTZ2}. For higher $M_{H_2}$, both the $Z$s have $p_T > 100$ GeV. For our cut-based 
analysis (CBA), we use these following selection cuts to separate signal from background to a good degree.

\item \underline{\textbf{Selection  Cuts (SC)}} \\
We use the following selection cuts: 
\begin{enumerate}
\item Invariant mass of the four lepton system: $M_{4l}$ to lie in the range, $ M_{H_2} \pm 10$ GeV 
\item Transverse momentum of leading lepton: $p_{T_{\ell_1}}> 90$ GeV
\item Transverse momentum of sub-leading lepton: $p_{T_{\ell_2}}> 70$ GeV
\item Transverse momentum of the other two leptons: $p_{T_{\ell_3}}> 50$ and $p_{T_{\ell_4}} > 20$  GeV
\item Invariant mass of the reconstructed $Z$ bosons: $M_{Z_1}$, $M_{Z_2} \in M_Z \pm 10$ GeV
\item Transverse momentum of the two reconstructed $Z$ bosons: $p_T(Z_1)$, $p_T(Z_2) > 100 $ GeV 
\end{enumerate}
\end{itemize}

\begin{figure}
\centering
\subfloat[]{\includegraphics[width=7.5cm,height=6.5cm]{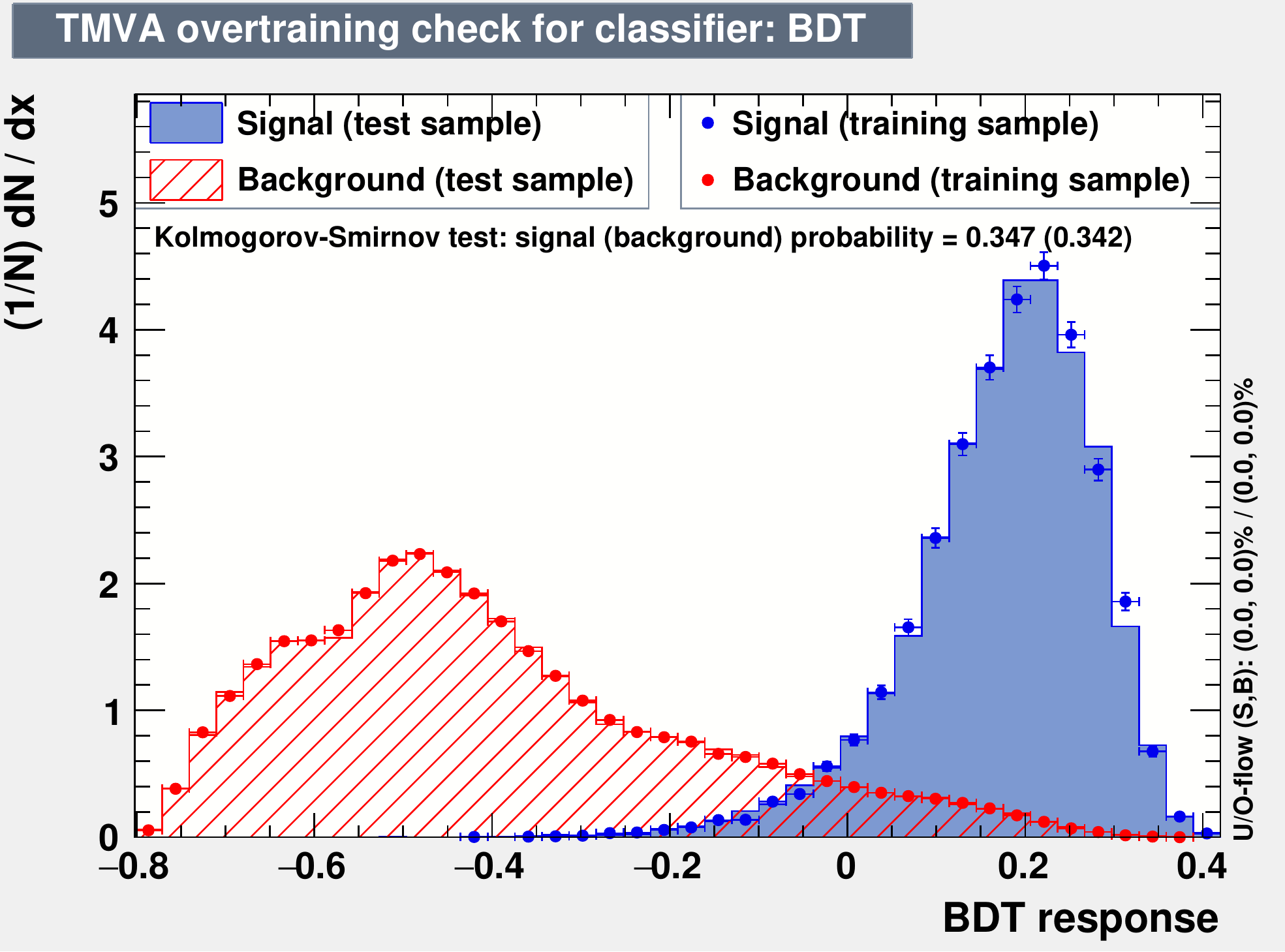}\label{fig:zzbdt250}}~~
\subfloat[]{\includegraphics[width=7.5cm,height=6.5cm]{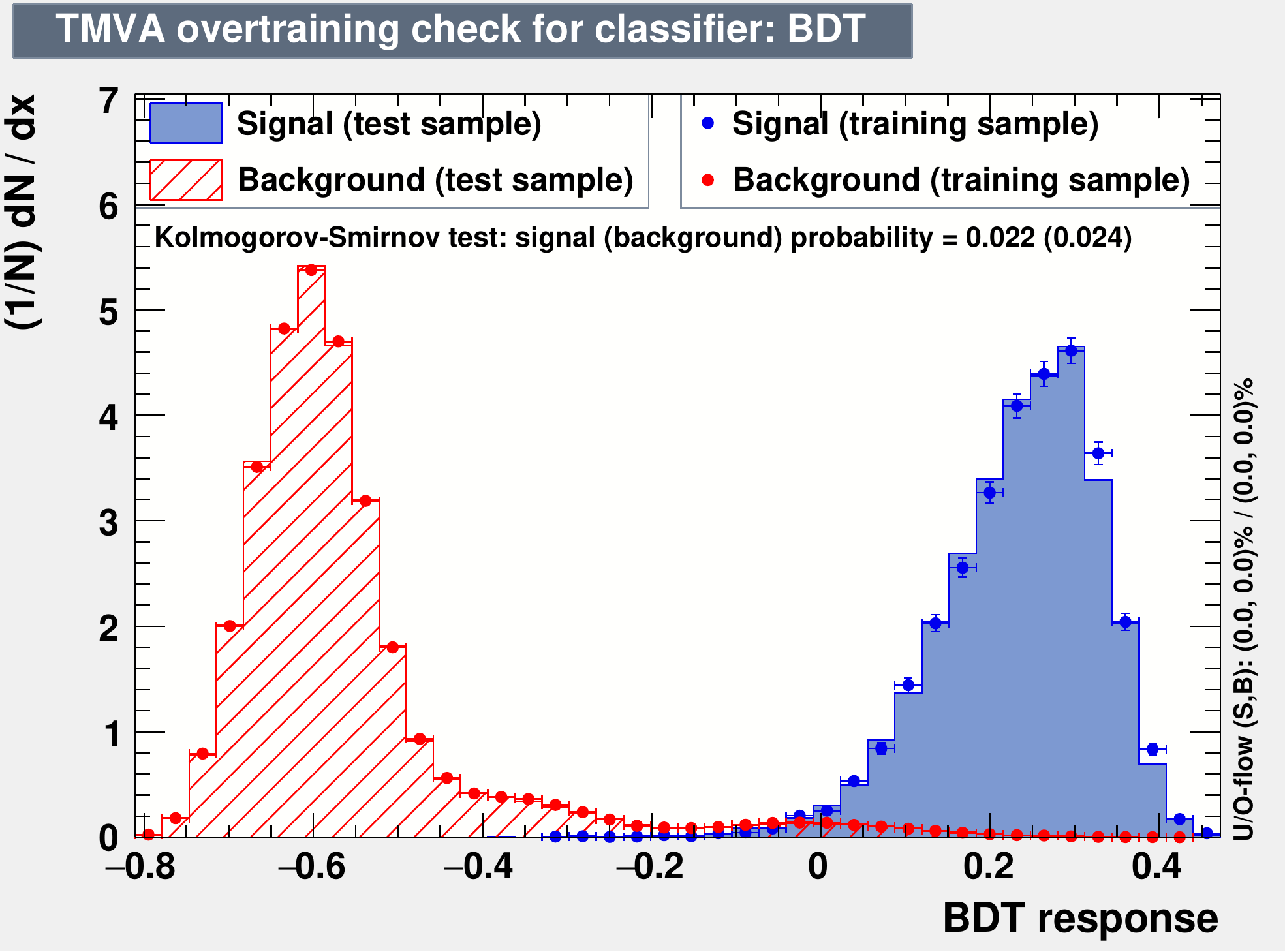}\label{fig:zzbdt500}}~~\\
\caption{Normalised signal and background distributions against BDT response for (a) $M_{H_2}=250$ GeV and (b) $M_{H_2}=500$ for the channel $pp\to H_2 \to ZZ \to 4 \ell$. }
\label{fig:zzbdt}
\end{figure}

\begin{figure}[ht!]
\centering
\subfloat[]{\includegraphics[width=7.5cm,height=6.5cm]{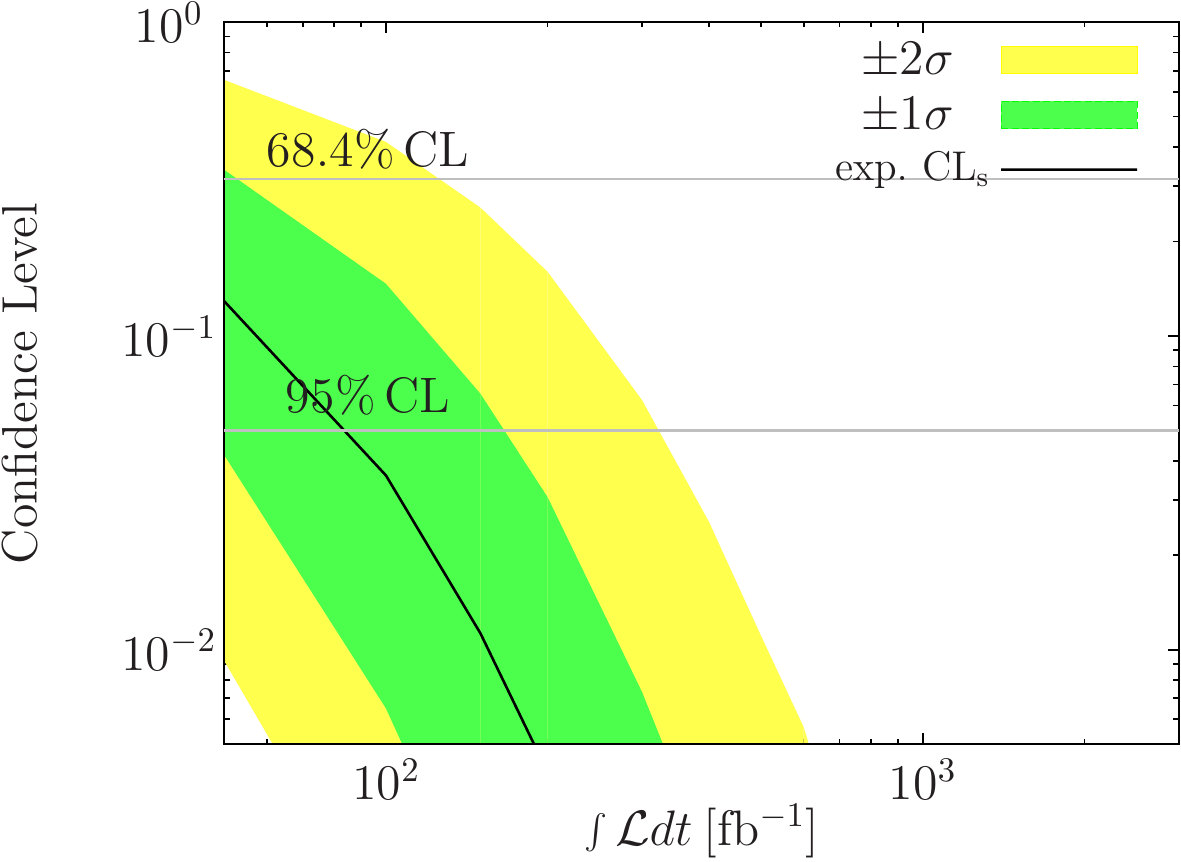}\label{fig:cls250}}~~
\subfloat[]{\includegraphics[width=7.5cm,height=6.5cm]{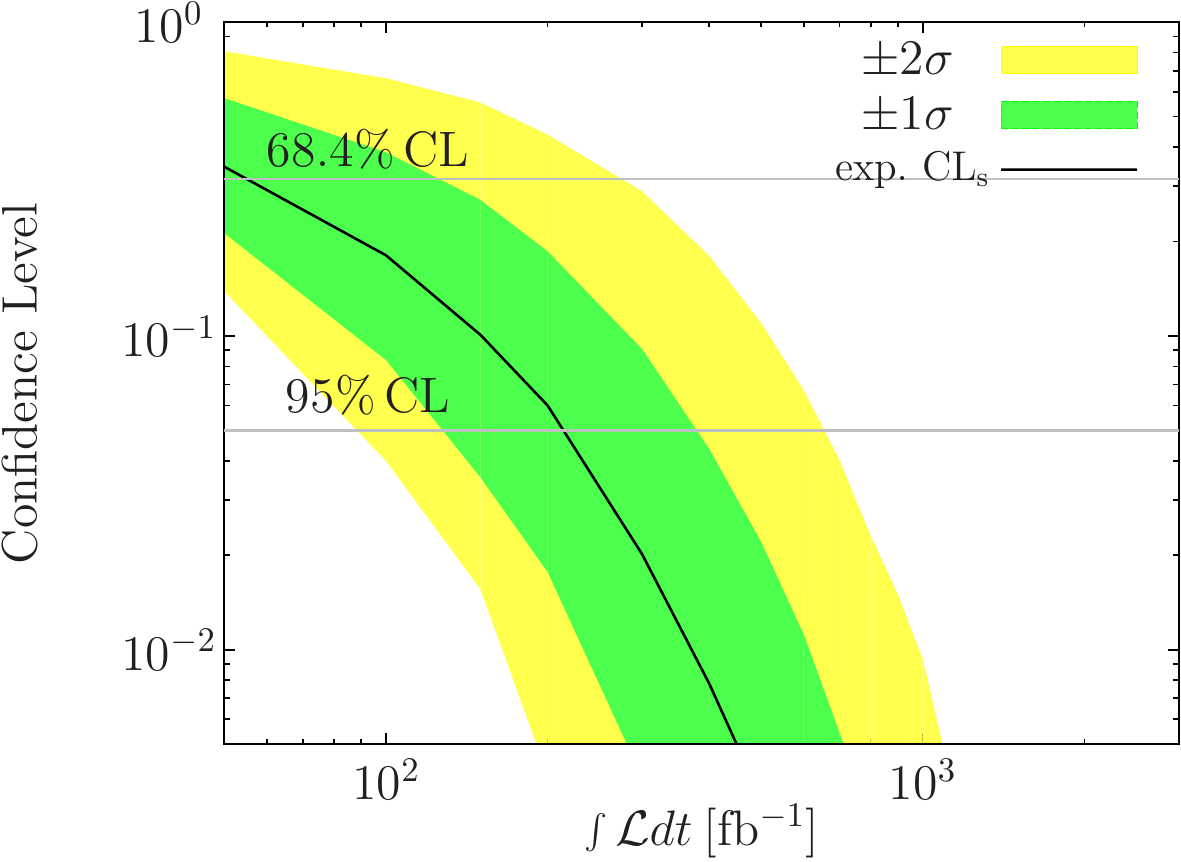}\label{fig:cls500}}~~\\
\subfloat[]{\includegraphics[width=7.5cm,height=6.5cm]{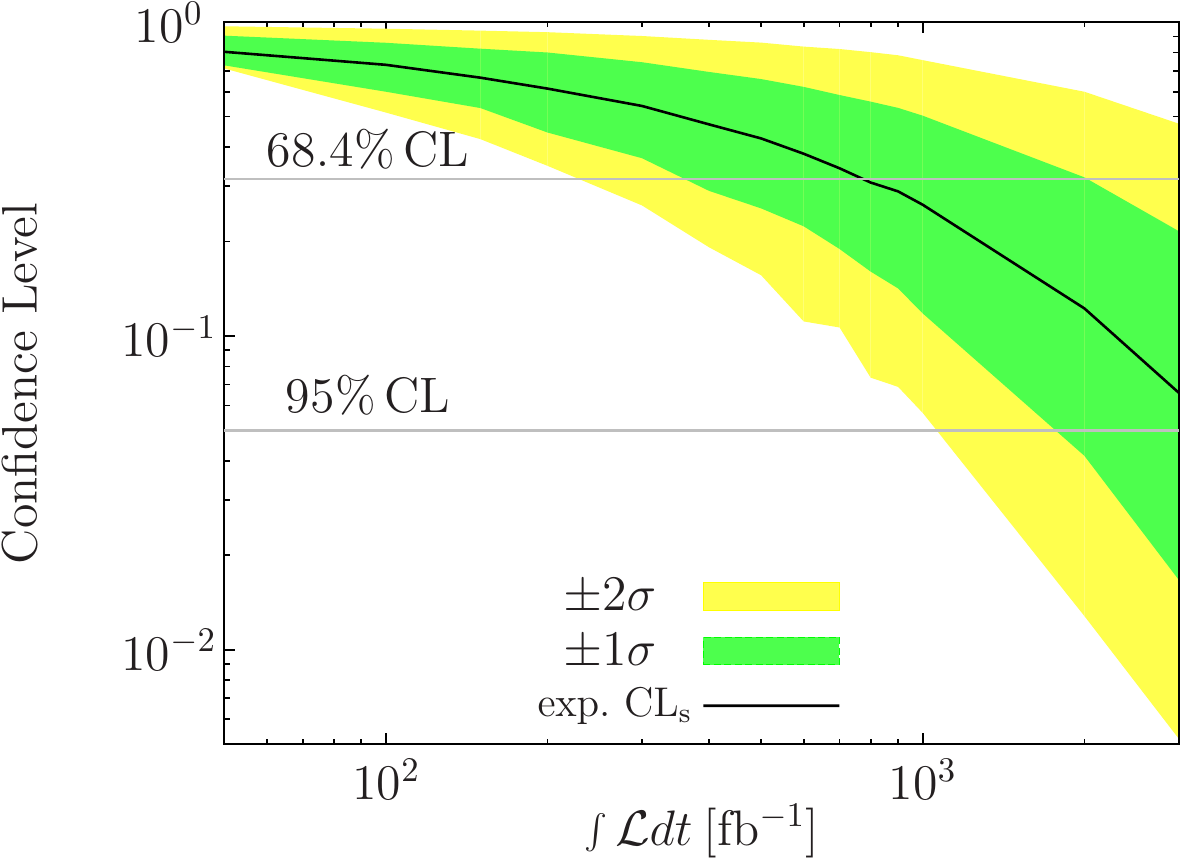}\label{fig:cls700}}~~
\caption{Confidence level contours for $M_{H_2}$ = (a) 250 GeV, (b) 500 GeV and (c) 700 GeV. We show results for integrated 
luminosities ($\int \mathcal{L} dt$) from 50 to 3000 $\mathrm{fb}^{-1}$. We assume a flat systematic uncertainty on the backgrounds of 
$10\%$.}
\label{fig:cls}
\end{figure}

For the MVA analysis, we choose a set of 18 kinematic variables with the maximum discriminating power, which are
$M_{4\ell}$, $p_{T_{\ell_i}}$, $\Delta R_{\ell_i \ell_j}$, $M_{Z_k}$, $p_{T}(Z_k)$, $\eta(Z_k)$ and $p_T(4\ell)$. Here $i,j=1,4$, $k=1,2$ and the leptons 
and two $Z$s are $p_T$ sorted. $p_T(4\ell)$ is the vector sum of $p_T$ of the four leptons.

We tabulate the signal and background cross-sections after the trigger cuts (TC) and the selection cuts (SC) in Table.~\ref{tab:MVA1}
for $M_{H_2}$ in the range $300-700$ GeV. We also list the number of signal ($\mathcal{N_S}$) and background ($\mathcal{N_B}$) events 
computed for an integrated luminosity of 3000 fb$^{-1}$ for the cut-based and BDT analyses after imposing the respective cuts.
Fig.~\ref{fig:zzbdt} shows the normalised distributions for the signal and the background against 
the BDT response for two benchmark masses,
$M_{H_2}=250$ GeV and 500 GeV. We find that with an increase in mass, the overlap between the signal and the background decreases. As a result, using a BDT, $S/B$ improves significantly.

The maximum significance is obtained for relatively small masses of $H_2$, where the cross-sections are sufficiently large. Note that, even with the cut-based analysis, the discovery prospect of the heavy Higgs is rich for 
3000 $\mathrm{fb}^{-1}$ integrated luminosity. As an example, with the cut-based analysis, a Higgs with mass $M_{H_2}=400$ GeV can be discovered with 
$8.6\sigma$ significance ($n_{CBA}$) at the HL-LHC with an integrated luminosity of 3000 $\mathrm{fb}^{-1}$. As expected, 
the BDT analysis is seen to improve the significance ($n_{BDT}$) by a considerable amount. In the entire mass spectrum, the maximum 
difference in $n_{CBA}$ and $n_{BDT}$ occurs for $M_{H_2}=300$ GeV, where the signal and background 
distributions mostly overlap, making it very difficult to impose rectangular cuts.

\begin{table}[!ht]
\centering
\begin{tabular}{|c|c|c|c|c|c|c|c|c|c|}
\hline 
$M_{H_2}$ &  $\sigma^{TC}$ & $\sigma^{SC}$ & $\mathcal{N}^{CBA}_{S}$ & $\mathcal{N}^{CBA}_{B}$ & $n_{CBA}$ & $\mathcal{N}^{BDT}_{S}$ & $\mathcal{N}^{BDT}_{B}$ & $n_{BDT}$   \\ 
(GeV) & (fb) & (fb) & & & & & & \\
\hline 
300 & 0.126 & 0.010 & 30  & 105  & 2.62 & 227 & 555 & 8.12 \\
350 & 0.132 & 0.042 & 125 & 162 & 7.37 & 262 & 419  & 10.03 \\
400 & 0.113 & 0.047 & 142 & 131 & 8.60 & 246 & 361  & 9.99 \\ 
450 & 0.078 & 0.034 & 101 & 101 & 7.14 & 168 & 243  & 8.29 \\
500 & 0.051 & 0.021 & 63  & 81  & 5.26 & 93  & 132  & 6.19 \\
550 & 0.034 & 0.013 & 40  & 48  & 4.23 & 54  & 70  & 4.82 \\
600 & 0.022 & 0.008 & 24  & 45  & 2.87 & 42  & 112  & 3.42 \\
650 & 0.015 & 0.005 &  14  & 32  & 2.12 & 23  & 60   & 2.54 \\
700 & 0.010 & 0.003 &  9   & 24  & 1.57 & 16  & 87    & 1.58 \\
\hline \hline 
SM & 28.626 & -     & -      & -     & -  &    & \\ 
\hline 
\end{tabular} 
\caption{NNLO cross sections after trigger cuts ($\sigma^{TC}$) and selection cuts ($\sigma^{SC}$). $\mathcal{N}_S$ and 
$\mathcal{N}_B$ represent the number of signal and background events, respectively, while the superscript and subscripts $CBA$ and 
$BDT$ represent the cut-based and BDT analysis. $n$ is the significance. The number of events have been computed for an integrated 
luminosity $3000 \, \rm{fb}^{-1}$. All the cross-sections include the higher order corrections 
to the NNLO level.}
\label{tab:MVA1}
\end{table}

 To estimate the necessary integrated luminosity to exclude the existence of the heavy Higgs boson, we use the BDT output, shown in 
 Fig.~\ref{fig:zzbdt}, weighted with the according cross section as input for a CLs likelihood ratio \cite{Junk:1999kv}, see 
 Fig.~\ref{fig:cls}. Conservatively, we assume a flat systematic uncertainty of $10 \%$ for each bin. While an $H_2$ with $M_{H_2}=250$ GeV can be 
 excluded at $95\%$ CL with $100~\mathrm{fb}^{-1}$ in this channel, excluding $M_{H_2}=700$ GeV requires $3000~\mathrm{fb^{-1}}$.

\subsection{$pp \to H_2 \to Z Z \to 2\ell + 2j$}

The channel $H_2 \to ZZ \to 2l 2j$ has been studied in \cite{Hackstein:2010wk} in the context of heavy SM Higgs boson searches. While the signal benefits from a larger branching ratio 
of the $Z$ boson to jets compared to leptons, the only major background in this channel remains continuum $ZZ$ production. 
Here, for convenience, we briefly describe the selection cuts discussed in full detail in Ref.~\cite{Hackstein:2010wk}. 
\begin{itemize}
 \item \textbf{Leptonic $\boldsymbol{Z}$ reconstruction :} We demand two isolated muons with $p_T > 15$ GeV and $|\eta|<2.5$. We 
 further demand an invariant mass window of $10$ GeV around $M_Z$.
 \item \textbf{Hadronic $\boldsymbol{Z}$ reconstruction :} We reconstruct the hadronic $Z$ following the algorithm given in 
 Ref.~\cite{Butterworth:2008iy}. Here also we require an invariant mass window of $10$ GeV around $M_Z$.
 \item \textbf{Heavy Higgs reconstruction :} If the previous two steps are successfully satisfied, then the invariant mass peaks 
 as $M_{H_2}^2=(p_{Z_{lep}}+p_{Z_{had}})^2$, where $p_{Z_{lep}}$ and $p_{Z_{had}}$ are respectively the four-momenta of the reconstructed $Z$ bosons in 
 the leptonic and hadronic channels. The Higgs mass windows used for the four benchmark masses are 
 $(300 \pm 30, 350 \pm 50, 400 \pm 50, 500 \pm 70, 600 \pm 100)$ GeV. These are found to optimise the results. Also because the Higgs 
 width increases significantly with $M_{H_2}$, we widen the windows for reconstruction purposes.
 \item \textbf{$\boldsymbol{ZZ}$ separation :} To further improve $S/B$, we require the leptonic and hadronic $Z$ bosons to be have a maximum 
 isolation of $\Delta R_{ZZ} < 3.2$, where $\Delta R = \sqrt{\Delta\eta^2+\Delta\phi^2}$ with $\Delta\eta$ and $\Delta\phi$ being the separation between 
 two objects in the pseudo-rapidity and azimuthal angle planes respectively. For $Z$ + jets, $\Delta R$ between the reconstructed $Z_{lep}$
 and the fake-$Z$ from the QCD jets often becomes large in order to account for the large Higgs invariant mass.
 \item \textbf{Pruning and trimming :} The pruning~\cite{Ellis:2009me,Ellis:2009su} and trimming~\cite{Krohn:2009th} algorithms are
 used to further reduce the QCD backgrounds because this technique helps in discriminating colour singlet resonances from QCD 
 jets~\cite{Soper:2010xk}. The details of this procedure are elucidated in Ref.~\cite{Hackstein:2010wk}.
\end{itemize}
After assuming 
$\sin \theta = 0.2$ and applying the reconstruction outlined in \cite{Hackstein:2010wk} we find the results shown in 
Tab.~\ref{tab:zz_2l2j}.

\begin{table}[!ht]
\centering
\begin{tabular}{|c|c|c|c|c|c|}
\hline 
$M_{H_2}$ & $\sigma^{ggF+VBF}_{SC}$ &  $\sigma^{bkg}_{SC}$ & $S/B$ & $S/\sqrt{S+B}_{100}$ & $S/\sqrt{S+B}_{3000}$  \\ 
(GeV) & (fb) & (fb) & & & \\
\hline 
300 & 0.048 & 2.10 & 0.023 & 0.331 & 1.811 \\
400 & 0.290 & 19.21 & 0.015 & 0.657 & 3.602 \\
500 & 0.223 & 18.01 & 0.012 & 0.522 & 2.858 \\
600 & 0.121 & 11.83 & 0.010 & 0.351 & 1.920 \\
\hline \hline
\end{tabular} 
\caption{$\sigma^{ggF+VBF}_{SC}$ is the production cross-section of $H_2$ from the $ggF$ and $VBF$ channels combined after 
employing the selection cuts discussed in Ref.~\cite{Hackstein:2010wk}. $\sigma^{bkg}_{SC}$ is the background cross-section for 
the same set of selection cuts. The table also shows the discovery potential of $H_2$ in this channel with the help of $S/B$ 
and $S/\sqrt{S+B}$. The subscripts $100$ and $3000$ imply the significance computed at the respective integrated luminosities 
in fb$^{-1}$.}
\label{tab:zz_2l2j}
\end{table}

Hence, the sensitivity in the $H_2 \to 2\ell 2j$ channel alone is fairly small for the $U(1)_{B-L}$ model, based on the 
reconstruction of boosted $Z$ bosons. However, this channel can be combined with the other channels in a global fit.
\\

\subsection{$pp \to H_2 \to W W \to \ell + \slashed{E}_T + \ge 2j$}

\begin{figure}
\centering
\subfloat[]{\includegraphics[width=7.5cm,height=6cm]{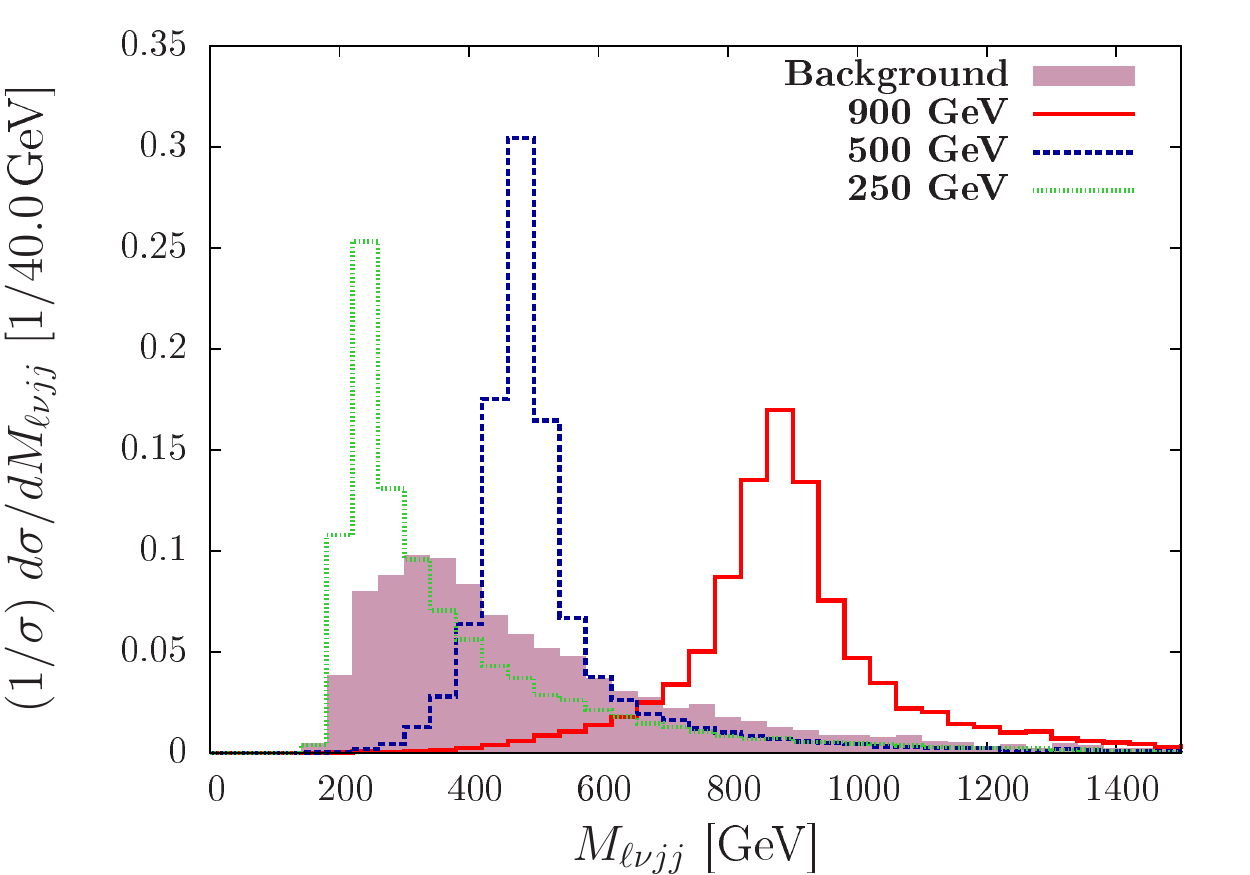}\label{fig:mh2}}~
\subfloat[]{\includegraphics[width=7.5cm,height=6cm]{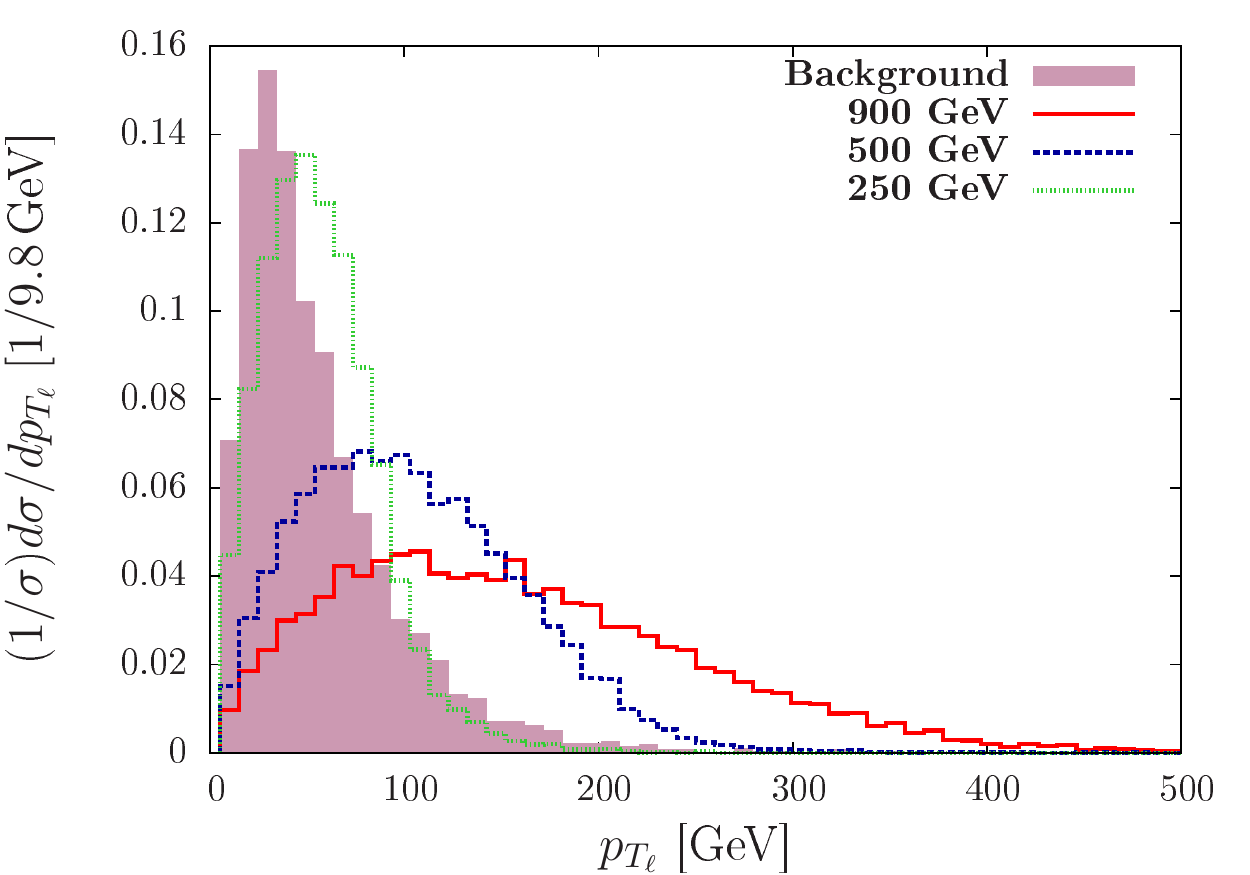}\label{fig:ptl}}~\\
\subfloat[]{\includegraphics[width=7.5cm,height=6cm]{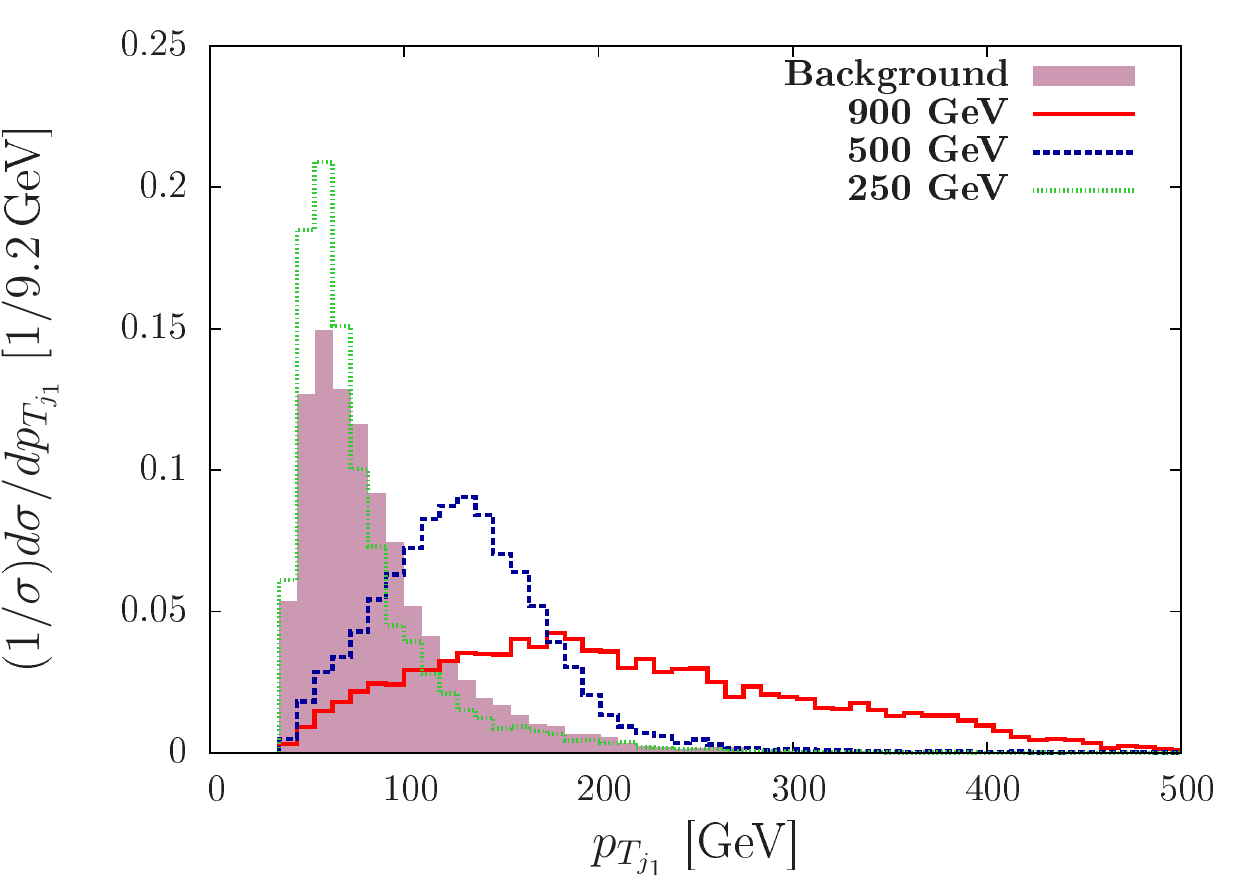}\label{fig:ptjt1}}~
\subfloat[]{\includegraphics[width=7.5cm,height=6cm]{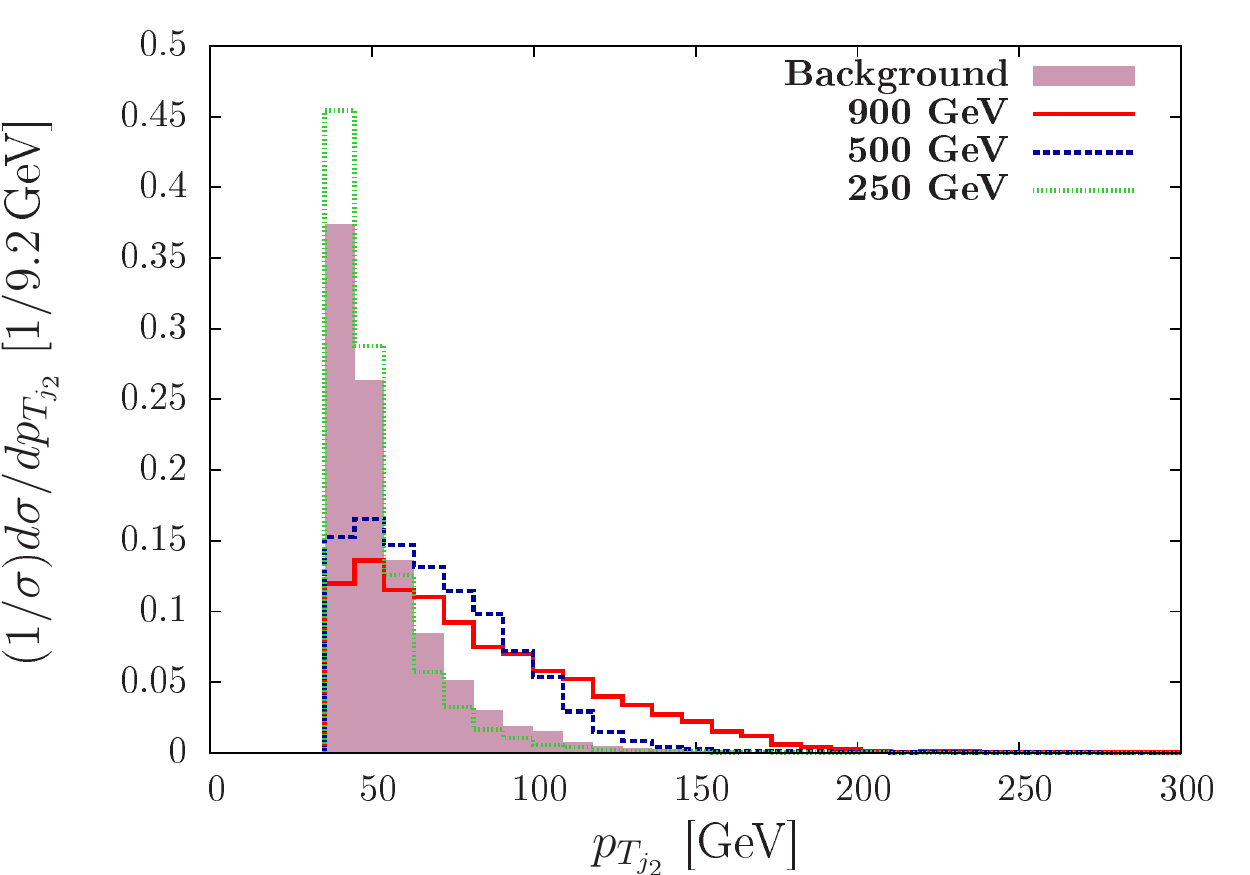}\label{fig:ptjt2}}
\caption{$pp\to\ WW \to \ell 2j \slashed{E}_T$ channel normalised distributions for $M_{H_2}$ = 250 GeV (green dotted line), 500 GeV (blue dashed line), 
900 GeV (red solid line) and background (purple solid): (a) Invariant mass of $\ell 2j \slashed{E}_T$, (b) $p_T$ of lepton, (c)-(d) $p_T$ of
the two tagged jets.}
\label{fig:distww1}
\end{figure}

\begin{figure}
\centering
\subfloat[]{\includegraphics[width=7.5cm,height=6cm]{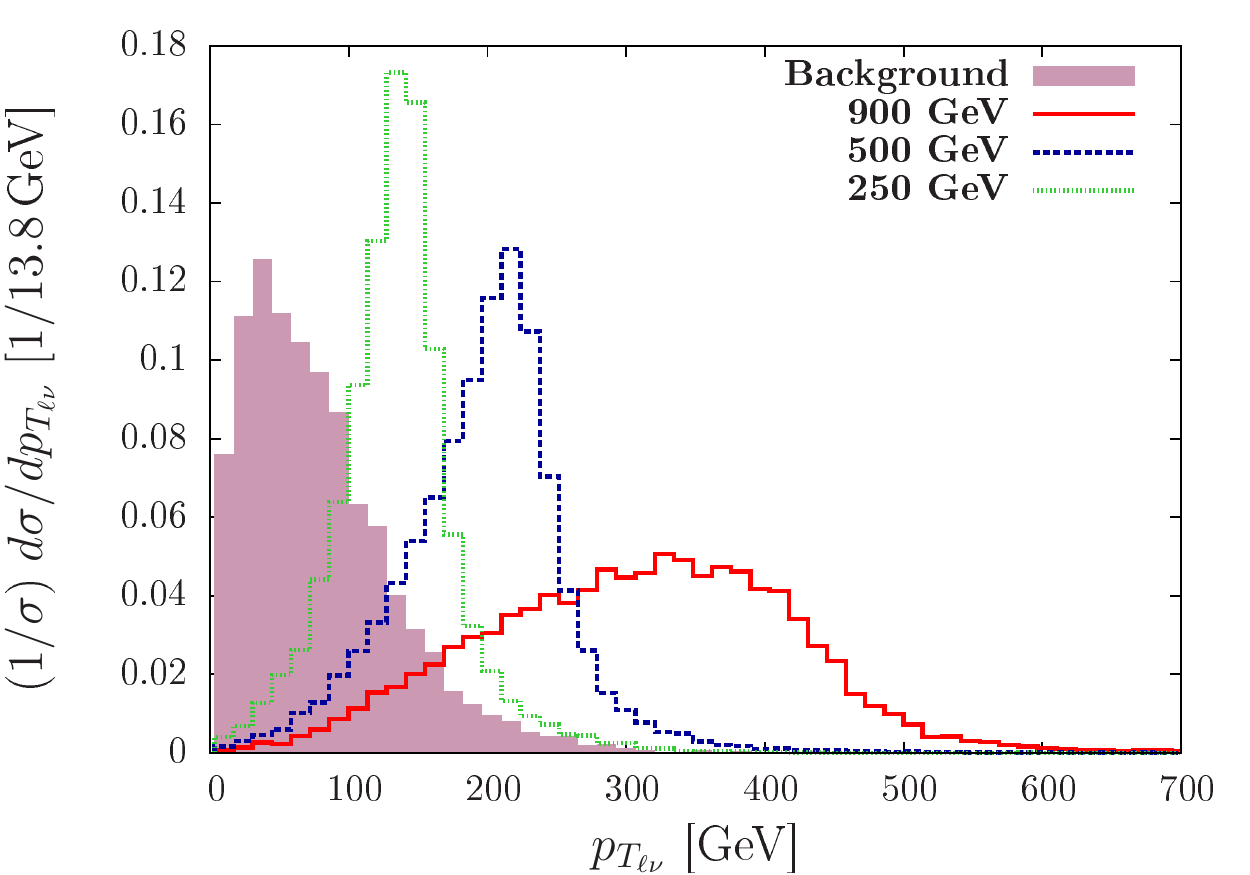}\label{fig:ptlnu}}~
\subfloat[]{\includegraphics[width=7.5cm,height=6cm]{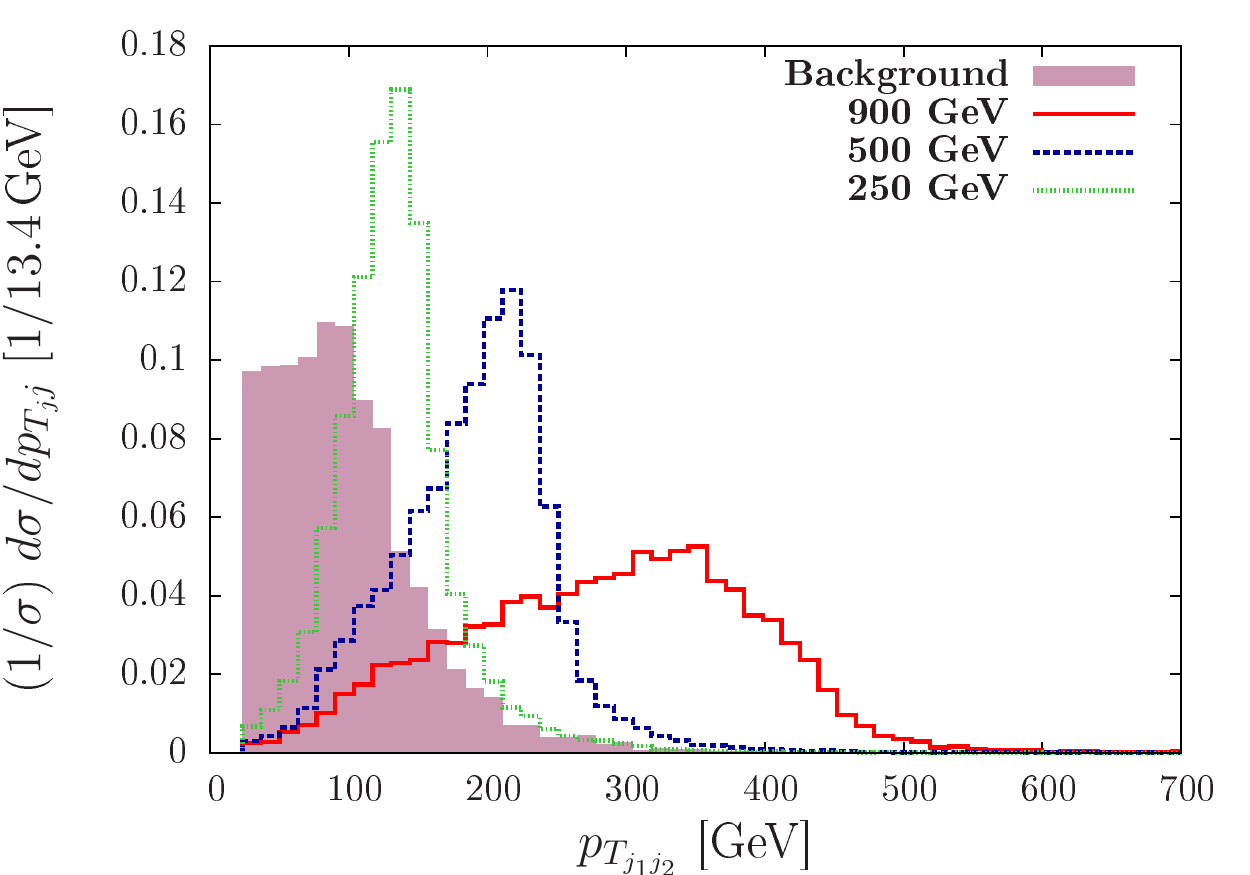}\label{fig:ptjj}}~\\
\subfloat[]{\includegraphics[width=7.5cm,height=6cm]{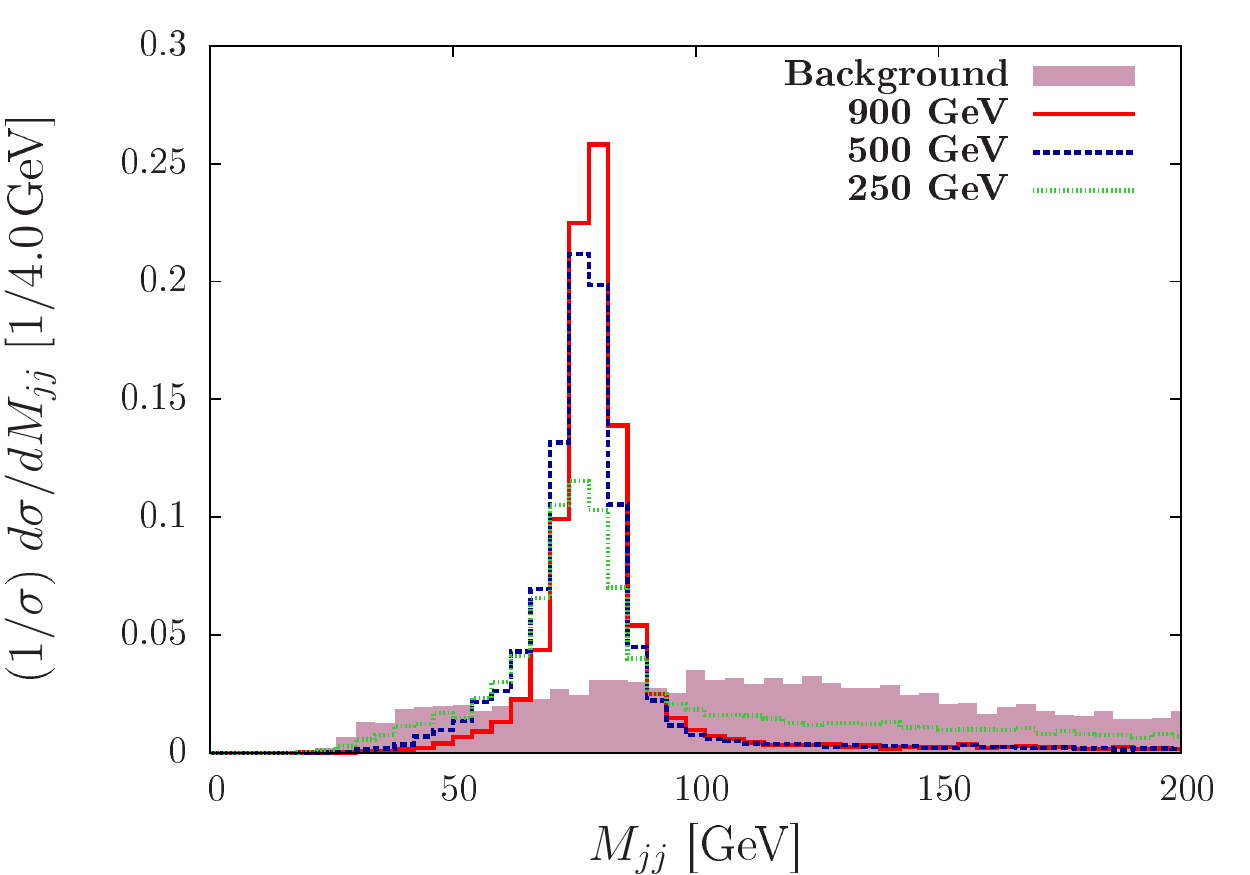}\label{fig:imjj}}~
\subfloat[]{\includegraphics[width=7.5cm,height=6cm]{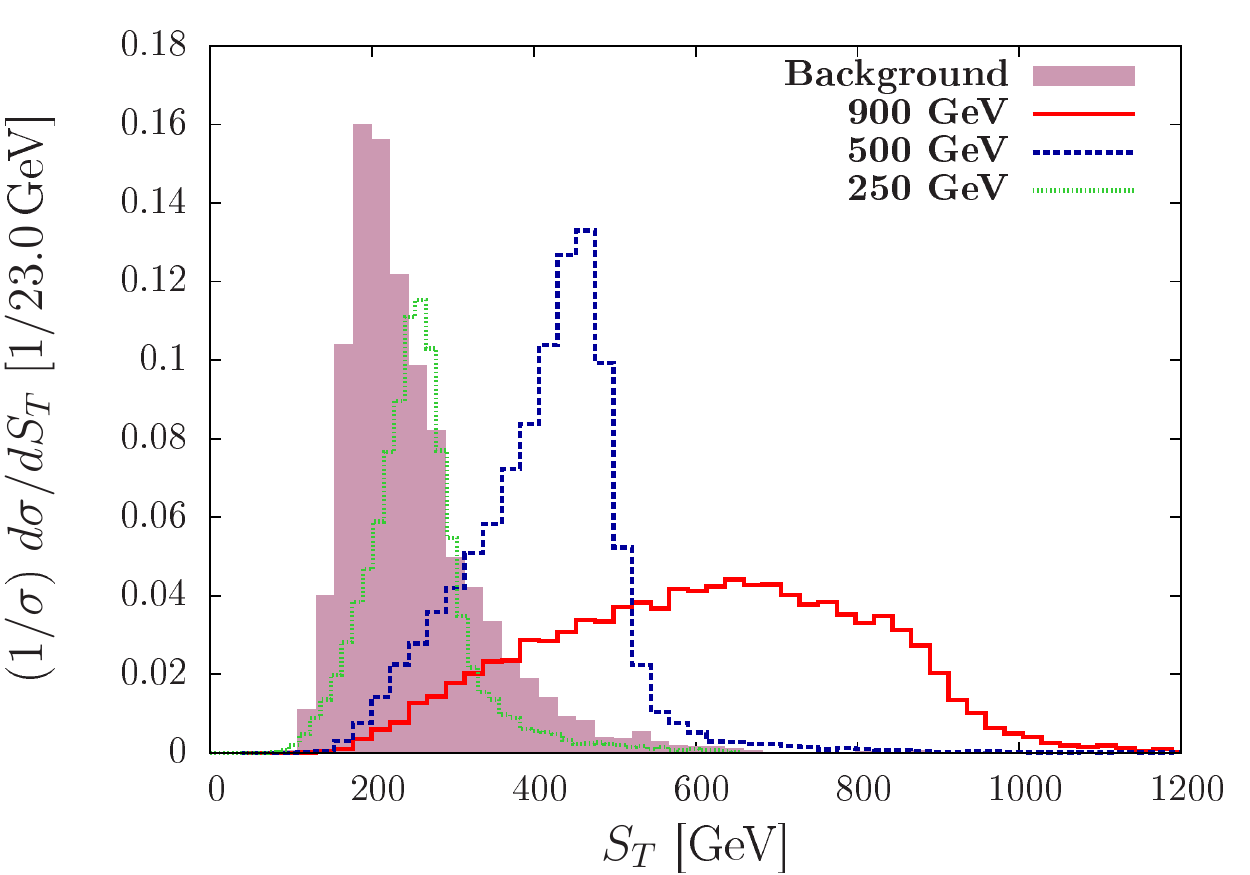}\label{fig:ptsum}}
\caption{$pp\to\ WW \to \ell 2j \slashed{E}_T$ channel normalised distributions for $M_{H_2}$ = 250 GeV (green dotted line), 500 GeV (blue dashed line), 
900 GeV (red solid line) and background (purple solid): (a) $p_T$ of the $\ell \nu$ system, (b) $p_T$ of the di-jet system which 
reconstructs the $W$ mass, (c) Invariant mass of di-jet system which reconstructs the $W$ mass, (d) scalar sum $p_T$ of lepton, 
two-tagged jets and $\slashed{E}_T$.}
\label{fig:distww2}
\end{figure}

\begin{figure}[!ht]
\centering
\subfloat[]{\includegraphics[width=7.5cm,height=6cm]{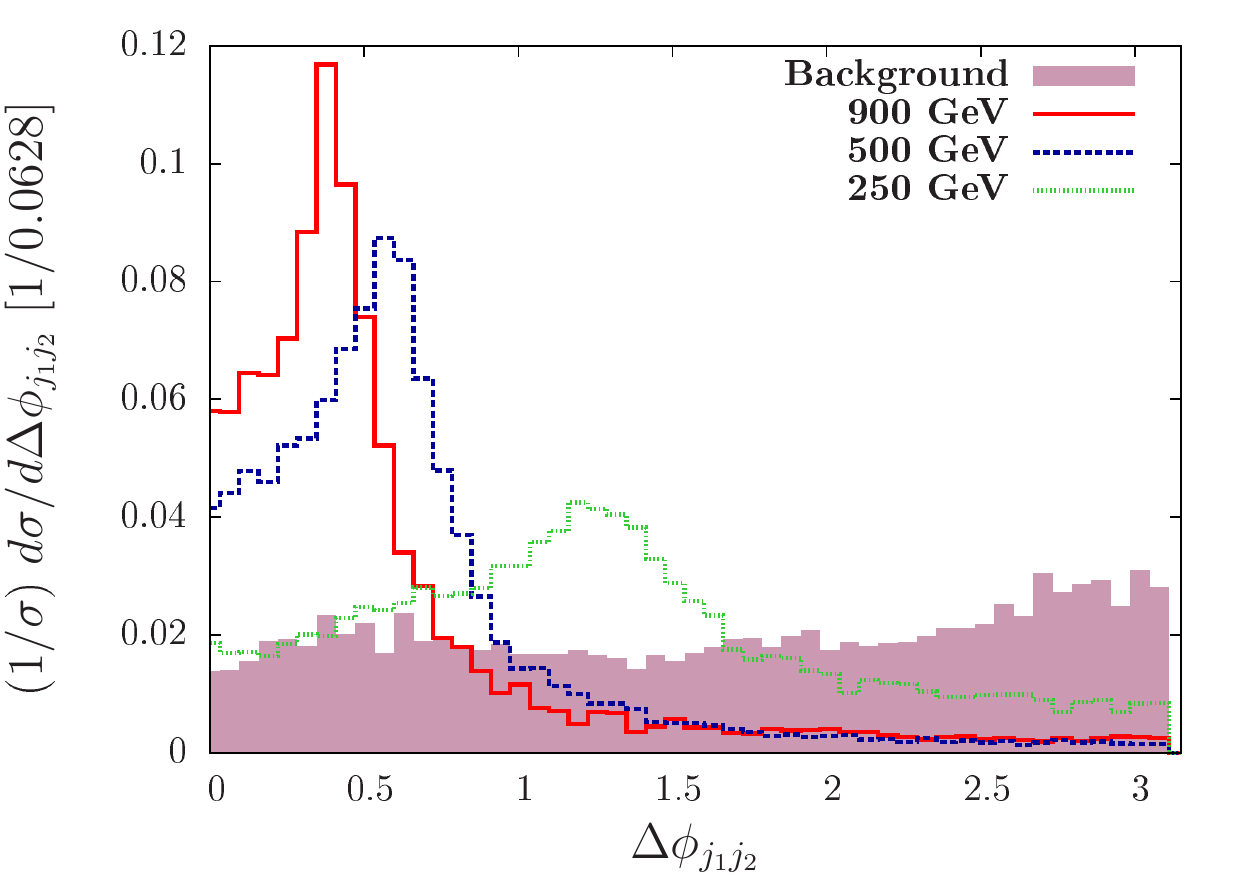}\label{fig:azij1j2}}~
\subfloat[]{\includegraphics[width=7.5cm,height=6cm]{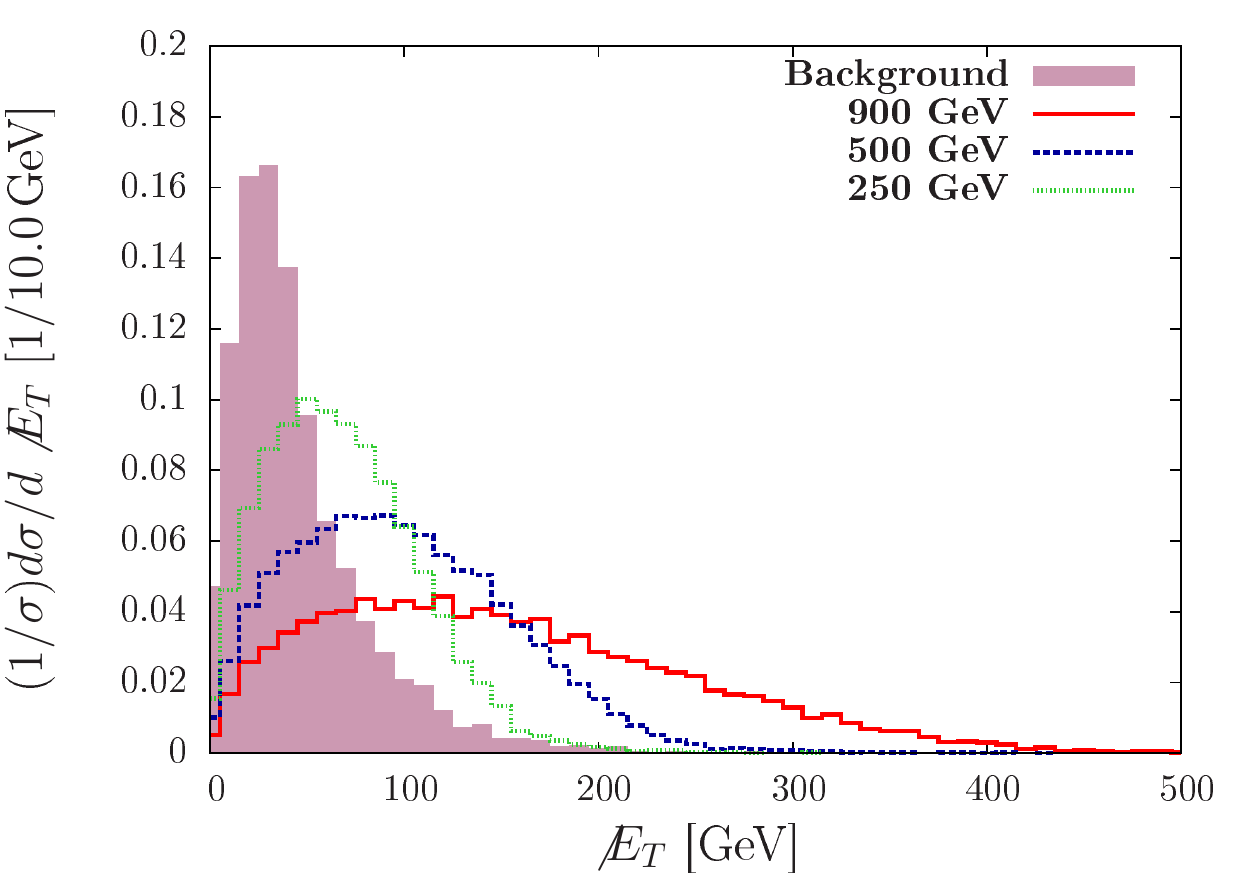}\label{fig:met}}~\\
\subfloat[]{\includegraphics[width=7.5cm,height=6cm]{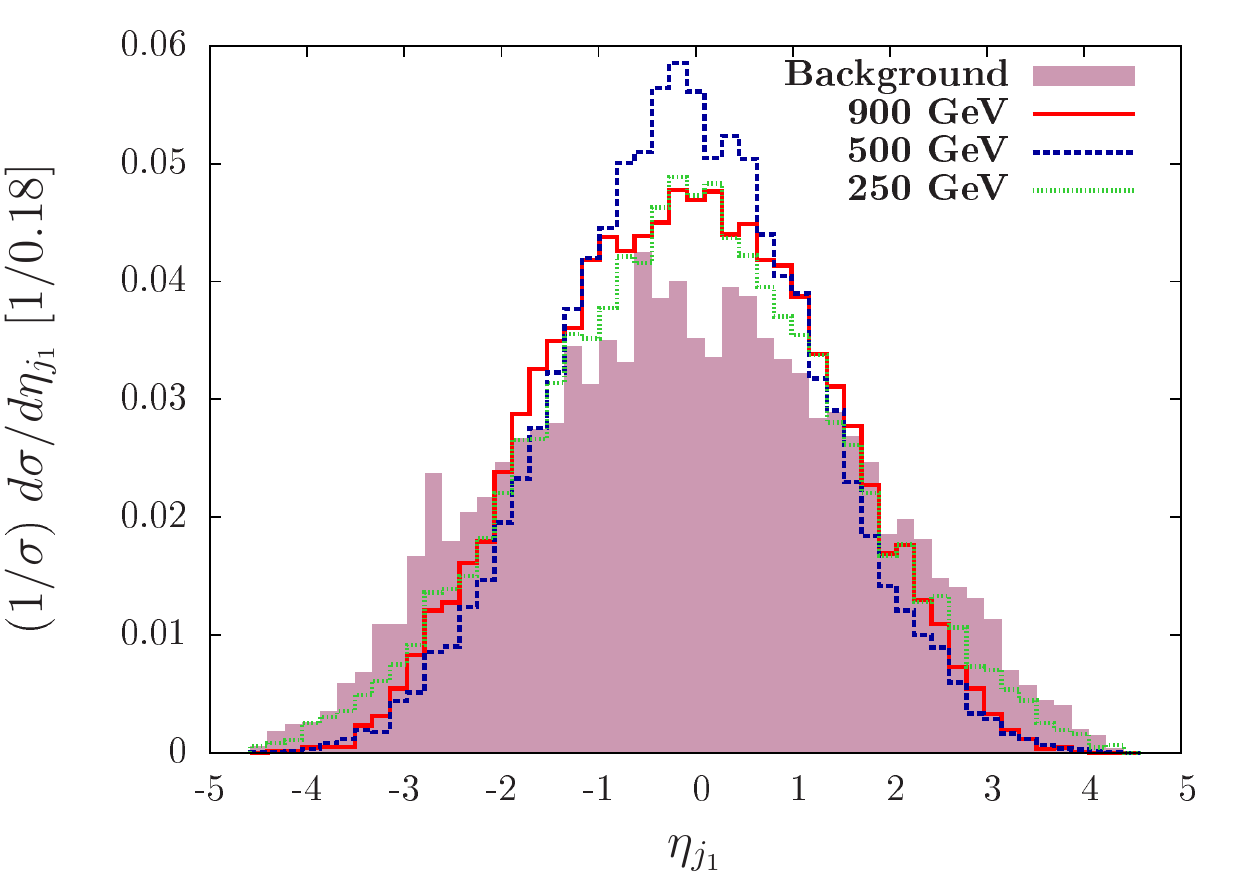}\label{fig:etaj1}}~ 
\subfloat[]{\includegraphics[width=7.5cm,height=6cm]{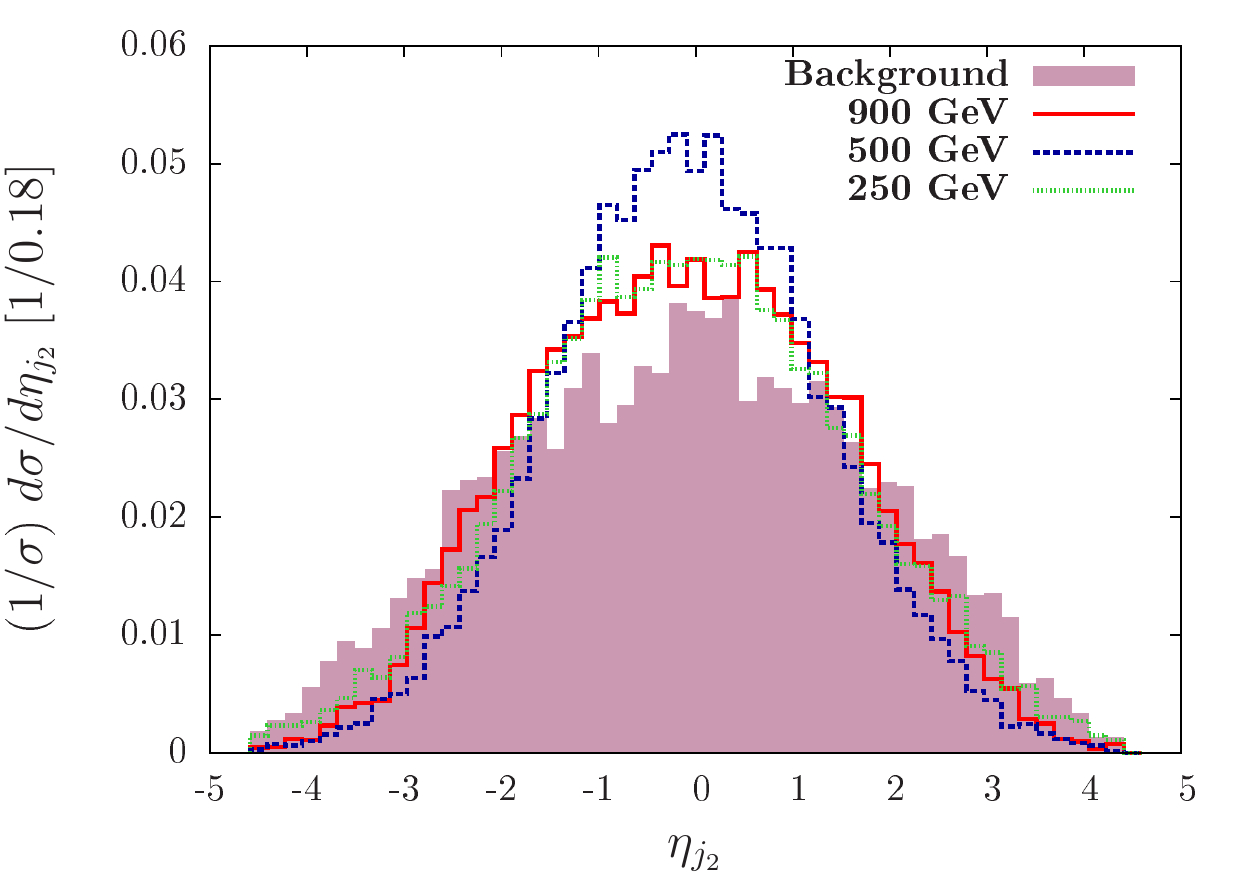}\label{fig:etaj2}}
\caption{$pp\to\ WW \to \ell 2j \slashed{E}_T$ channel normalised distributions for $M_{H_2}$ = 250 GeV (green dotted line), 500 GeV (blue dashed line), 
900 GeV (red solid line) and background (purple solid): (a) $\Delta \phi_{j_1 j_2}$ between the two tagged jets, 
(b) $\slashed{E}_T$, (c) pseudorapidity of the hardest $p_T$ jet, (d) 
pseudorapidity of the second hardest $p_T$ jet and (e) pseudorapidity of the lepton.}
\label{fig:distww3}
\end{figure}

\begin{figure}[!ht]
\centering
\includegraphics[width=7.5cm,height=6cm]{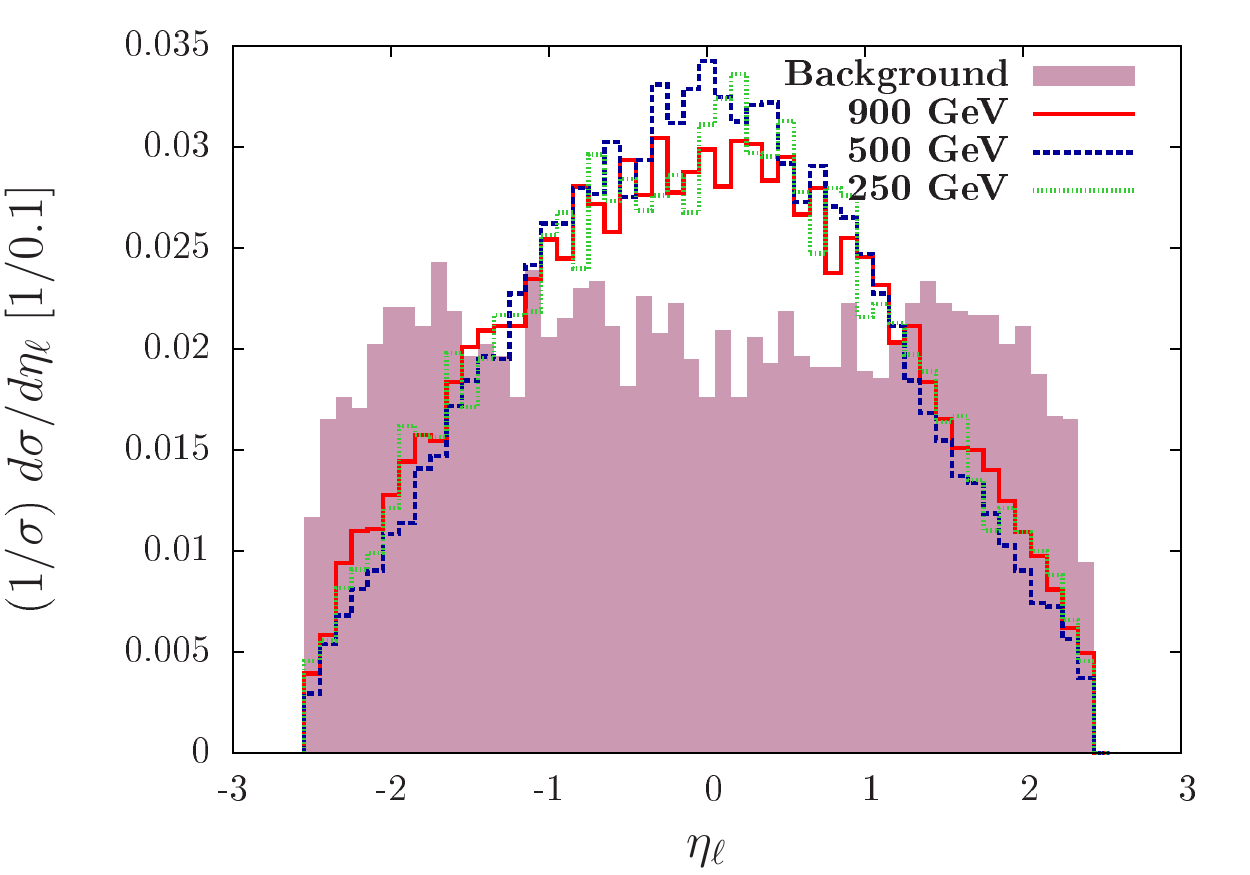}\label{fig:etal}
\caption{$pp\to\ WW \to \ell 2j \slashed{E}_T$ channel normalised distributions for $M_{H_2}$ = 250 GeV (green dotted line), 
500 GeV (blue dashed line), 
900 GeV (red solid line) and background (purple solid): pseudorapidity of the lepton.}
\label{fig:distww4}
\end{figure}

In this scenario, $H_2$ decays to $W^{+} $ and $ W^{-}$, followed by the subsequent decay of one of the $W$s to 
a lepton and missing energy and the other one to a pair of jets. 

Recently, the ATLAS collaboration has searched for $WW/WZ$ resonances decaying to a lepton, neutrino and jets 
\cite{Aad:2015ufa}, that mimic our signal. With a $p_T(W)>400$ GeV cut, used in Ref.~\cite{Aad:2015ufa}, our signal cross-sections will be extremely small. In Ref.~\cite{Aad:2015ufa}, the authors have considered two 
benchmark models, \textit{viz.,} (i) a spin 2 Kaluza-Klein Graviton for the $WW$ resonance and (ii) a spin-1 SSM $W^{\prime}$ decaying 
to a $WZ$ pair. From  Fig.~\ref{fig:figcrossallx}, we find that the production cross-section of 
$pp \to H_2 \to W^{+} W^{-}$ is approximately $\sigma \le 0.098$ pb, that is way below the exclusion limit, as given in Fig.~2 
of Ref.~\cite{Aad:2015ufa}. Hence, we adopt different sets of cuts which are suitable for our analysis.

For a heavy Higgs boson, the intermediate $W$s are expected to have large $p_T$. Because the
$W$ bosons are highly boosted, the leptons and jets are expected to have a large isolation of $\Delta R(l, j_i)$. We show 
the normalised distributions of various kinematic variables in Figs.~\ref{fig:distww1},~\ref{fig:distww2}, ~\ref{fig:distww3}
and ~\ref{fig:distww4}, for the masses $M_{H_2}=250,\, 500, \, 900$ GeV and for the SM background. 

To identify the leptons and jets, we apply the following minimal cuts:
\begin{itemize}
\item \underline{\textbf{Trigger Cuts (TC)}}
\begin{enumerate}
\item {Transverse Momentum of the jets :} $p_T(j_i) > 30$ GeV, where $i$ is the jet-index
\item {Transverse Momentum of the lepton :} $p_T(l) > 20$ GeV 
\item {Pseudo-rapidity of the lepton :} $|\eta(l)| < 2.5$
\item {Pseudo-rapidity of the jets :}  $|\eta(j_i)| < 5.0$
\item {Radial Distance between jets $i$ and $j$:} $\Delta R(j_i,j_j) > 0.4$ 
\item {Radial Distance between lepton and the $i^{th}$ jet:} $\Delta R(l,j_i) > 2.0$ 
\end{enumerate}
\end{itemize}

In the above, $j_1$ and $j_2$ denote two leading jets, sorted according to their transverse momentum. 
For the  Higgs masses, $M_{H_2}$ of our interest, both the intermediate gauge bosons are on-shell and this allows us to fully reconstruct them.
For the hadronically decaying $W$, we use the jet four-momentum, while for the leptonically decaying $W$, we fix the longitudinal component of the neutrino momentum, $p_z(\nu)$ by imposing 
the constraint $M^2_W=(p_l+p_{\nu})^2$. 

The major part of the background originates from non-resonant $W^{+} W^{-}$ production, with one $W$ decaying hadronically and the other 
decaying leptonically. From the different $p_T$ distributions in Figs.~\ref{fig:distww1} and~\ref{fig:distww2}, it is evident that for low masses, the signal and 
background has large overlap, making the discrimination a difficult task. In addition, for the invariant mass, $l jj \slashed{E}_T$, 
the signal and background show a large overlap for $M_{H_2} \approx 250$ GeV (see Fig.~\ref{fig:mh2}), 
whereas for larger masses, the overlap decreases. $p_T$ of the lepton and the leading-jet are large for 
$500 ~ \textrm{GeV} < M_{H_2} < 900 ~ \textrm{GeV}$. We also show the transverse momentum of the two reconstructed $W$s in 
Figs.~\ref{fig:ptlnu} and~\ref{fig:ptjj}. For the signal, they  peak  at  $p_T(W)> 100$ GeV, while for the background, most of the 
intermediate $W$s have lower $p_T$. In addition, the signal has a large missing transverse energy ($\slashed{E}_T$), as evident from 
Fig.~\ref{fig:met}. In Fig.~\ref{fig:imjj}, we show the invariant mass distribution of the two hardest jets and in Fig.~\ref{fig:azij1j2} we 
show the distribution of their azimuthal angle separation.
 
Note that the partonic signal cross-section, $\sigma_{\rm{sig}}$ varies from few tens of fb to $\mathcal{O}(0.1)$ fb, for $M_{H_2}$ 
varying between 300 GeV and 900 GeV. However, the background for this process is extremely large $\sigma_{\rm{bkg}} \sim 3380$ pb. 
Hence, to extract the signal from background in a statistically viable fashion, we categorise the signals into four separate regions 
and implement stringent cuts both at the generation level as well as at the detector level.

\begin{table}[!ht]
\centering
\begin{tabular}{|c|c|c|c|c|}
\hline 
$M_{H_2}$ & $p_T(l/j_1/j_2)$ &  $\Delta R(j_1,j_2)_{\rm{min}}$ & $\Delta R(j_1,j_2)_{\rm{max}}$  & $\slashed{E}_T $ \\
(GeV) & (GeV) & & & (GeV) \\
\hline 
350    &   30   &  0.4 & 1.4 &  50 \\
500 &  40   &  0.2 & 1.0 &   60 \\
700 & 50  &  0.2 & 0.8  &  70  \\
900 & 70 & 0.2 & 0.6 &   90 \\
\hline \hline 
\end{tabular} 
\caption{Basic trigger cuts used to separate the signal from background in the 
$pp \to H_2 \to W W \to \ell + \slashed{E}_T + \ge 2j$ channel.}
\label{tab:mg5cuts}
\end{table}

\begin{table}[!ht]
\begin{tabular}{|c|c|c|c|c|c|c|c|}
\hline 
$M_{H_2}$  & $p_{T,l/j_{1,2}}$ & $p_{T,W_{1,2}}$  &  $\Delta R_{j_1,j_2}^{\rm{max}}$  & $\slashed{E}_T $  & $S_T$  & $|M_{ljj\slashed{E}_T}-M_{H_2}|$  &
$|M_{jj}-M_{W}|$  \\
(GeV) & (GeV) & (GeV) & & (GeV) & (GeV) & (GeV) & (GeV) \\
\hline 
350    &   35   &  100 &  1.35 &  55  & 225 & 50 & 20\\
500 &  45   & 100 &  0.9 &   70  & 250 & 50 & 20 \\
700 & 55  & 100 &  0.75 &  75 & 250 & 50 & 20  \\
900 & 75 & 100& 0.58 &  95 & 600 & 50 & 20 \\
\hline \hline 
\end{tabular} 
\caption{Selection cuts to separate out signal from the background in the $pp \to H_2 \to W W \to \ell + \slashed{E}_T + \ge 2j$ channel. }
\label{tab:seleccuts}
\end{table}
\begin{figure}[!ht]
\centering
\subfloat[]{\includegraphics[width=7.5cm]{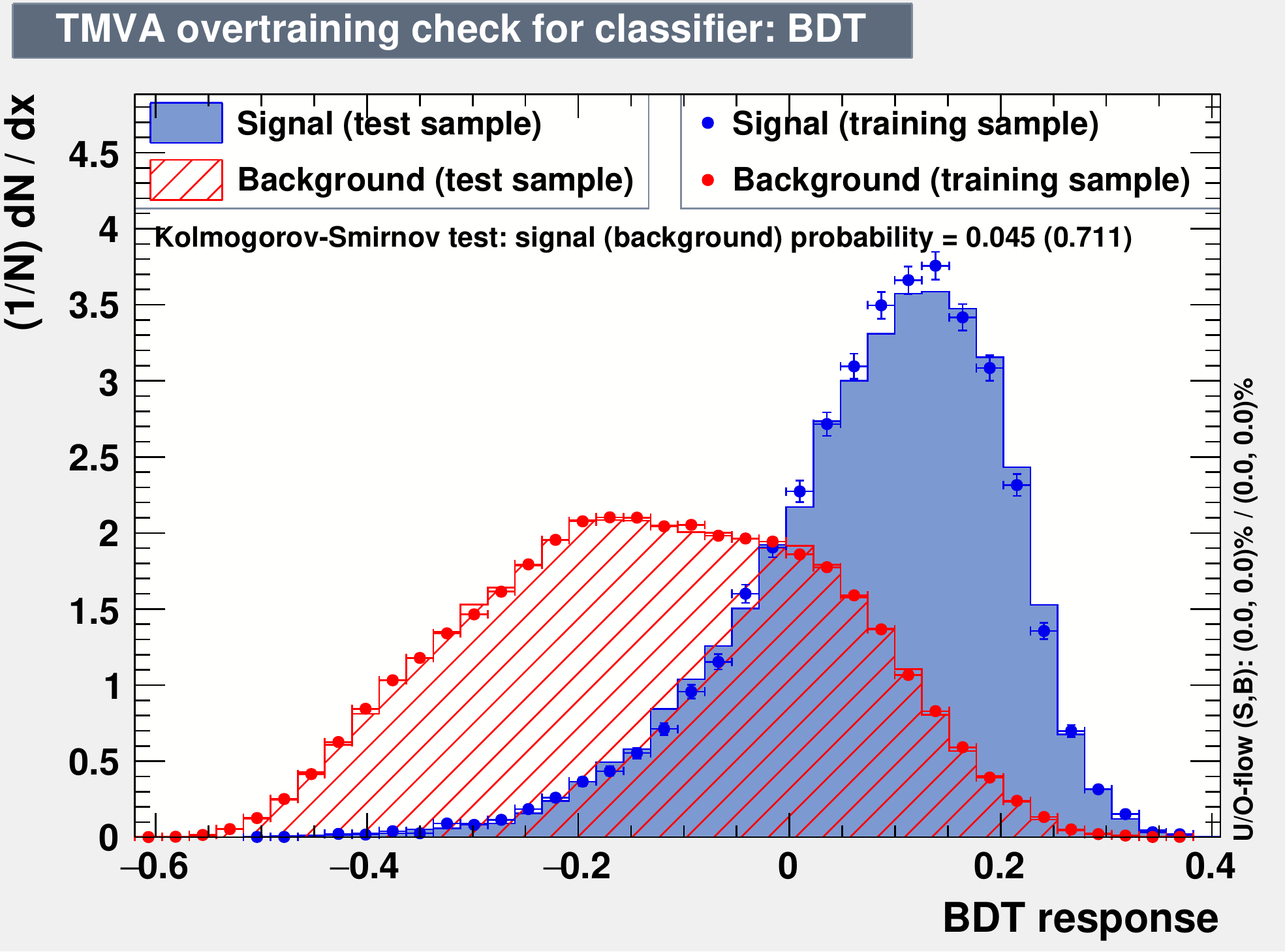}\label{fig:wwbdt350}}~~
\subfloat[]{\includegraphics[width=7.5cm]{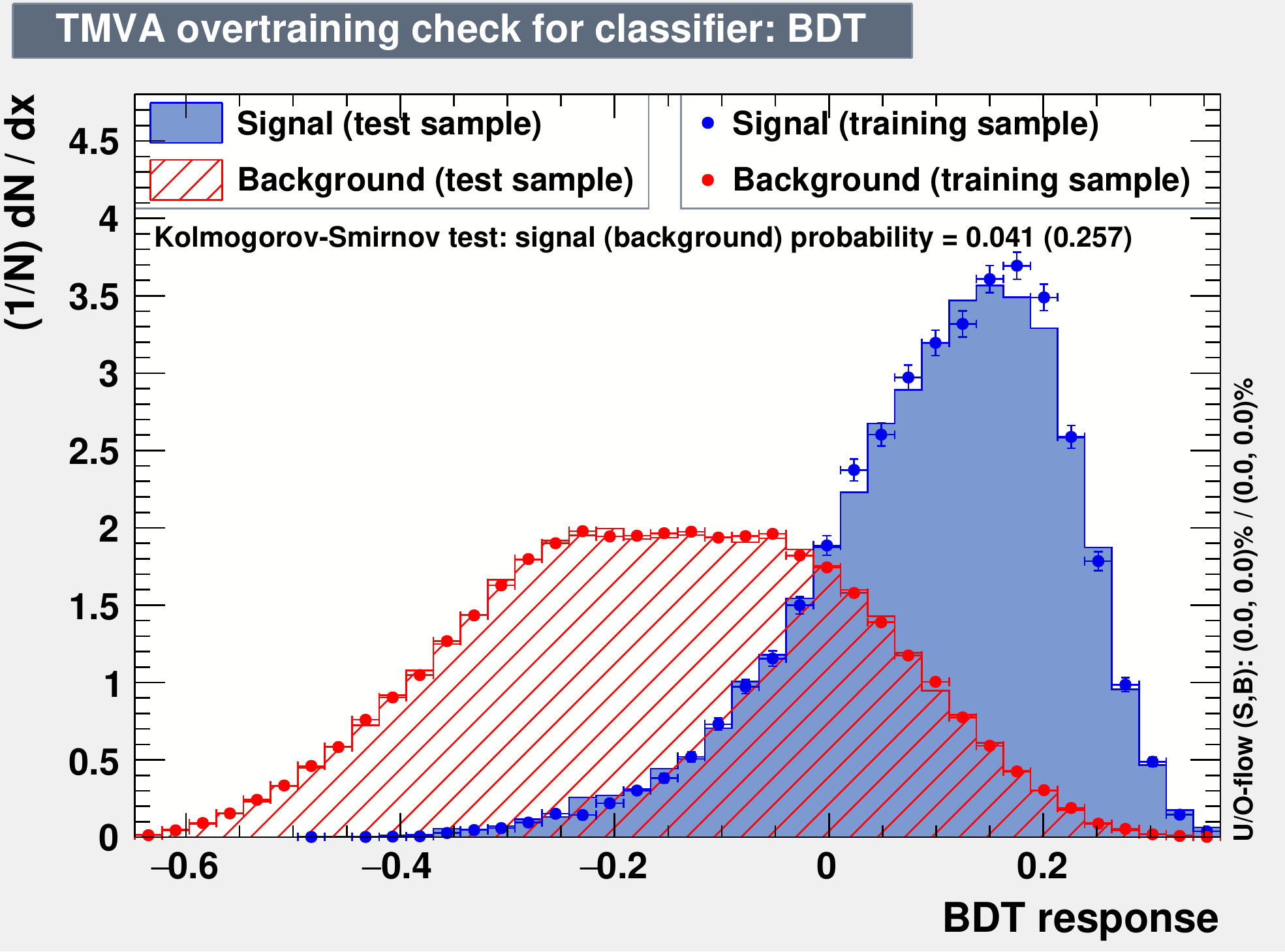}\label{fig:wwbdt500}}~~\\
\caption{Normalised signal and background distributions against BDT response for (a) $M_{H_2}=350$ GeV and (b) $M_{H_2}=500$ for the channel $pp\to H_2 \to WW \to \ell \nu j j$.}
\label{fig:wwbdt}
\end{figure}
\begin{itemize}
\item {{The low mass region: ${M_{H_2} \sim 350}$ GeV case:}} For this case the signal and background has a large 
overlapping region. The transverse momentum variable for the lepton and jets peak around 30 GeV for the background. Hence, to reduce 
the background, we implement selection cuts, such as, $p_T> 35$ GeV, missing transverse energy $\slashed{E}_T > 55$ GeV and 
the scalar sum of transverse momentum of the $l \nu jj$ system, $S_T > 225$ GeV. In addition, the selection cuts on the invariant mass 
of the $l \nu jj$ system, $|M_{l \nu jj}-M_{H_2}|=50$ GeV reduces the background significantly.

\item {{Intermediate mass range: $M_{H_2}{\sim 500}$ GeV: }}
For the intermediate mass range, such as, $M_{H_2}=500$ GeV, the signal and background has a relatively smaller overlap, as is 
evident from Figs.~\ref{fig:ptl},~\ref{fig:ptjt1} and~\ref{fig:ptjt2}. In addition, the reconstructed  $W$s have large 
transverse momenta, as shown in Figs~\ref{fig:ptlnu} and~\ref{fig:ptjj}. The invariant mass of the $l \nu jj$ system peaks 
around 500 GeV, while for background the invariant mass peaks at much lower values. The larger cuts on the transverse momenta 
of the $W$s and the cuts on the invariant mass efficiently reduces the background cross-section from $\sim 3380$ pb to $\sim 6.69$ pb.

\item {{Large masses: $M_{H_2}{\sim 700}$ GeV: }}
For large masses, such as, $M_{H_2}=700$ GeV, the signal cross-section after the selection cuts is relatively small $\sigma \sim 0.84$ fb. However, the 
background is well separated. Hence, the stringent cuts on the kinematic variables improves the sensitivity. We use larger cuts on 
the transverse momenta of the lepton, jets and the reconstructed $W$s, as shown in Tables~\ref{tab:mg5cuts} and~\ref{tab:seleccuts}. 
The background cross-section after the different trigger and selection cuts reduce to $\sim 2.56$ pb. 

\item {{Very large masses: $M_{H_2}{\sim 900}$ GeV: }}
For $M_{H_2}$ as large as 900 GeV, the signal cross-section after imposing the selection cuts becomes extremely low, 
$\sigma \sim 0.06$ fb. Here, we thus use higher trigger and selection cuts as is given in Table.~\ref{tab:mg5cuts} and 
Table.~\ref{tab:seleccuts}, in order to reduce the background. 
\end{itemize}
\begin{table}
\begin{tabular}{ccccccccc}
\hline
$M_{H_2}$  & $\sigma_s$ & $\sigma_{\rm{bkg}}$ & $\sigma^{SC}_s$ & $\sigma^{SC}_{\rm{bkg}}$ & $\mathcal{N}^{CBA}_{S_{100}}$ & $\mathcal{N}^{CBA}_{B_{100}}$  &  $\mathcal{N}^{CBA}_{S_{3000}}$ & $\mathcal{N}^{CBA}_{B_{3000}}$\\
(GeV) & (fb) & (pb) & (fb) & (fb) & & & & \\
\hline
350 & 16.85   & 24.22  & 5.49  & 1666.29  & 549 & 166629 & 16474 & 4998880 \\
500 & 9.06 & 6.69 & 3.44 & 360.26   & 344 & 36026 & 10305 & 1080769 \\ 
700 & 2.39  & 2.56  & 0.84  & 77.63  & 84 & 7763 & 2509 & 232887 \\
900 & 0.57  & 0.65 & 0.06  & 6.19  & 6 & 619 & 193 & 18564 \\
\hline 
\end{tabular}
\caption{The NNLO signal ($\sigma_s$) and background ($\sigma_{\rm{bkg}}$) cross-sections and the number of events for different Higgs masses. The $\mathcal{N}^{CBA}_{S_{100}/B_{100}}$,  
$\mathcal{N}^{CBA}_{S_{3000}/B_{3000}}$ are the signal and background events with 100 and 3000\,$\rm{fb}^{-1}$ integrated luminosities 
for the cut based analysis.}
\label{tab:tabevent}
 \end{table}

\begin{table}
\centering
\begin{tabular}{ccccc}
\hline
$M_{H_2}$ &  $\mathcal{N}^{BDT}_{S_{100}}$ & $\mathcal{N}^{BDT}_{B_{100}}$  &  $\mathcal{N}^{BDT}_{S_{3000}}$ & $\mathcal{N}^{BDT}_{B_{3000}}$\\
(GeV) & & & & \\
\hline
 350 & 591 & 116793 & 17731 & 3503792 \\
500 & 309 & 18761 & 9270 & 562838 \\
700 &  68 & 3718 & 2055 & 111544  \\
900 & 12& 1358 & 366 & 40735 \\
\hline 
\end{tabular}
\caption{The signal  and background  events after BDT cut for different Higgs masses. The $\mathcal{N}^{BDT}_{S_{100}/B_{100}}$,  
$\mathcal{N}^{BDT}_{S_{3000}/B_{3000}}$ are the signal and background events with 100 and 3000\,$\rm{fb}^{-1}$ integrated luminosities.}
\label{tab:bdtww}
 \end{table}
 
In order to perform the multivariate analysis, we choose a set of 27 kinematic variables with excellent discriminating power, which are 
$M_{\ell j j \nu}$, $p_T(\ell)$,  $\eta(\ell)$, $p_T(j_i)$, $\eta(j_i)$, $\slashed{E}_T$,  
$\phi(\slashed{E}_T)$, $p_T(\ell,\slashed{E}_T)$, $p_T(j_1,j_2)$, $|\Delta \phi (W_1, W_2)|$, $|\Delta \phi (\ell, j_1)|$, $\Delta \eta (\ell ,j_2)$, $\Delta \eta (\ell, j_i)$, $|\Delta \phi (j_1, j_2)|$, $\Delta \eta (j_1, j_2)$, $|\Delta \phi (j_i, \slashed{E}_T)|$, $S_T$, $M_{j_i \ell}$,  $M_{j_1 j_2 l}$, $\Delta R(\ell, j_i)$ and $\Delta R(j_1 j_2)$.
In the above, $i=1,2$ and the jets and the reconstructed $W$s are $p_T$ sorted.

The results for the cut-based and the multivariate analyses are shown in Tables~\ref{tab:tabevent} and~\ref{tab:bdtww}. As 
expected, we find the MVA to be performing better with respect to the cut-based analysis. 
Fig.~\ref{fig:wwbdt} shows the normalised distributions of the signal and background as a function of the BDT response 
for two benchmark masses, i.e. $M_{H_2}=350$ GeV and 500 GeV. We see that for both cases, there is a large region of overlap 
between the signal and background. This implies that even with a BDT, the separation of signal and background is challenging, resulting in a very small $S/B \lesssim 1/100$.
Assuming zero systematic uncertainties we quote the statistical significance in 
Table~\ref{tab:sensitivity}. The discovery potential of $H_2$ in this channel is however bleak.
\begin{table}[!ht]
\centering
\begin{tabular}{|c|c|c|c|}
\hline 
$M_{H_2}$ GeV   & $\mathcal{L}$~$[\mathrm{fb}^{-1}]$ & $n_{CBA}$  &  $n_{BDT}$ \\
\hline 
\multirow{2}{1cm}{350}  & 100 & 1.34 & 1.73 \\ \cline{2-4} 
&   3000 & 7.36 & 9.45 \\
\hline
\multirow{2}{1cm}{500} &  100 & 1.80 & 2.24 \\ \cline{2-4} 
&  3000 & 9.86 & 12.26 \\
\hline
\multirow{2}{1cm}{700} & 100 & 0.94 & 1.11 \\  \cline{2-4} 
&  3000 & 5.17 & 6.10 \\
\hline
\multirow{2}{1cm}{900}  & 100  & 0.26 & 0.33 \\ \cline{2-4} 
&  3000 & 1.41  & 1.81 \\
\hline \hline 
\end{tabular} 
\caption{The significance  for cut-based and multivariate analysis for integrated luminosity $100\, \rm{fb}^{-1}$ and $3000\, \rm{fb}^{-1}$. }
\label{tab:sensitivity}
\end{table}

\begin{figure}
\centering
\subfloat[]{\includegraphics[width=7.0cm,height=7.0cm]{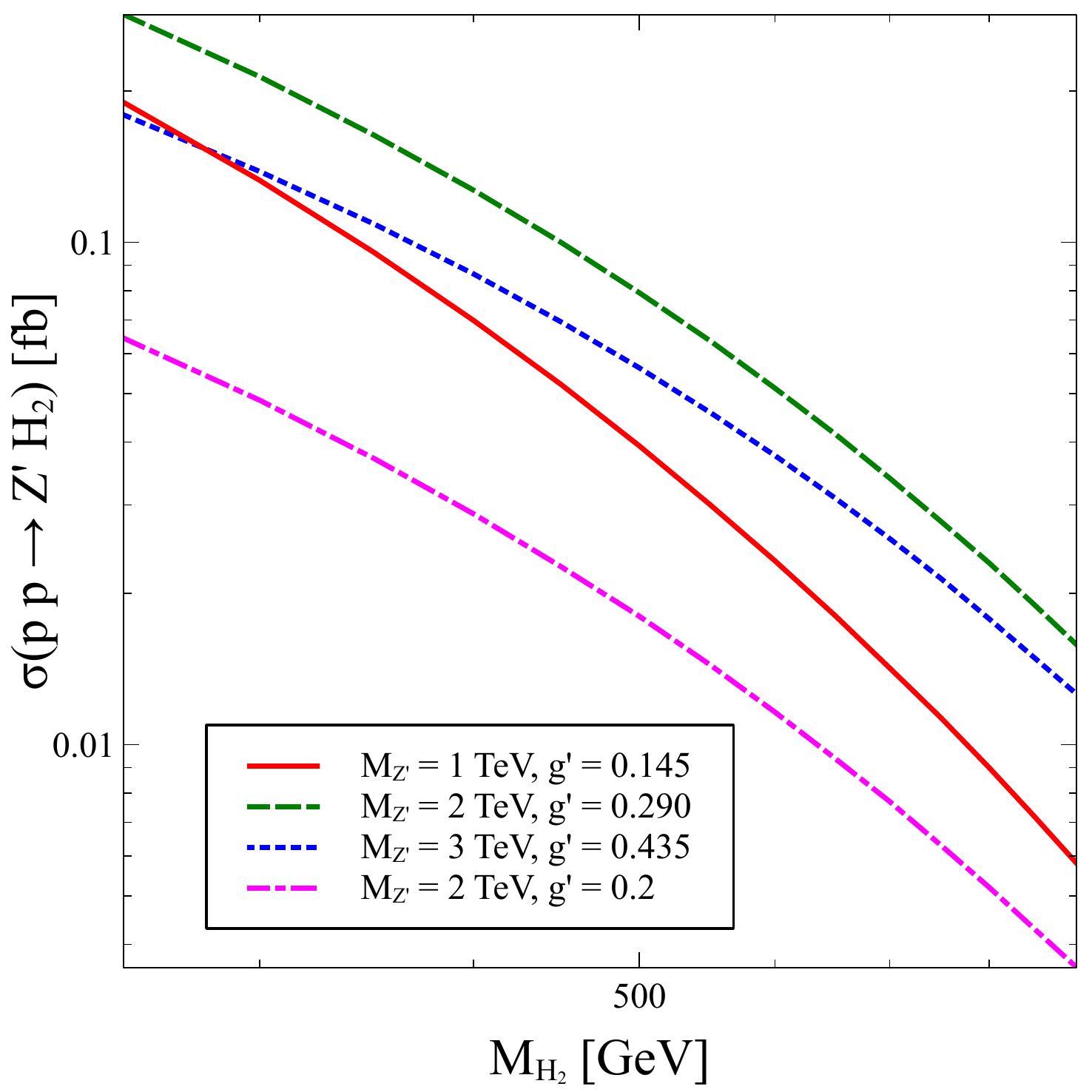}\label{fig:zph2a}}~~
\subfloat[]{\includegraphics[width=7.0cm,height=7.0cm]{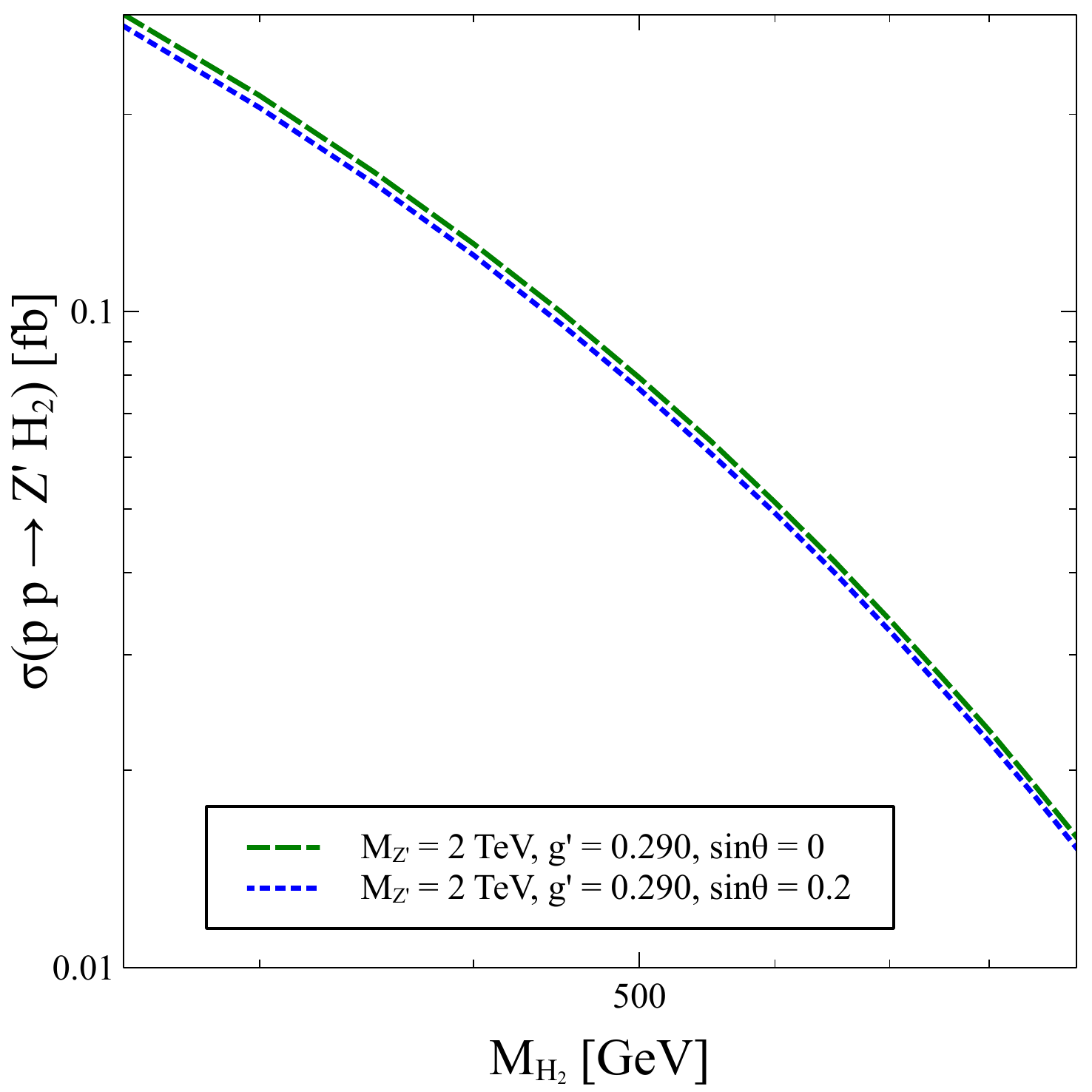}\label{fig:zph2b}}
\caption{Left panel: The LO cross-section for the associated production  $p p \to Z^{\prime} H_2$ for mixing $\theta=0$ and different 
values of $M_{Z'}$ and $g^{\prime}$ (solid red, dashed green, dotted blue and dot-dashed magenta for [$M_{Z'}=1$ TeV, 
$g^{\prime}=0.145$], [$M_{Z'}=2$ TeV, $g^{\prime}=0.290$], [$M_{Z'}=3$ TeV, $g^{\prime}=0.435$] and [$M_{Z'}=2$ TeV, $g^{\prime}=0.2$] 
respectively) such that $\frac{M_{Z'}}{g^{\prime}} \ge 6.9$ TeV (Eq.~\ref{indirect}). Right panel: Comparison between the associated 
production $p p \to Z^{\prime} H_2$ between $\sin \theta=0$ and $\sin \theta=0.2$ (dashed green and dotted blue are for [$M_{Z'}=2$ 
TeV, $g^{\prime}=0.290$, $\sin{\theta}=0$] and [$M_{Z'}=2$ TeV, $g^{\prime}=0.290$, $\sin{\theta}=0.2$] respectively).}
\label{associated}
\end{figure}

\section{Non-standard heavy Higgs production \label{non-standard-prod}}

In addition to the standard production processes, i.e. gluon fusion, $VBF$, $V H_2$ ($V=W,Z$), $t \bar{t} H_2$, 
the heavy Higgs $H_2$ can also be produced  in association with a $Z^{\prime}$ \cite{Pruna:2011me, Basso:2008iv} {\footnote{For a 
complete list of   production processes, see \cite{Emam:2007dy, Basso:2010yz,Pruna:2011me}.}}. In the decoupling limit, the mixing  
$\theta \sim  0$ and hence the gluon fusion contribution would be negligible. However,  the vertex factor $Z^{\prime}Z^{\prime}H_2$ is 
proportional to $\cos \theta$ (see Eq.~\ref{int}). Therefore, even in the decoupling limit, the process $p p \to Z^{\prime} H_2$  will give a 
non-zero contribution.  We show the  production cross-section of this process  in Fig.~\ref{fig:zph2a} for the decoupling limit. Note 
that, for a non-zero value of $\theta$, both   the  $s$-channel ( $ p p \to {Z^{\prime}}^* \to H_2 Z^{\prime}$ ) {\footnote{Note that, 
$Z' H_2 \gamma$  contribution is negligible because the vertex is loop suppressed and also because in the minimal $B-L$ scenario, the 
$Z'W^{+}W^{-}$ coupling is absent. Hence, we neglect this contribution in our study.}} and $t$-channel diagrams mediated by quarks will 
contribute (see Fig.~\ref{fig:zph2b}). However, the $t$-channel contribution will be small.

\section{Decay to heavy neutrinos \label{decay-neutrino}}

\begin{figure}
\centering
\subfloat[]{\includegraphics[width=7.0cm,height=7.0cm]{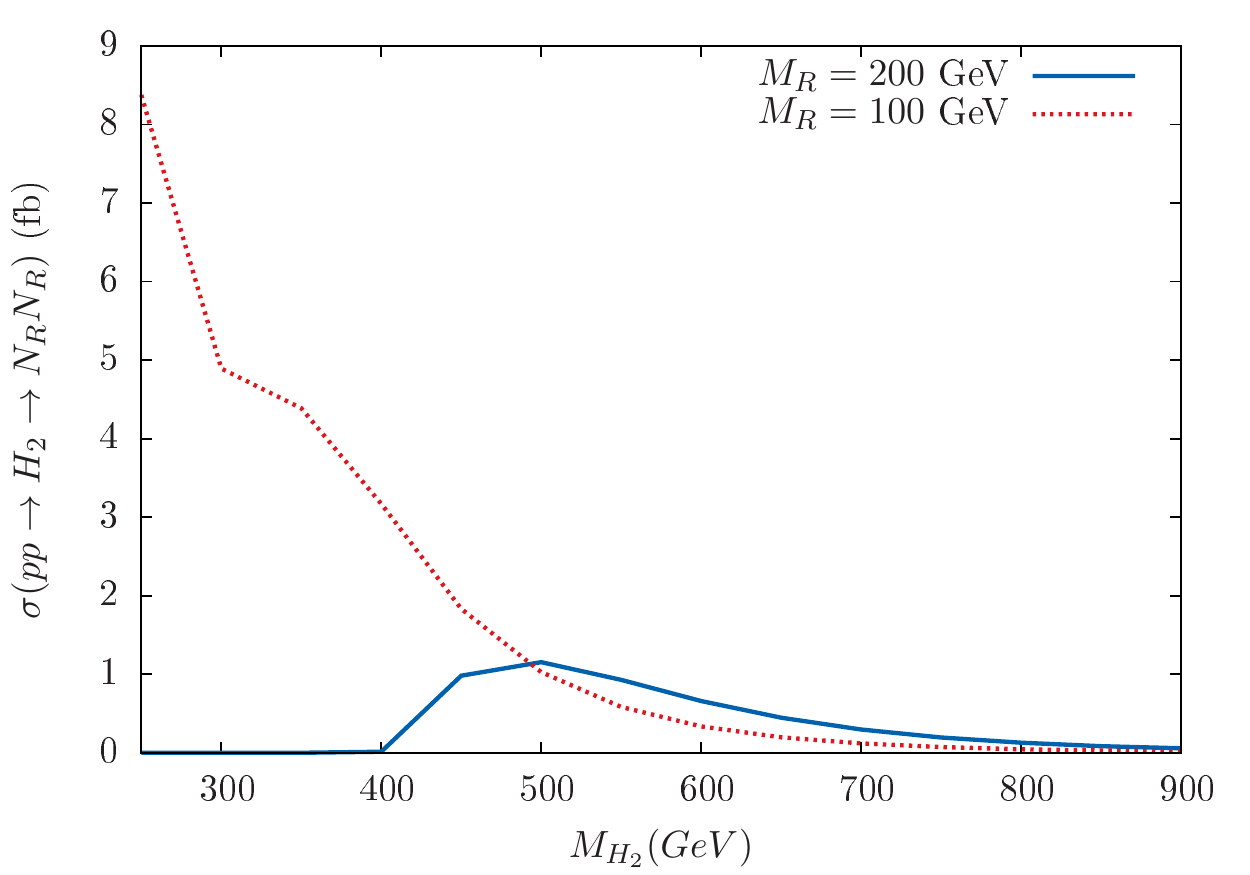}\label{figh2nrnr}}~~
\subfloat[]{\includegraphics[width=7.0cm,height=7.0cm]{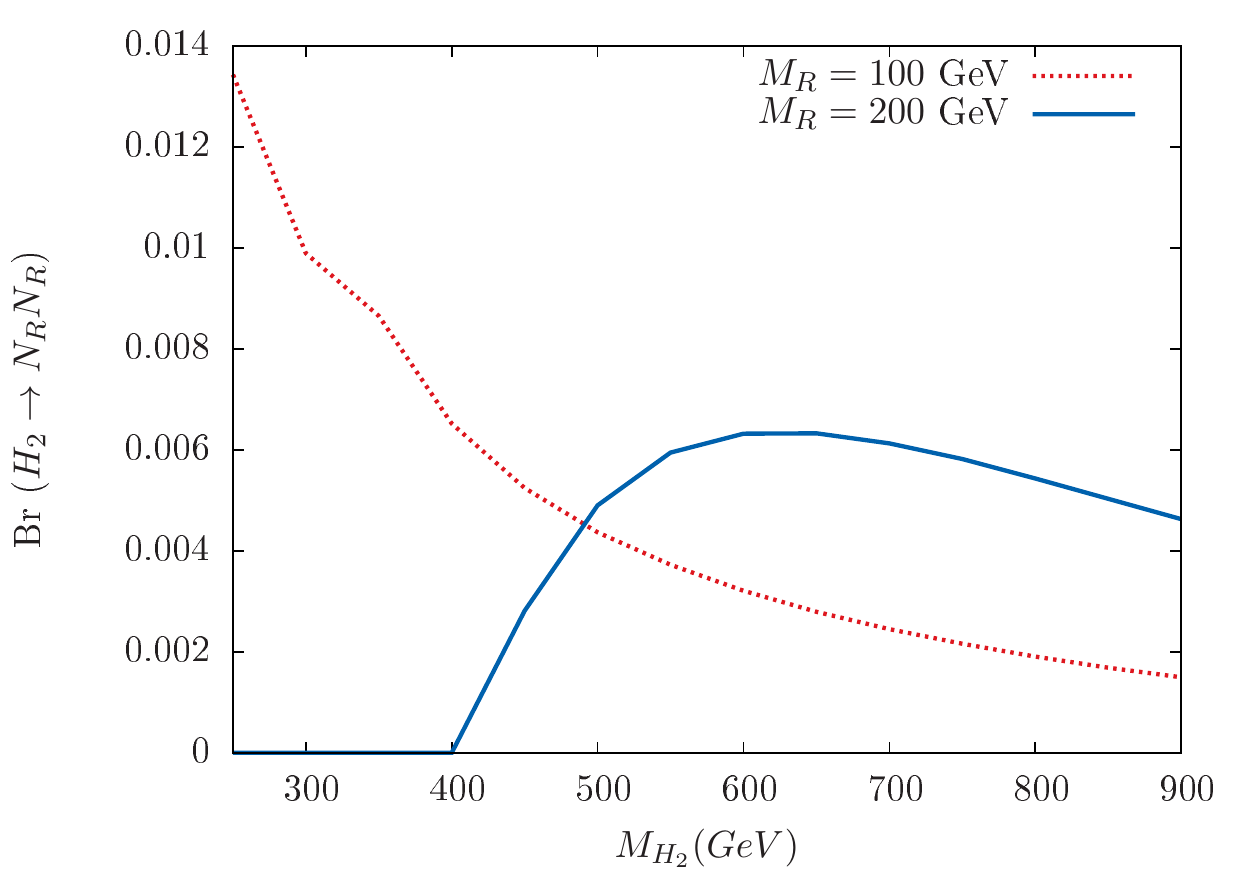}\label{h2nrnrbr}}
\caption{Left panel: The NNLO production cross-section of $p p \to H_2 \to N_R N_R$. Right panel: The branching ratios of 
$H_2 \to N_R N_R$ (solid blue : $M_R=200$ GeV and dotted red : $M_R=100$ GeV).}
\label{fig:nrnr}
\end{figure}

Here  we  briefly  discuss
the decay of $H_2$ to two heavy neutrino states, {\it i.e.}, $p p \to H_2 \to N_R N_R$. The unique feature of the $B-L$ model is that both 
 $Z^{\prime}$ and $H_2$ can decay to a pair of  heavy neutrinos. These produced heavy neutrinos can  further decay to 
$l W$, $\nu Z$ and $H_1 \nu$ states, producing  same as well as opposite sign leptonic signatures. The same sign leptonic signature 
confirms the Majorana nature of the heavy neutrinos, as well as, the presence of $B-L$ breaking {\footnote{Same sign dileptonic 
signature can also arise from a Higgs triplet Type-II seesaw scenario \cite{Perez:2008ha}}}. In Ref.~\cite{Perez:2009mu} the authors 
have studied the same sign dilepton+4-jets signature from $p p \to Z^{\prime} \to N_R N_R$ channel. In Ref.~\cite{Huitu:2008gf}, 
multilepton final states have been studied. In Fig.~\ref{figh2nrnr} we show the NNLO cross-section of $p p \to H_2 \to N_R N_R$, which 
is only few fb for $M_N =100 $ GeV.  In addition, we also show the branching ratio of $H_2\to N_R N_R$  in Fig.~\ref{h2nrnrbr}. Note that, 
this branching ratio is small. Hence, even for lower $N_R$ masses, e.g. 100 or 200 GeV, the previous analysis of 
$p p \to H_2 \to W^{+} W^{-}$ and $p p \to H_2 \to Z Z$ will remain practically unchanged. A detailed study for 
$p p \to h_2 \to N_R N_R$ has been presented in Refs.~\cite{Basso:2010yz, Pruna:2011me}. This channel also offers other final states, e.g.  $l^{\pm} l^{\pm}+ 4j$, $l^{\pm} l^{\mp} l^{\pm} + 2j+ \slashed{E}_T$, $4l+ \slashed{E}_T$,  
$4b+\slashed{E}_T$ and $l+\slashed{E}_T+b\bar{b}+2j$.

\section{Summary and conclusions \label{conclu}}
The gauged $B-L$ model is well-motivated and phenomenologically  rich. Three heavy neutrinos, required for anomaly cancellation, 
participate in the $seesaw$ mechanism and generate viable light neutrino masses and mixings. In addition to the heavy neutrinos, 
this model has an extra $Z'$ gauge boson and an additional SM-singlet Higgs state. In this work, we studied in detail the discovery 
prospect of the gauged $B-L$ model through the Higgs window. We considered the $B-L$ breaking scale to be few TeV, for which the $Z'$ and 
the heavy neutrinos are naturally of TeV scale, and we considered heavy Higgs masses starting from 250 GeV up to 900 GeV. 

The SM-singlet Higgs state mixes with the SM Higgs with a mixing angle $\theta$, that is constrained by direct and indirect searches, 
such as, vacuum stability, electroweak precision data and $M_W$ mass measurement. The precise determination of the SM Higgs masses and 
the compatibility of its couplings with the SM prediction bounds the mixing angle to be $\sin^2 \theta \lesssim 0.3$. However, the most 
stringent limit  on the mixing angle arises from the $M_W$ mass measurement. In section~\ref{mix}, we reviewed the different constraints and 
for our subsequent analysis we consider the mixing angle $\sin \theta =0.2$, compatible with $M_W$ mass measurement. 

During future LHC runs, Higgs coupling measurements can provide strong indirect constraints on the mixing angle between the SM-like Higgs boson and the heavy neutral Higgs, independent of the mass of the heavy state. The predicted bound on $\sin{\theta}$ from the 14 TeV run of LHC with $\int{\mathcal{L} dt} = 300$ fb$^{-1}$ is $\sin{\theta} \sim 0.36$ from the $H\to WW^*$ coupling measurement. However, if $m_{H_2} \lesssim 500$ GeV direct searches will be more sensitive. 

The heavy Higgs of this model can be produced via gluon fusion, VBF, associated Higgs production, out of which the gluon fusion offers 
the highest cross-section. We considered $pp \to H_2$ folded with $H_2$ decaying into different SM states. For our parameters of 
interest,
$H_2$ does not decay to any additional invisibles state. The produced $H_2$ decays predominantly to $W^{+} W^{-}$, $H_1 H_1$ and $ZZ$ 
final states. The decays of the gauge bosons lead to different final states, such as $4\ell$ and $ljj \slashed{E}_T$. We studied
the discovery prospect of a heavy Higgs $H_2$ in the $4\ell$, $2\ell 2j$ and the $\ell jj \slashed{E}_T$ channels at the LHC (with 
$\int \mathcal{L} dt=100$ fb$^{-1}$) and HL-LHC ($\int \mathcal{L} dt=3000$ fb$^{-1}$), where we employed a boosted decision tree to separate signal from background.

The channel with four leptons was found to be the cleanest. The signal and background cross-sections for these processes are 
$\sigma_S \simeq 0.1$ fb and $\sigma_B \simeq 42$ pb, respectively. Using the  cuts on the i) invariant mass of $4\ell$ and on the reconstructed 
$Z$ bosons, ii) the $p_T$ cuts on the momenta of four leptons, as well as, the reconstructed $Z$ bosons, we found that for a mass 
$M_{H_2} \le 500 $ GeV, the $H_2$ can be discovered with a significance of $\sim 5 \sigma$ at HL-LHC with $3000 \, \rm{fb}^{-1}$. 
 
The $ZZ\to 2 \ell 2j$ final state has a larger cross section than the $4 \ell$ final state. Particularly for heavy $H_2$ masses an increased cross section is important to extend the LHC's reach. However, we find that for small mixing angles ($\sin \theta = 0.2$) this channel has small $S/B$ and sensitivity.
 
$pp \to H_2 \to W^{+}W^{-} \to \ell jj \slashed{E}_T$ offers an even larger cross-section 
for the signal, i.e.
$\sigma \simeq \mathcal{O}(10)\, \rm{fb}$. Unfortunately, huge backgrounds $\sigma_B \simeq \mathcal{O}(10^3)$ pb make this channel 
extremely difficult to deal with. We explored the different mass regions of the heavy Higgs $M_{H_2}$ and applied exclusive trigger and 
selection cuts. We found that in the most optimistic scenario, assuming zero systematic uncertainties, a heavy Higgs search for 
$350 \leq M_{H_2} \leq 700$ GeV has a statistical significance of $2.24 \sigma$ with $100 \, \rm{fb}^{-1}$ and up to $5 \sigma$ with 
$3000 \, \rm{fb}^{-1}$. Since entirely negligible systematic uncertainties are not realistic in this fairly complex final state with 
jets and missing-transverse energy, even using data-driven methods only,  the $4l$ channel proves to be superior in discovering a heavy 
Higgs boson in the $B-L$ model.

The heavy Higgs $H_2$ decays to $H_1 H_1 $ with a  branching ratio $\sim 20\%$ for $M_{H_2}\gtrsim 400$ GeV.
Due to the small production cross section for $H_2 \to H_1 H_1 \to b\bar{b}b\bar{b}$ or $H_2 \to H_1 H_1 \to b\bar{b}\gamma\gamma$ 
searching for $H_2$ in these final states is very challenging. However, we showed that for $M_{H_2}$ up to $500$ GeV, the production
cross-section for $p p \to H_1 H_1$ is significantly enhanced with respect to the SM expectation. Hence, it might be interesting to 
look for this channel in this region of the parameter space in more details.

In this analysis, we also briefly considered the $Z'$ searches. We used the ATLAS search at 8 TeV for $Z'$ in the dileptonic channel to recast $Z'$ mass constraints for various values of the $U(1)_{B-L}$ coupling $g'$ and the heavy neutrino masses, $M_N$.

\section*{Acknowledgements}
S.B. and M.M. would like to thank Satyaki Bhattacharya, Shilpi Jain, Arghya Choudhury, Tanumoy Mandal and Saurabh Niyogi for some helpful discussions regarding
{\tt TMVA} and {\tt FastJet}. The work of S.B. was partially supported by funding available from the Department of Atomic Energy, 
Government of India for the Regional Centre for Accelerator-based Particle Physics (RECAPP), Harish-Chandra Research Institute. 
S.B. and M.M would also like to acknowledge the hospitality of IPPP, Durham, where the work began. The work of M.M was partially supported by
DST-Inspire project grant. This research was supported in part by the European Commission through the 'HiggsTools' Initial Training Network PITN-GA-2012-316704.

\end{document}